\documentclass[12pt,a4paper]{article}
\pdfoutput=1
\usepackage{jheppub}
\usepackage{dsfont}
\usepackage{tikz}
\newcommand{\cA}{\ensuremath \mathcal A}
\newcommand{\cB}{\ensuremath \mathcal B}
\newcommand{\cC}{\ensuremath \mathcal C}
\newcommand{\cF}{\ensuremath \mathcal F}
\newcommand{\cM}{\ensuremath \mathcal M}
\newcommand{\cG}{\ensuremath \mathcal G}
\newcommand{\cW}{\ensuremath \mathcal W}
\def\cZ{{\cal Z}}

\def\mf{{\mathfrak F}}

\DeclareMathOperator{\vol}{vol}
\DeclareMathOperator{\Vol}{Vol}

\def\RR{\ensuremath{\mathbb R}}
\def\ZZ{{\mathds{Z}}}

\def\Im{\mathop{\rm Im}}
\def\Re{\mathop{\rm Re}}

\usepackage{subfigure}

\numberwithin{equation}{section}

\newcommand{\be}{\begin{equation}} \newcommand{\ee}{\end{equation}}
\newcommand{\bea}{\begin{equation} \begin{aligned}} \newcommand{\eea}{\end{aligned} \end{equation}}

\def\tilde{\widetilde}

\def\hat{\widehat}

\def\bar{\overline}
\def\a{\alpha}

\def\rt2{\sqrt{2}}

\def\Re{\mathop{\rm Re}}
\def\Im{\mathop{\rm Im}}

\def\det{\mathop{\rm det}}

\DeclareMathOperator{\Li}{Li}

\def\1{{\ds 1}}

\def\O{\mathrm{O}}

\def\repa{\raise4pt\hbox{$\square$}\mkern-14mu\raise-4pt\hbox{$\square$}}
\def\repab{\overline{\raise4pt\hbox{$\square$}\mkern-14mu\raise-4pt\hbox{$\square$}\mkern-1mu}}

\def\smileface{\ensuremath{\hbox{\large$\bigcirc$}\mkern-15mu\raise-1pt\hbox{\scriptsize$\smallsmile$}%
\mkern-10mu\raise4pt\hbox{..}\mkern4mu}}
\def\frownface{\ensuremath{\hbox{\large$\bigcirc$}\mkern-15mu\raise-1pt\hbox{\scriptsize$\smallfrown$}%
\mkern-10mu\raise4pt\hbox{..}\mkern4mu}}

\newcommand{\ba}{\begin{array}}
\newcommand{\ea}{\end{array}}
\newcommand{\bi}{\begin{itemize}}
\newcommand{\ei}{\end{itemize}}

\def\bea#1\eea{\allowdisplaybreaks \begin{align}#1\end{align}}
 \newcommand{\ben}{\begin{enumerate}}
\newcommand{\een}{\end{enumerate}}
\newcommand{\bean}{\begin{eqnarray*}}
\newcommand{\eean}{\end{eqnarray*}}

\newcommand{\comment}[1]{}

\def\aup#1 {\overset{#1}{\uparrow} \, \overset{\tilde{#1}}{\downarrow}}

\newcommand{\AdS}{\mathrm{AdS}}
\newcommand{\Mink}{\mathrm{Mink}}
\newcommand{\drm}{\mathrm{d}}
\newcommand{\Srm}{\mathrm{S}}

\title{Non-supersymmetric AdS$_6$ and the swampland}
\author[a,b]{Fabio Apruzzi,}
\author[c]{G.~Bruno De Luca,}
\author[d,e]{Gabriele Lo Monaco,}
\author[f]{Christoph F.~Uhlemann}

\affiliation[a]{Mathematical Institute, University of Oxford, \\
	Andrew-Wiles Building,  Woodstock Road, Oxford, OX2 6GG, UK}
\affiliation[b]{Albert Einstein Center for Fundamental Physics, Institute for Theoretical Physics,\\  University of Bern,
Sidlerstrasse 5, CH-3012 Bern, Switzerland}
\affiliation[c]{Stanford Institute for Theoretical Physics, Stanford University,\\
382 Via Pueblo Mall, Stanford, CA 94305, United States}
\affiliation[d]{Universit\'e Paris-Saclay, CNRS, CEA, Institut de Physique Th\'eorique, \\ 91191, Gif-sur-Yvette, France}
\affiliation[e]{Department of Physics, Stockholm University, AlbaNova, 10691 Stockholm, Sweden}
\affiliation[f]{Leinweber Center for Theoretical Physics, Department of Physics\\
	University of Michigan, Ann Arbor, MI 48109-1040, USA}

\emailAdd{apruzzi@itp.unibe.ch}
\emailAdd{gbdeluca@stanford.edu}
\emailAdd{gabriele.lomonaco@ipht.fr}
\emailAdd{uhlemann@umich.edu}

\preprint{LCTP-21-27}

\abstract{
We discuss infinite families of non-supersymmetric AdS$_6$ solutions in Type IIB string theory. They are siblings of supersymmetric solutions which are associated with $(p,q)$ 5-brane webs and holographically dual to 5d SCFTs engineered by those brane webs. The non-supersymmetric backgrounds carry identical 5-brane charges and are connected to the supersymmetric ones by RG flows. We study the stability of the non-supersymmetric solutions, identifying perturbative and non-perturbative decay channels for all the backgrounds explicitly available. We also identify likely decay mechanisms for solutions that have not been constructed explicitly but may be expected to exist based on brane web considerations. Finally, we exclude scale separation by constructing universal spin 2 modes with masses comparable to the mass-scale of the cosmological constant. 
}

\begin{document}

\setcounter{tocdepth}{2}
\maketitle

\section{Introduction and summary}

In this paper we investigate the stability of non-supersymmetric AdS$_6$ solutions in Type IIB string theory. 
Potentially stable non-supersymmetric AdS$_6$ solutions would be of great interest as they would provide holographic duals for non-supersymmetric 5d CFTs. The understanding of interacting UV-complete QFTs in $d>4$ that are believed to exist so far relies strongly on supersymmetry. Whether non-supersymmetric higher-dimensional CFTs exist is an interesting question.
Based on a generalization of the weak gravity conjecture \cite{Arkani-Hamed:2006emk}, non-supersymmetric AdS solutions supported by fluxes were generally conjectured to be unstable in \cite{Ooguri:2016pdq}. In line with this conjecture, non-supersymmetric AdS$_7$ solutions in massive Type IIA were shown to be unstable in \cite{Apruzzi:2019ecr} and an AdS$_6$ solution in massive Type IIA was shown to be unstable in \cite{Suh:2020rma}.
On the other hand, results consistent with the existence of non-supersymmetric 5d CFTs were found in \cite{Fei:2014yja,Nakayama:2014yia,Bae:2014hia,Chester:2014gqa} and \cite{DeCesare:2021pfb}, and AdS$_8$ duals for non-supersymmetric 7d CFTs were proposed in \cite{Cordova:2018eba}. An RG flow ending in a non-supersymmetric 5d CFT was conjectured in \cite{BenettiGenolini:2019zth}, though an instability was identified in \cite{Bertolini:2021cew}. 
As we discuss below, Type IIB string theory admits infinite families of non-supersymmetric AdS$_6$ vacua. We initiate an investigation of their stability to further test the conjecture of \cite{Ooguri:2016pdq} and gain intuition for potentially stable non-supersymmetric AdS$_6$ solutions.\footnote{The stability of vacua in other dimensions was studied e.g.\ in \cite{Maldacena:1998uz,Gaiotto:2009mv,Narayan:2010em,Antonelli:2019nar,Bena:2020xxb, Bena:2020qpa, Basile:2021vxh}; a family of AdS$_4$ solutions whose stability is open at this point was discussed in \cite{Guarino:2020jwv,Guarino:2020flh}.}

The non-supersymmetric solutions we will consider are associated with 5d SCFTs constructed using $(p,q)$ 5-brane webs \cite{Aharony:1997ju,Aharony:1997bh}, which can involve additional ingredients like 7-branes \cite{DeWolfe:1999hj,Benini:2009gi} and orientifold planes \cite{Bergman:2015dpa,Zafrir:2015ftn}.
Supergravity solutions describing the near-horizon geometry of these brane webs were constructed in \cite{DHoker:2016ujz,DHoker:2016ysh,DHoker:2017mds,DHoker:2017zwj} (for earlier attempts see \cite{Apruzzi:2014qva,Kim:2015hya}).
The geometry is a warped product of AdS$_6$ and $S^2$ over a Riemann surface $\Sigma$, and the solutions are characterized by a pair of locally holomorphic functions $\cA_\pm$ on $\Sigma$.
The external 5-branes of the brane webs are represented as poles in the differentials $\partial\cA_\pm$ on the boundary of $\Sigma$, while 7-branes correspond to punctures with $SL(2,\ZZ)$ monodromy in the interior of $\Sigma$.
Each of these supergravity solutions admits a consistent truncation to 6d gauged supergravity \cite{Hong:2018amk,Malek:2018zcz,Malek:2019ucd}.
The details of the 10d seed solution are encoded in the uplift, i.e.\ for each choice of  functions $\cA_\pm$ on $\Sigma$ there is one uplift to Type IIB.
The supersymmetric 10d AdS$_6$ solutions give rise to the supersymmetric AdS$_6$ vacuum in 6d.
6d gauged supergravity also admits a non-supersymmetric AdS$_6$ vacuum, which is connected to the supersymmetric one by an RG flow \cite{Gursoy:2002tx,Karndumri:2012vh}. The ratio of central charges between the supersymmetric and non-supersymmetric solutions is $27/25$.
The entire RG flows and in particular their non-supersymmetric IR fixed points can be uplifted to Type IIB, with one distinct uplift for each supersymmetric 10d seed solution.
The resulting 10d non-supersymmetric AdS$_6$ solutions have the same brane sources as the supersymmetric seed solutions, and we expect them to be related to non-supersymmetric deformations of the brane webs associated with the seed solution.
General features of the RG flows will be discussed in the main part along with a field theoretic interpretation for the supersymmetry-breaking deformation.

We will show that the non-supersymmetric counterparts of all explicitly constructed supersymmetric solutions available so far are (at best) part of the swampland \cite{Vafa:2005ui}: they may or may not be stable within Type IIB supergravity (we do not discuss the complete KK spectrum), but they are certainly unstable with respect to brane creation and polarization processes in string theory.
In other words, among the uplifts we consider we find no embedding of the non-supersymmetric AdS$_6$ solution of 6d gauged supergravity which is stable in 10d. 
This is in line with the conjecture of \cite{Ooguri:2016pdq} that there are no stable non-supersymmetric AdS solutions supported by fluxes.

A general feature of the non-supersymmetric AdS$_6$ solutions is that the 5-brane sources on the boundary of $\Sigma$ are unstable; they exhibit an instability in which the 5-branes move into the interior of $\Sigma$.
Concretely, we show that probe D5-branes wrapping AdS$_6$ near a D5-brane pole are unstable in the non-supersymmetric solutions and are expelled into the interior of $\Sigma$. 
The arguments only use the behavior of the solutions near the 5-brane sources; they generalize to generic $(p,q)$ 5-brane poles by the $SL(2,\ZZ)$ symmetry of Type IIB and apply regardless of the global form of $\Sigma$.
Since all solutions constructed explicitly so far involve  5-brane sources (solutions with only 5-branes have at least 3 poles, solutions with mutually local $[p,q]$ 7-branes have at least two \cite{Chaney:2018gjc} and solutions with O7 planes can have as little as one 5-brane pole \cite{Uhlemann:2019lge}), this instability applies to the non-super\-symmetric siblings of  all solutions that are explicitly available at this point.
We study the non-Abelian action for multiple D5-branes as well, to investigate configurations where the 5-branes are polarized into 7-branes,
and derive the fluctuation spectrum in the supersymmetric and non-supersymmetric solutions.
We find that 5-branes polarized into 7-branes are stable in the super\-symmetric solutions, but unstable in the non-supersymmetric ones, due to tachyons below the BF bound which transform non-trivially under the $SU(2)$ R-symmetry.
In brane web terms the tachyons correspond to fluctuations out of the plane of the web.

The natural next step is to study 7-brane punctures. 
Solutions with 5-brane poles and arbitrary numbers of (mutually local) 7-brane punctures were constructed in \cite{DHoker:2017zwj}. To assess the stability of punctures in the associated non-supersymmetric solutions we add probe 7-branes and study their stability. The 7-branes wrap AdS$_6\times S^2$ and are localized at points in $\Sigma$ which lie along a curve. In the non-supersymmetric solutions they can be embedded at the same points on $\Sigma$ as in the supersymmetric ones.
For a sample of AdS$_6$ solutions we study fluctuations in the embeddings and worldvolume fluxes of such probe 7-branes. 
The sample comprises the $T_N$ and $+_{N,M}$ solutions, associated with the brane webs in fig.~\ref{fig:plus} and \ref{fig:TN}, respectively,
and the $+_{N,M,j,k}$ solutions associated with the web in fig.~\ref{fig:plus-D7} as an example with backreacted D7-brane punctures. 
The free energies and various local and non-local operators were studied and matched to field theory computations for the supersymmetric solutions in \cite{Bergman:2018hin,Fluder:2018chf,Uhlemann:2019ypp,Uhlemann:2020bek,Gutperle:2020rty}; recent studies  include a reformulation that connects to gauge theory deformations \cite{Legramandi:2021uds}, Penrose limits \cite{Gutperle:2021nkl}, integrability \cite{Roychowdhury:2021oiq,Alencar:2021ljc} and compactifications \cite{Legramandi:2021aqv}.
For the probe D7-branes we find that the masses of fluctuations are independent of the location of the D7-branes on $\Sigma$ and, moreover, identical among the aforementioned solutions.
For the supersymmetric solutions they match brane web expectations.
For fluctuations preserving the $S^2$ isometries the masses are above the BF bound for the supersymmetric as well as the non-supersymmetric solutions.
Among the modes that transform non-trivially under $SU(2)$, however, we find tachyons signaling instabilities in the non-supersymmetric solutions.
For 7-branes approaching the boundary of $\Sigma$ the fluctuation spectrum smoothly connects to the results obtained for 5-branes before. 
The universality of these results suggests that also 7-brane punctures may generally be unstable in the non-supersymmetric solutions.

We study two further types of instabilities. The first is associated with domain walls which describe flows onto the moduli space in the supersymmetric theories. These can turn into instabilities in the non-supersymmetric solutions.
The second type is the nucleation of brane bubbles, through which a solution can decrease its brane charges by nucleating bubbles of new vacua with lesser charge. 
These channels are closely related for branes that are pointlike in the internal space \cite{Apruzzi:2019ecr}. 
The relevant branes in the Type IIB AdS$_6$ solutions, however, are strings in the internal space.
As a result they have richer dynamics.
To anchor the discussion and highlight the differences we start for both channels with the Brandhuber-Oz solution in Type IIA, which describes the near-horizon limit of D4-branes probing a D8/O8 system \cite{Brandhuber:1999np}, and then discuss 5-brane webs.

We start with domain wall instabilities and the Brandhuber-Oz solution. The moduli space of the dual 5d SCFT corresponds to configurations where the D4 branes are separated. It can be explored holographically with (stable) D4-brane embeddings that extend along $\RR^{1,4}$ slices in AdS$_6$ while being pointlike in the internal space.
In the non-supersymmetric siblings there are no analogous static embeddings; D4-branes always feel a force that generically expels them towards the conformal boundary of AdS$_6$.
This signals an instability in the sense of \cite{Gaiotto:2009mv,Apruzzi:2019ecr,Bena:2020xxb} and was discussed previously for this solution in~\cite{Suh:2020rma}.

For the 5d SCFTs with duals in Type IIB and their non-supersymmetric siblings we focus on Coulomb-branch domain walls, which preserve the $SU(2)$ R-symmetry.
Moving onto the Coulomb branch of the 5d SCFTs corresponds to resolving the associated intersection of $(p,q)$ 5-branes to a web of 5-branes with closed faces, without moving the external 5-branes. An example for the $+_{N,M}$ theory is shown in fig.~\ref{fig:plus-CB}.
In the AdS$_6$ supergravity solutions such states can be realized by configurations of 5-brane segments that are embedded along $\RR^{1,4}$ in AdS$_6$ and extend along curves in the internal space, rather than being pointlike. 
Embeddings preserving the $SU(2)$ R-symmetry extend along the boundary of $\Sigma$, where the $S^2$ in the geometry collapses, and connect to the 5-brane poles.
We discuss these embeddings and their brane web interpretation in the supersymmetric solutions and derive a general analytic form.
In the non-supersymmetric backgrounds we can still find static embeddings along the boundary of $\Sigma$. 
However, in contrast to the supersymmetric case, the probe 5-branes diverge towards the conformal boundary of AdS$_6$ before reaching the 5-brane poles.
We interpret this as an instability in which 5-branes can be pulled out of the Poincar\'e horizon of AdS$_6$ and get partly expelled towards the conformal boundary near the 5-brane poles.
In the brane webs this might be understood as 5-brane segments bending out towards infinity along the external 5-branes of the web. 
This is different from the D4-brane case, where the entire branes were expelled to the conformal boundary.
We also discuss solutions where some of the 5-brane poles have been Higgsed and replaced by 7-brane punctures. We find that the domain wall 5-branes in the non-supersymmetric siblings are not expelled towards the conformal boundary when crossing the branch cuts.

For the brane bubble nucleation channel the relevant process in the Brandhuber-Oz solution is D4-brane nucleation. 
We construct the instanton describing the tunneling as a D4-brane wrapping $S^5$ in Euclidean AdS$_6$, in the spirit of Coleman-De Luccia \cite{Coleman:1980aw}.
Upon changing coordinates, this becomes an expanding bubble with $SO(1,5)$ symmetry in global Lorentzian AdS$_6$.
These highly symmetric bubble configurations can be found at those points in the internal space where a domain wall D4-brane is expelled to the conformal boundary of AdS$_6$ without moving in the internal space.
This is in line with the conclusion of \cite{Apruzzi:2019ecr} that the criteria for the existence of the two instabilities are equivalent for branes that are pointlike in the internal space.

For the Type IIB solutions corresponding to 5-brane webs  we seek probe 5-branes with a similar symmetry, i.e.\ branes wrapping $S^5$ in Euclidean AdS$_6$.
One could consider a variety of bubble nucleation channels, corresponding to different embeddings in $\Sigma$.
To be consistent with the domain wall discussion we seek probe 5-branes along the same curve in $\Sigma$ as the Coulomb-branch domain wall 5-branes. 
However, we do not find bubble nucleation solutions, i.e.\ Euclidean configuration in which the size of the $S^5$ stays finite along the curve that the 5-branes wrap in $\Sigma$, so that they do not reach the conformal boundary and have finite action.
The only bubble nucleation embeddings we find reach to the conformal boundary near the 5-brane poles and have infinite action.
We match this feature with the brane web perspective, where charge conservation prevents 5-brane segments of finite extent from nucleating without attached semi-infinite 5-branes.
This distinguishes the 5-brane web setups from the D4-brane configuration, where D4 branes can be created or eliminated without altering branes with more than 5 non-compact dimensions.

As for the domain wall embeddings before, one can study what happens if 5-brane poles are replaced by 7-brane punctures. We find, similarly to the domain wall embeddings, that  bubble 5-brane embeddings can transition over the branch cut associated with the puncture continuously. This can again be understood from the brane web perspective. 

This discussion reveals qualitative differences between the  pointlike D4-branes in the Brandhuber-Oz solution and the analogous 5-brane configurations in the Type IIB solutions. In the Type IIB solutions the 5-brane poles on the boundary of $\Sigma$ drive the domain wall instabilities and obstruct bubble nucleation instabilities for 5-branes embedded along the boundary of $\Sigma$ -- in both cases by expelling the embeddings towards the conformal boundary. For D4-branes that do not move in the internal space, on the other hand, domain wall and bubble nucleation instabilities exist at the same points in the internal space -- the domain wall embeddings reach the conformal boundary while bubble nucleation embeddings do not.

We note that the 5d $USp(N)$ theory dual to the Brandhuber-Oz solution can also be engineered by a 5-brane web in Type IIB \cite{Bergman:2015dpa}. There is no contradiction with our results, as the brane web for the $USp(N)$ theory has no large numbers of external 5-branes and would correspond to a supergravity solution with no 5-brane poles.

Based on our results we may also speculate on the stability of solutions where all 5-brane poles are replaced by 7-brane punctures, corresponding to brane webs where all external 5-branes are terminated on 7-branes in large groups. Such solutions have not been constructed explicitly, but they should be captured by the general local solution of \cite{DHoker:2016ujz}. With no 5-brane poles to begin with, the 5-brane pole instability discussed above would not affect the non-supersymmetric siblings of these solutions. However, our 7-brane discussion suggests a universal instability for 7-brane punctures in the non-supersymmetric solutions, which may render these solutions unstable as well. The absence of 5-brane poles would also open the possibility to nucleate closed 5-brane loops along the boundary of $\Sigma$. We leave more detailed studies for the future.

We conclude our analysis by looking at the problem of \emph{scale separation}. Starting from a higher-dimensional gravitational theory such as type IIB supergravity, there are different mass scales that arise upon compactification to lower-dimensional vacua, such as the mass of the cosmological constant $m_\Lambda= \sqrt{|\Lambda|}$, the mass of the first Kaluza-Klein (KK) mode $m_{KK}$ and the reduced Planck mass $m_P$. 
The vast majority of known AdS vacua of ten- end eleven- dimensional supergravities have no separation among these scales, and in particular \mbox{$m_{KK}\sim m_\Lambda$}. This property has been named absence of \emph{separation of scales}, and the challenges in constructing controlled scale-separated vacua in supergravity led to conjectures that these might not exist as vacua of UV complete theories, relegating them to the swampland \cite{lust-palti-vafa}.

Explicitly computing the Kaluza-Klein spectrum can be challenging, especially for warped backgrounds. To date there are no known complete KK spectra of warped compactifications, with only some recent results obtained in the context of ExFT \cite{malek-samtleben-kk, malek-nicolai-samtleben}. Luckily, knowledge of the full spectrum is not required to rule out scale-separation, since it is enough to show that $m_{KK}\sim m_\Lambda$ even for a single mode. We employ this strategy to show that all the non-supersymmetric AdS$_6$ backgrounds we are considering are not scale separated.
As shown in \cite{csaki-erlich-hollowood-shirman,bachas-estes}, the spin 2 fluctuations always decouple and their masses are given by a universal operator only depending on the internal metric and the warping.\footnote{Recently, \cite{REC, deluca-deponti-mondino-tomasiello} showed that this operator is natural in the mathematical framework of \emph{Bakry-\'Emery geometry}, which allows to exploit known mathematical results on the spectrum of the Bakry-\'Emery Laplacian to rigorously bound $m_{KK}$ for general warped compactifications.}
By generalizing the analysis of \cite{spin2-susy} we show that the entire spectrum of spin-2 fluctuations of the non-supersymmetric solutions can be obtained from that of the supersymmetric solutions by a simple mass shift.

When supersymmetry is present, there are general arguments to expect low-lying modes corresponding to protected operators \cite{polchinski-silverstein}.
We show in this paper that also the large class of non-supersymmetric AdS$_6$ backgrounds we are considering is not scale separated. This situation is similar to the AdS$_7$ case, where \cite{Apruzzi:2019ecr} showed that all supersymmetric AdS$_7$ vacua and their non-supersymmetric siblings are not scale-separated, even in presence of orientifold planes.
While it has been suggested that O-planes are an important ingredient for scale separation \cite{gautason-schillo-vanriet-williams, REC} and they appear explicitly in some proposed scale-separated vacua (see e.g. \cite{dgkt, Petrini:2013ika, Cribiori:2021djm, Demirtas:2021nlu}), they do not improve the situation in the non-supersymmetric AdS$_7$ and AdS$_6$ partners of the supersymmetric backgrounds. Technically, this happens because one can construct universal classes of low-lying modes which are unaffected by the presence of the orientifold planes.

We conclude by urging the reader to interpret the results presented here, both on the instability and the absence of scale separation, \emph{cum grano salis}. Though they hold for the large class of non-supersymmetric partners of AdS$_6$ (and AdS$_7$) solutions, these vacua are very similar to the original supersymmetric ones. Since to date a classification of more general non-supersymmetric vacua (both AdS$_6$ and AdS$_7$) is not available, there is no indication on whether these results are representative of the full higher-dimensional AdS landscape or not. It would be important then to construct more general non-supersymmetric backgrounds to better address this issue.

\medskip

\textbf{Outline:} In Sec.~\ref{sec:intro} we discuss the Type IIB AdS$_6$ solutions, their non-super\-symmetric siblings and the RG flows.
In Sec.~\ref{sec:poles} we discuss the stability of 5-brane poles in the non-supersymmetric solutions and the polarization of 5-branes into 7-branes. In Sec.~\ref{sec:probe-D7} the fluctuation spectrum of probe 7-branes is studied for example solutions. 
In Sec.~\ref{sec:DW} and \ref{sec:D5-instanton} we discuss, respectively, domain wall instabilities and bubble nucleation.
In Sec.~\ref{sec:spin2} we discuss the spectrum of spin-2 fluctuations and scale separation.

\section{Non-supersymmetric \texorpdfstring{AdS$_6$}{AdS6} in Type IIB}\label{sec:intro}

The non-supersymmetric solutions we consider owe their existence to infinite families of supersymmetric Type IIB AdS$_6$ solutions, constructed in \cite{DHoker:2016ujz,DHoker:2016ysh,DHoker:2017mds,DHoker:2017zwj}. Each supersymmetric solution has a consistent truncation to 6d gauged supergravity \cite{Hong:2018amk,Malek:2018zcz}.
In addition to the supersymmetric AdS$_6$ vacuum, 6d gauged supergravity also has a non-supersymmetric AdS$_6$ vacuum \cite{Romans:1985tw}.
The 6d non-supersymmetric AdS$_6$ vacuum can be uplifted to Type IIB using the uplift associated with any of the supersymmetric seed solutions, leading to infinite families of non-supersymmetric AdS$_6$ solutions of Type IIB.
In the following we give explicit expressions for the non-supersymmetric solutions in parallel with the supersymmetric ones.
More details on the supersymmetric solutions and the uplifts can be found in the references.

The geometry of the solutions is a warped product of AdS$_6$ and $S^2$ over a Riemann surface $\Sigma$.
The solutions are defined by a pair of locally holomorphic functions $\cA_\pm$ on $\Sigma$.
The metric is given by
\begin{align}\label{eq:metric-gen}
	ds^2&=f_6^2 ds^2_{AdS_6}+f_2^2ds^2_{S^2}+4\rho^2 |dw|^2~,
\end{align}
where $w$ is a complex coordinate on $\Sigma$, and $ds^2_{Ads_6}$ and $ds^2_{S^2}$ are, respectively, the line elements for unit-radius AdS$_6$ and $S^2$. The warp factors are
\begin{align}\label{eq:functionsExpl}
	f_6^2&=\frac{20}{V_0 X}\sqrt{6\cG T}~,
	&
	f_2^2&= \frac{X}{9}\sqrt{6\cG}\,T ^{-\tfrac{3}{2}}~,
	&
	\rho^2&=\frac{X\kappa^2}{\sqrt{6\cG}} T^{\tfrac{1}{2}}~,
\end{align}
where, respectively for the supersymmetric and non-supersymmetric solutions,
\begin{align}
	X_{\rm susy}&=1~, & X_{\rm non-susy}&=3^{-\tfrac{1}{4}}~,
\end{align}
and $V_0=-X^{-6}+12 X^{-2}+9X^2$. The remaining quantities are defined in terms of $\cA_\pm$ as
\begin{align}\label{eq:kappa2-G}
	\kappa^2&=-|\partial_w \cA_+|^2+|\partial_w \cA_-|^2~,
	&
	\partial_w\cB&=\cA_+\partial_w \cA_- - \cA_-\partial_w\cA_+~,
	\nonumber\\
	\cG&=|\cA_+|^2-|\cA_-|^2+\cB+\bar{\cB}~,
	&
	T^2&=\left(\frac{1+R}{1-R}\right)^2=X^4+\frac{2|\partial_w\cG|^2}{3\kappa^2 \, \cG }~.
\end{align}
The axion-dilaton scalar $B=(1+i\tau)/(1-i\tau)$ is given by
\begin{align}
	B&=-\frac{(T/X^2+1)\partial\cA_+ \overline{\partial\cG}-(T/X^2-1)\overline{\partial\cA_-}\partial\cG}{(T/X^2+1)\partial\cA_- \overline{\partial\cG}-(T/X^2-1)\overline{\partial\cA_+}\partial\cG}~.
\end{align}
The two-form potentials take the same general form for the supersymmetric and non-supersymmetric solutions,
\begin{align}\label{eq:cC}
	B_2+iC_2&= \cC \vol_{S^2}~,
	&
	\cC&=\frac{2i}{3}\left(
	\frac{\partial_{\bar w}\cG\partial_w\cA_++\partial_w \cG \partial_{\bar w}\bar\cA_-}{3\kappa^{2}T^2} - \bar{\mathcal{A}}_{-} - \mathcal{A}_{+}  \right).
\end{align}
However, $T$ depends on $X$ and so does as a result $\cC$.
As we will discuss below, the 5-brane charges are nevertheless identical, suggesting that the non-supersymmetric solutions may be related to non-supersymmetric 5-brane configurations similar to their supersymmetric counterparts.
We will also need the dual six-form potentials, which also depend on $X$.
Expressions for the field strength can be found in \cite{Gutperle:2018vdd} and the potential was derived in \cite{Corbino:2017tfl}.
As extension to include the non-supersymmetric solutions we find 
\begin{align}\label{eq:C6-cM}
	B_6-i C_6&=\cM\vol_{AdS_6}\,, &
	\cM=4X^4(2&-X^4)^2\Big[
	20(\cW_++\overline \cW_-)-3(2-X^4)\cG U
	\nonumber\\ &&&
	+5(\cA_++\bar \cA_-)(2|\cA_+|^2-2|\cA_-|^2-3\cG)\Big],
\end{align}
where $\partial_w\cW_\pm=\cA_\pm\partial_w\cB$.

Concrete solutions are characterized by a choice of $\cA_\pm$, which is constrained by regularity conditions.
For supergravity solutions associated with $(p,q)$ 5-brane webs without 7-branes, $\Sigma$ can be taken as the upper half plane and the functions  $\cA_\pm (w)$ are characterized by poles in $\partial\cA_\pm$ on the boundary of $\Sigma$.
The poles represent the external 5-branes of the associated brane configuration, with the residues encoding the 5-brane charges.
Concretely, $\cA_\pm$ are given in terms of pole positions $r_{\ell}$ and the respective residues $Z_\pm^\ell$ by
\begin{align}\label{eqn:cA}
	\cA_\pm (w) &=\cA_\pm^0+\sum_{\ell=1}^L Z_\pm^\ell \ln(w-r_\ell)~,
	&
	Z_\pm^\ell&=\frac{3}{4}\alpha^\prime \left(\pm q_\ell+ ip_\ell\right)~.
\end{align}
The locations $r_\ell$ and constants $\cA_\pm^0$ are determined in terms of the residues by regularity conditions,
enforcing e.g.\ that the $S^2$ in (\ref{eq:metric-gen}) collapses on the boundary of $\Sigma$ to close off the geometry smoothly.
Details can be found in \cite{DHoker:2017mds}; concrete examples will be discussed below.
The real function $\cG$ can be expressed in a compact form using the Bloch-Wigner function as \cite{Uhlemann:2020bek}
\begin{align}
	\label{eq:GBloch}
	{\cal G}&=\sum_{\ell,k=1}^{L}2iZ^{[\ell,k]}\text{D}\left(\frac{r_\ell-w}{r_\ell-r_k}\right)\,,& \text{D}(u)&=\Im(\text{Li}_2(u))+\text{arg}(1-u)\ln|u|\,,
\end{align}
where $Z^{[\ell,k]}=Z_+^\ell Z_-^k-Z_+^k Z_-^\ell$.
Generalizations incorporating [p,q] 7-branes can be found in \cite{DHoker:2017zwj} and O7-planes in \cite{Uhlemann:2019lge}. 
We will discuss those cases but will not need explicit expressions for $\cA_\pm$.

We close this part with a discussion of the 5-brane charges.
Around each pole on the boundary of $\Sigma$ there is a 3-cycle, formed out of a curve connecting the boundary components to either side of the pole and the $S^2$. The $S^2$ collapses on both ends of the curve. 
The 5-brane charges at the pole are given by the integral of the complex 3-form along this cycle, which reduces to the discontinuity of $\cC$ in (\ref{eq:cC}) at the pole. This discontinuity is independent of the first term in the expression for $\cC$ in (\ref{eq:cC}), and thus independent of $X$.
The 5-brane charges are therefore identical for the supersymmetric and non-supersymmetric solutions.
Moreover, the non-supersymmetric solutions have the same $SU(2)$ symmetry as the supersymmetric solutions, where it corresponds to the R-symmetry of the dual SCFT. We will occasionally take the freedom to refer to this symmetry as R-symmtery also in the non-supersymmetric solutions. This suggests that, similar to the supersymmetric setups, the non-supersymmetric solutions describe planar 5-brane configurations with an $SU(2)$ symmetry corresponding to rotations in the directions transverse to the plane.

\subsection{RG-flow solutions}

We briefly discuss the RG flows that lead to the non-supersymmetric solutions described above and the nature of the deformation. 
The flow is constructed in 6d $F(4)$ gauged supergravity \cite{Gursoy:2002tx,Karndumri:2012vh} and involves only the metric and the real scalar $X$ of the 6d supergravity multiplet.
The solution can be obtained from the ansatz
\begin{align}
	ds^2_6&=e^{2A(r)} ds^2_{1,4}+dr^2~, & X=X(r)=e^{-\frac{\sigma(r)}{2\sqrt{2}}},
\end{align}
where $\sigma$ is the dilaton field in $F(4)$ gauged supergravity. The solution interpolates between AdS$_6$ in the UV and AdS$_6$ of smaller radius in the IR,
\begin{align}
	r\rightarrow  -\infty:&\quad  A(r) \sim \frac{3r}{L_{\rm UV}}~, &
	r\rightarrow  \infty: &\quad A(r) \sim \frac{3r}{L_{\rm IR}}~.
\end{align}
While the fixed points are known analytically, an explicit solution for the RG flow was obtained numerically in \cite{Gursoy:2002tx}.
The metric in the uplifted 10d solutions takes the form
\begin{equation}
	ds^2=f_6(w,\bar w, r)^2 (e^{2A(r)} ds^2_{1,4}+dr^2)+f_2(w,\bar w, r)^2  ds^2_{S^2}+4\rho(w,\bar w, r)^2 |dw|^2~,
\end{equation}
where $f_6,f_2,\rho$ are as in \eqref{eq:functionsExpl} with the difference that now $X=X(r)$. The remaining 10d fields are obtained accordingly.

The 6d gauged supergravity fields are holographically dual to the stress tensor multiplet of the 5d SCFTs.
The flows to the non-supersymmetric vacuum are thus triggered by deforming the SCFT with the scalar operator in the stress tensor multiplet, which is an $SU(2)$ R-symmetry singlet. This  scalar has $m^2L = -6$ and therefore could be dual to an operator with scaling dimension $\Delta=3$ or $\Delta=2$ \cite{Gursoy:2002tx,Karndumri:2012vh}. The dual operator is the R-symmetry neutral scalar operator in the stress energy tensor multiplet, denoted as $\mathbf{B_2}[0,0]_3^{(0)}$ in \cite{Cordova:2016emh}, and therefore has $\Delta=3$. 
The entire 10d flow solutions preserve the $SU(2)$ symmetry of the $S^2$, corresponding to the R-symmetry in the dual SCFTs, while breaking the isometries of AdS$_6$ to the Poincar\'e symmetry of the Minkowski slices.
The flows are similar in those regards to flows triggered by turning on finite gauge coupling as a relevant deformation or Coulomb branch flows. Crucially, they break supersymmetry.
In terms of a brane web picture, the universal nature of the flows and their relation to the stress tensor multiplet  suggest that they correspond to deformations that act uniformly across the brane web, while the preserved $SU(2)$ indicates that the deformation acts in the plane of the web.

It would be interesting to understand the supersymmetry-breaking deformation in more detail from the brane web perspective. The flows are similar to flows of 6d SCFTs described by 7d gauged supergravity, for which a brane interpretation was proposed in \cite{Apruzzi:2016rny}.

\section{Stability of 5-brane poles}\label{sec:poles}
The external 5-brane stacks in the 5-brane junction associated with a given supergravity solution are represented by the poles in $\partial\cA_\pm$.
To investigate the stability of the brane configurations we consider the generic behavior of the solutions near a D5-brane pole.
We will show that a probe D5 brane added at the location of the pole is tolerated in the supersymmetric solutions but is expelled into the interior of $\Sigma$ for the non-supersymmetric solutions.
The analysis generalizes to generic $(p,q)$ 5-branes by $SL(2,\ZZ)$ duality, and indicates an instability where the 5-brane stacks represented by the poles on the boundary of $\Sigma$ disintegrate.

\subsection{5-brane polarization}\label{eq:D5-Abelian}
In this section we work in Einstein frame, where it is possible to write an action for generic $(p,q)$ 5-branes \cite{Bergshoeff:2006gs}.\footnote{This form of the action holds if the four-form gauge potential vanishes and $C_{(2)}^2=B^2=B\wedge C_{(2)}=0$.}
Namely,
\begin{equation}
\label{eq:(p,q)action}
S^{(5)}_{(p,q)}\,=\,-T_5\int d^6\xi \sqrt{\mathfrak q^t\,\mathfrak{ M}\,\mathfrak q}\sqrt{-\det \left( g_{ab}\,+\,\frac{\mathcal{F}_{ab}} {\sqrt{\mathfrak q^t\,\mathfrak{M}\,\mathfrak q}}\right)}+T_5\int\!p\,C_{6}+q \,B_{6}\,,
\end{equation}
where, with $\mathfrak{q}=(q,p)$, fluxes $F_1$, $F_2$ and dilaton convention $\tau=\chi+ie^{-2\phi}$,
\begin{align}\label{eq:pq5-action}
\mathfrak{ M}\,&=\,e^{2\phi}\left(
\begin{matrix}
\chi^2+e^{-4\phi}\,\,\, & \chi\\
\chi & 1
\end{matrix}
\right)\,,&
\mathcal{F}\,&=\,p(F_{1}-B_{2})-q(F_{2}-C_{2})\,.
\end{align}
With this action we could uniformly treat poles corresponding to generic 5-branes.
However, we will focus here on D5-branes, corresponding to $(p,q)=(0,\pm 1)$, for simplicity, noting that the results generalize.
Then
\begin{equation}
\label{eq:D5action}
S_{D5}\,=\,-T_5\int d^6\xi e^{\phi}\sqrt{-\det \left( g_{ab}+e^{-\phi}\,\mathcal{F}_{ab}\right)}+T_5\int\!C_{6}\,.
\end{equation}
In order to study the behavior of the D5-branes at the boundary of the Riemann surface $\Sigma$, we first consider the action for a single D5-brane sitting close to a pole. The probe is thus extended along the external $\text{AdS}_6$ space and we will consider the worldvolume flux to be vanishing. We derive the required behavior of the background supergravity solutions near poles from the expressions for $\cA_\pm$ in \eqref{eqn:cA}, but the near-pole behavior is identical for solutions with 7-brane punctures, so the results extend straightforwardly.
The D5-brane location is taken as
\begin{equation}
w_n\,=\,r_{n}+\delta\,e^{i\alpha}\,,
\end{equation}
where $0<\alpha<\pi$ and the radial coordinate $\delta$ is small, {\it i.e.} $0<\delta \ll |r_n-r_\ell|\,\forall \ell$. We want to evaluate the probe action in this limit of small $\delta$. Close to the pole, all the functions characterizing the background have a simple behavior. Let us provide some examples: 
The Bloch-Wigner function in (\ref{eq:GBloch}) can be approximated in the following way for small (modulus of) the argument:
\begin{equation}
\begin{split}
\text{D}\left( \delta \,e^{i\alpha} \right)\,&=\,-\delta (\ln\delta -1)\sin\alpha-\frac{1}{2}\delta ^2\ln\delta \sin(2\alpha)\,+\,O(\delta ^2)\,,\\
\text{D}\left(r_n+\delta \,e^{i\alpha}\right)\,&=\,-\frac{1}{2}\left( \frac{\ln r_n^2}{1-r_n} +\frac{\ln(1-r_n)^2}{r_n}\right)\delta \sin\alpha+O(\delta ^2)\,.
\end{split}
\end{equation}
Plugging the expansions above into \eqref{eq:GBloch} leads to the asymptotic behavior of $\cG$ at the pole,
\begin{equation}
{\cal G}\,=\,\sum^L_{\ell=1}2 \kappa_n^2\,\delta |\ln \delta |\,\sin \alpha+2 \lambda_n\,\delta\,\sin\alpha\,+O(\rho^2)\,, 
\end{equation}
where we defined the constants
\begin{align}
\label{eq:kldef}
\kappa_n^2&=\sum_{\ell=1}^L \frac{2iZ^{[\ell, n]}}{p_n-p_\ell}\,, & Z^{[\ell,n]}&=\,Z^{\ell}_+Z^n_--Z^n_+Z^\ell_-
~.
\end{align}
The expression for $\lambda_n$ is more involved and not relevant in the following. Similar expansions can be performed for the other functions like $f_6,\,f_2$. The action for a D5-brane extended along AdS$_6$ can then be written as
\begin{equation}
S_{D5}\,=\,-T_5\int d^6\xi\,\sqrt{-g_6}\left(\, 2\,f_6^4\,\rho^2\,e^{\phi}\,\nabla_\beta \delta \nabla^\beta\delta\,+\,f_6^6\,e^{\phi}\,+\,\Im \cM \right)\,,
\end{equation}
where $\delta$ has been promoted to be a field on AdS$_6$, whose covariant derivative has been denoted by $\nabla$. $g_6$ is the determinant of the AdS$_6$ factor. We only require the following expansions:
\begin{align}
f_6^6\,e^{\phi}&\approx \frac{20^3}{V_0^3}\frac{8\,\kappa^4_n \delta^2|\ln \delta|^2}{X^2|Z^n_+|^2}\,\left\{ 1-\frac{1}{\ln \delta}\left(\frac{(1-3X^4)^2+(1+6X^4-9 X^8)\cos 2\alpha}{12X^4}+\frac{2\lambda_n}{\kappa^2_n}\right) \right\},
\nonumber\\
\Im\cM&\approx -\frac{20^3}{V_0^3}\frac{8\,\kappa^4_n \delta^2|\ln \delta|^2}{X^2|Z^n_+|^2}\,\left\{1+\frac{1}{\ln \delta}\left(\frac{1-9X^4+(3X^4-1)\cos2\alpha}{6X^4}-\frac{2\lambda_n}{\kappa^2_n}\right) \right\}.
\end{align}
The two leading terms cancel, which is the zero force condition that must hold for the probe brane. We also need the normalization factor of the kinetic term,
\begin{equation}
4f_6^4\,e^{\phi}\rho^2\,=\,\left(\frac{20}{V_0}\right)^2\frac{4\kappa_n^4}{|Z^n_+|}|\ln \delta|\,.
\end{equation}
We can plug the previous expression into the probe action. In doing so, we also perform the change of variable
\begin{align}\label{eq:near-pole-coord}
	t=\frac{20}{V_0}\frac{2\kappa^2_n}{|Z^n_+|}\delta\sqrt{|\ln \delta|}~,
\end{align}
obtaining, at the lowest order in $t$, the following normalized Lagrangian for a massive scalar field propagating in AdS$_6$:
\begin{equation}
\label{eq:actionAbelian}
\frac{L_{D5}}{\sqrt{-g_6}}=-\frac{1}{2}\nabla_\beta t \nabla^\beta t-5\,t^2 \frac{-1+6X^4-27 X^8+(-1+18 X^4-27 X^8)\cos2\alpha)}{X^6V_0}+\dots
\end{equation}
From this Lagrangian we can read off the masses for two different modes: one of them corresponds to the motion of the probe 5-brane along the boundary ($\alpha=0$) and coincides with the fluctuation mode $\delta x$. The other one ($\alpha=\frac{\pi}{2}$) describes the motion of the 5-brane toward the interior of $\Sigma$, {\it i.e.} fluctuations $\delta y$. The mass matrices are thus
\begin{equation}\label{eq:D5-mass}
	M^{\rm susy}_{\rm D5}=\begin{pmatrix} -4 & 0 \\ 0 & -6\end{pmatrix}\,,\quad\qquad 
	M^{\rm non-susy}_{\rm D5}=\begin{pmatrix} 0 & 0 \\ 0 & -10\end{pmatrix}\,,
\end{equation}
where the first entry corresponds to $\delta x$. As expected, the two modes in the supersymmetric case are both above the BF bound ($m^2\geq-\frac{25}{4}$), indicating that the 5-branes are stable at the poles. 
The $\delta x$ fluctuations are dual to an operator with $\Delta=4$, consistent with a fermion mass term, and preserve the $SU(2)$ R-symmetry.
The $\delta y$ fluctuations are dual to an operator with $\Delta=2$ or $\Delta=3$, depending on whether standard or alternative quantization is used. 
Moving a single D5 brane into the interior of $\Sigma$ would break the $S^2$ isometries corresponding to the $SU(2)$ R-symmetry.
This will be studied in more detail below.
In the non-supersymmetric case the $\delta x$ fluctuations are massless and dual to marginal operators. 
The $\delta y$ modes have a tachyonic mass below the BF bound. This means that, in general, in the non-supersymmetric backgrounds the 5-branes at the poles tend to enter into $\Sigma$. This tachyonic mode is universal for all backgrounds with 5-brane poles and destabilizes them.

\subsection{Non-Abelian modes}\label{sec:D5-nonAbelian}

As D5-branes move into the interior of $\Sigma$ they may polarize into 7-branes.
We now study this process in more detail, using the non-Abelian action for a stack of $N$ $D5$-branes \cite{Myers:1999ps}. We will mostly follow the conventions of \cite{McGuirk:2012sb} and start from
\begin{equation}
S^N_{\text{D5}}\,=\,-T_{D5}\int d^6\xi\,\text{Str}\left\{e^{\phi/2}\sqrt{-\det M}\sqrt{\det Q}-e^{\iota^2_{\Phi}} \mathrm{P}\left(C\wedge e^{-B_2+F}\right)\right\},
\end{equation}
where $F$ is the world-volume field strength (vanishing in the case at hand), $P$ denotes the pullback and we defined the metrics
\begin{equation}
\begin{split}
M_{\mu\nu}&=\mathrm{P}\left(g_{\mu\nu}+e^{\phi}g_{\mu i}(Q^{-1}-\delta)^{ij}g_{j\nu}\right)+e^{-\phi}\mathcal{F}_{\mu\nu}\,,\\
 {Q^i}_j\,&=\,{\delta^i}_j+[\Phi^i,\Phi^k]e^{\phi}E_{kj}\,,\\
 E_{MN}\,&=\,g_{MN}-e^{-\phi}B_{2\,MN}\,,
\end{split}
\end{equation}
where $M,N,\dots$ are ten-dimensional indices, $\mu,\nu,\dots$ are world-volume indices and $i,j,\dots$ are transverse indices. The transverse coordinates are promoted to adjoint fields $\Phi^i$ and the contraction $\iota^2_{\Phi}$ has been defined as
\begin{equation}
\iota^2_{\Phi}\left(\frac{1}{2}F_{MN}dx^{M}\wedge dx^N\right)\,=\,\frac{1}{2}[\Phi^i,\Phi^j]F_{ij}\,,
\end{equation}
for any two-form $F$. The Wess-Zumino part of the Lagrangian simplifies to
\begin{equation}
T_{D5}\int C_6\,+\,\iota^2_{\Phi}\left(C_8-B_2\wedge C_6 \right)\,.
\end{equation}
In oder to evaluate the action we thus need the leading-order behavior for $B_2$ and $C_8$, the latter needing the asymptotic behavior of the axion $\chi$ to be computed.
Following the same procedure as above, we find
\begin{align}
B_2&=\left(-\frac{\kappa^2_n}{3}\frac{3X^4-1}{|Z^n_+|}\,\frac{r}{\log r}\sin^3\alpha+\mathcal O(r/(\log r)^2\right) \text{vol}_{\text{S}^2}\,,
\nonumber\\
\chi&= -\frac{4 |Z_+^n|^2}{\kappa_n^2}\,\frac{3X^4-1}{3X^4}\,\frac{\cos\alpha}{r}+\mathcal O(1)\,,
\end{align}
where we took a gauge where $B_2$ vanishes at the D5 pole. Both expressions are vanishing in the non-supersymmetric case due to the factor $3X^4-1$. However, we do not need to go further in the expansion, since such corrections would be subleading in the expansion of the non-Abelian action. Using the fact that
\begin{equation}
\text{d}C_8\,=\,e^{2\phi}\star \text{d}\chi-\text{d}B_2\wedge C_6\,,
\end{equation}
we can compute the the leading order of $C_8$ as
\begin{equation}
C_8=8 \frac{3X^4-1}{3X^2}\left(\frac{20}{V_0}\right)^3\frac{\kappa^6_n}{|Z^n_+|^2}\,r^3\,\log r\,\sin^3\alpha+\dots\,.
\end{equation}
The previous expressions can be rewritten in terms of the coordinates $x$ and $y$ on the Riemann surface. 
To deal with the logarithmic behavior of the background solution near the pole one could use a coordinate transformation similar to (\ref{eq:near-pole-coord}). However, we will use a more pragmatic procedure and displace the stack of D5-branes slightly from the pole.
That is, we place the stack of D5-branes at a position $(\delta x,\delta y)$ with $\delta y\ll \delta x\ll |r_n-r_\ell |\,\,\,\forall \ell$. 
We will find that the fluctuation spectrum is independent of $\delta x$ and expect a smooth limit of the fluctuation spectrum as $\delta x\rightarrow 0$.
In this way, we can approximate $\log \delta\approx \log|x|$ in the background solution and find for the non-Abelian action
\begin{align}
\label{eq:polAction}
S^{N}_{\text{D5}}\,\approx\,-\tau_{\text{D5}}\int&d^6\xi\frac{\kappa^2_n}{\zeta_n^2}\left(\frac{20}{V_0}\right)^2|\log x |\,\text{Str}\left\{ \frac{1}{2}\nabla_\mu\Phi^i \nabla^\mu\Phi_i+\left(\frac{20}{V_0}\right)\frac{3(1-6X^4+3 X^8)}{2X^6}\,\Phi^i\,\Phi_i \right.
\nonumber\\
&\,\,\left.+ \left(\frac{20}{V_0}\right)X^2[\Phi^i,\Phi^j][\Phi_i,\Phi_j]-\frac{2}{3}\frac{(3X^4-1)}{X^2}\left(\frac{20}{V_0}\right)\epsilon_{ijk}\Phi^i[\Phi^j,\Phi^k]\right\}\,.
\end{align}
In doing so we also performed a rescaling $\Phi\rightarrow \frac{2|Z^n_+|}{\kappa^2_n}\Phi$. The coordinates $\Phi^i$ are the coordinates on the $\mathbb{R}^3$ given by
\begin{equation}
4 \rho^2 \text{d}y^2+f_2^2\,\vol_{\text{S}^2}\,\approx 4\rho^2(\text{d}y^2+y^2 \vol_{\text{S}^2})=4\rho^2\,\vol_{\mathbb{R}^3}.
\end{equation}
We use the Lagrangian in the parenthesis of \eqref{eq:polAction} as the effective action describing the polarization of the branes for fixed $x$. Varying this action with respect to $\Phi^i$ leads to the equations of motion
\begin{equation}
\label{eq:radiusConstr}
\frac{3(1-6X^4+3 X^8)}{X^4} \Phi^i-2(3X^4-1){\epsilon^i}_{jk}[\Phi^j,\Phi^k]-4 X^4[[\Phi^i,\Phi^l],\Phi_l]\,=\,0\,.
\end{equation}
Polarization is described by configurations of the type $\Phi^i=R\sigma^i$, where $\sigma$ are $SU(N)$ matrices satisfying the $SU(2)$ algebra $[\sigma^i,\sigma^j]\,=\,\epsilon^{ijk}\sigma_k$. Equation \eqref{eq:radiusConstr} thus becomes a constraint on the value of the ``radius" $R$
\begin{equation}
\frac{3(1-6X^4+3 X^8)}{X^4}R-4(3X^4-1)R^2+8 X^4\,R^3\,=\,0~.
\end{equation}
For both the supersymmetric and non-supersymmetric case there is one positive-radius solution,
\begin{align}
	R&=\frac{3}{2}~,
\end{align}
and the trivial solution $R=0$. 
In the non-supersymmetric backgrounds the $R=0$ solution is unstable, as shown in the Abelian D5 analysis above.
The value of the potential for the $R=\frac{3}{2}$ solution is lower than for the $R=0$ solution, suggesting that the 5-branes in the non-supersymmetric solutions could form a non-commutative $S^2$ configuration as they move into the bulk and polarize into 7-branes.
We will discuss this in more detail shortly.
The vacuum defined by the $\sigma^i$ breaks the $SU(N)$ flavor symmetry and the $SU(2)$ acting on the $\Phi_i$, but a diagonal $SU(2)$ subgroup is preserved, which we can identify with the $SU(2)$ R-symmetry corresponding to the symmetry of the $S^2$ in the 10d solution.

To further study the polarization into 7-branes we now study the stability of the $R=\frac{3}{2}$ vacuum.
That is, fluctuations of the form $\Phi_i=\frac{3}{2}\sigma_i+\phi^i$. The second-order variation of the effective polarization Lagrangian is
\begin{align}
-\delta^2 L=&\nabla_{\mu}\phi^i\,\nabla^\mu\phi^i+\frac{20}{V_0}\cdot \frac{3(1-6X^4+3X^8)}{X^6}\phi^i\phi_i\,+\,
\nonumber\\
&+\frac{20}{V_0}\cdot 9X^2\left([\phi^i,\sigma^j][\phi_i,\sigma_j]-[\phi^i,\sigma^j][\phi_j,\sigma_i]\right)-\frac{20}{V_0}\cdot \frac{9X^4-6}{X^2}[\sigma^i,\sigma^j][\phi_i,\phi_j]\,.
\end{align}
The mass matrix can be evaluated using the techniques explained in \cite{DeLuca:2018zbi, Apruzzi:2019ecr}. In particular, we can decompose the $SU(N)$ generators into representations of $SU(2)$ such that
\begin{equation}
[\sigma^i,T^a]=j^i_{ab}T^b, 
\end{equation}
and the mass matrices can be rewritten as follows,
\begin{equation}
\begin{split}
(M^{ij}_{\text{susy}})_{ab}\,&=\,-6\left(\delta^{ij}\left(1+\frac{3}{2}j^kj^k\right)-j^j j^i-\frac{1}{2}j^ij^j\right)_{ab}\,,\\
(M^{ij}_{\text{non-susy}})_{ab}\,&=\,-10\left(\left( \delta^{ij}+\frac{1}{2}j^{k}j^{k}\right)-j^j j^i+\frac{1}{2}j^{i}j^j\right)_{ab}\,.
\end{split}
\end{equation}
The $\sigma^i$ are the generators of an $\mathfrak{su}(2)$ algebra, and can be decomposed into irreducible representations,
\begin{align}\label{eq:sigma-decomp}
\sigma^i&=\mathbf{d_1}\oplus \mathbf{d_2}\oplus\dots \mathbf{d}_p~.
\end{align}
One can prove (using the Jacobi identity) that the matrices $j^{i}$ are also $\mathfrak{su}(2)$ generators, {\it i.e.}  $[j^i, j^j]=\epsilon^{ijk}j^k$ and our goal is then determining their representation, i.e. the number and dimensions of the blocks appearing in the $j^i$. The irreducible representations contained in the matrices $j^i$ can be obtained if one understands those as tensor products of two copies of \eqref{eq:sigma-decomp},
\begin{equation}\label{eq:j-rep}
\begin{split}
j^i: \ (\mathbf{d_1}\oplus \mathbf{d_2}\oplus\dots \mathbf{d}_p)&\otimes (\mathbf{d}_1\oplus \mathbf{d_2}\oplus\dots \mathbf{d}_p)\,=\,\\
\,=\, &\oplus_a(\mathbf{2d_a-1}\oplus \mathbf{2d_a-3}\oplus\dots \oplus\mathbf{1})\oplus\\
&2\oplus_{a>b}(\mathbf{d_a+d_b-1}\oplus\mathbf{d_a+d_b-3}\oplus\dots \oplus\mathbf{d_a-d_b+1})
\end{split}
\end{equation}
subtracting at the end a singlet. 
There is a total of $3 d$ modes for each $\mathbf{d}$ appearing in \eqref{eq:j-rep}. and they are organized in the multiplets $\mathbf{d}\otimes  \mathbf{3}= \mathbf{d}\,\oplus\,\mathbf{d+2}\,\oplus \,\mathbf{d-2}$ for $d>1$. 
The masses associated to these three representations are given in Table \ref{tab:masses}.

\begin{table}
	\centering
	\begin{tabular}{|cc|c|c|}
		\hline
		\multicolumn{2}{|c|}{$\mathrm{SU}(2)_{\mathcal R}$ rep.} &  $m^2_\text{susy}$ & $m^2_\text{non-susy}$\\
		\hline\hline
		$\mathbf{d}$   & ($d\ge 2$)& $0 $			& $0$  \\
		$\mathbf{d-2}$ & ($d\ge 3$)& $\frac{3}{4}(d-1)(3d+7)$ & $\frac{5}{4}(d-1)(d+5)$ \\
		$\mathbf{d+2}$ & 		  & $\frac{3}{4}(d+1)(3d-7)$ & $\frac{5}{4}(d+1)(d-5)$ \\
		\hline
	\end{tabular}
	\caption{\small Masses for the supersymmetric and non-supersymmetric vacua, and their R-symmetry representation. For supersymmetric solutions all masses are above the BF bound. The non-supersymmetric solutions have tachyonic fluctuations below the BF bound for $d=1,2,3$.}
	\label{tab:masses}
\end{table}

The fluctuations in the $\mathbf{d}$ series are dual to Nambu-Goldstone bosons associated with broken symmetries. These operators only exist for $d\geq 2$. In the series $\mathbf{d-2}$ and $\mathbf{d+2}$ the masses in the non-supersymmetric solutions are decreased by $(d-1)^2$ and $(d+1)^2$, respectively, with respect to the supersymmetric solutions.
The series $\mathbf{d-2}$ with $d=3$ is a singlet fluctuation and has mass $m^2_{\rm susy}=24$ and $m^2_{\rm non-susy}=20$, in agreement with the $\delta y$ fluctuations of probe D7-branes to be discussed in the next section.
The series $\mathbf{d+2}$ with $d=1$ is a triplet with $m^2_{\rm susy}=-6$ and $m^2_{\rm non-susy}=-10$, in agreement with the Abelian D5-brane fluctuations discussed in Sec.~\ref{eq:D5-Abelian}.
The $\mathbf{d+2}$ series contains for $d=5$ fluctuations that are massless in the non-supersymmetric solutions and dual to marginal operators.

The masses in the supersymmetric backgrounds are all above the BF bound. In that case the configuration $\Phi_i=\frac{3}{2}\sigma_i$ is stable (as is $\Phi_i=0$). The masses obtained for the special case $N=2$ perfectly agree with the results on the spectrum of matter-coupled $F(4)$ gauged supergravity in \cite{Karndumri:2012vh}. This suggest that matter-coupled $F(4)$ gauged supergravity in six dimensions captures some modes associated to 5-brane modes in ten dimensions.

For the non-supersymmetric solutions we have modes violating the BF stability bound in the $\mathbf{d+2}$ series for $d=1, d=2$ and $d=3$, with masses $m^2=-10$, $m^2=-45/4$ and $m^2=-10$, respectively. This means that the $\Phi_i=\frac{3}{2}\sigma_i$ configuration, though it has lower potential than $\Phi_i=0$, still is not a stable configuration and the D5-branes do not form a stable non-commutative sphere configuration.
All tachyons below the BF bound break the $SU(2)$ R-symmetry. This can be related to the brane web perspective, where the $SU(2)$ symmetry corresponds to rotations in the directions transverse to the brane web. 
We will discuss this below. The stability of 7-branes will be discussed in Sec.~\ref{sec:probe-D7}.

The multiplets of even dimension ({\it i.e.} half-integer spin) in Table \ref{tab:masses} represent interaction modes between different stacks of branes. In fact, let us assume that $\Phi^i$ is in an irreducible representation of spin $s$ (and dimension $d=2s+1$) corresponding to a configuration with a single stack of branes. Following \eqref{eq:j-rep}, the irreducible representations in the matrices $j^i$ are $\mathbf{2s+1},\,\mathbf{2s-1},\dots,\mathbf{3}$, {\it i.e.} blocks of integer spin only. Thus, in order to have half-integer spin multiplets, $\Phi^i$ has to be in a reducible SU$(2)$ representation, with at least one block of integer spin and one block of half-integer spin.  In particular, this implies that the tachyonic modes in the $\mathbf{d+2}$ series for $d=2$, in the non-supersymmetric case, are related to interactions among stacks of branes. In contrast, any integer-spin mode can be generated also if $\Phi^i$ is in an irreducible $\mathrm{SU}(2)$ representation. 

Given an arbitrary representation of $\Phi^i$, there is always a tachyonic mode of mass $m^2=-10$, that belongs to the $\mathbf{d+2}$ series for $d=1$ or $d=3$. This implies that there is not actually a stable configuration with localized branes. One can thus wonder what the endpoint of the decay of a solution with localized sources may be. In order to address such question, let us observe that the tachyons mass $m^2=-10$ are also present in the Abelian analysis of the previous section (see the discussion around \eqref{eq:actionAbelian}). This means that the 5-branes can also move into the interior of $\Sigma$ without polarizing into 7-branes. A possible end point could be a configuration where the 5-branes are smeared over $\Sigma$.

\subsection{Brane interpretation}

The results of the previous section show that the brane configurations described by the non-supersymmetric AdS$_6$ solutions are unstable to processes where 5-branes change their position in the brane configuration while breaking the $SU(2)$ symmetry. This corresponds to moving out of the plane of the brane web. To develop intuition for this process we start by discussing analogous deformations in the supersymmetric seed solutions.
These deformations are not instabilities in the supersymmetric solutions, but our results indicate that similar processes are instabilities for the  non-supersymmetric partner solutions.

\begin{figure}
	\subfigure[][]{\label{fig:plus-inst-1}
	\begin{tikzpicture}
		\foreach \i in {-3/8,-1/8,1/8,3/8}{
			\draw (-5/3,\i) -- (5/3,\i);
			\draw (\i,-5/3) -- (\i,5/3);
			\draw[fill=black] (-5/3,\i) circle (1.5pt);
			\draw[fill=black] (5/3,\i) circle (1.5pt);
			\draw[fill=black] (\i,-5/3) circle (1.5pt);
			\draw[fill=black] (\i,5/3) circle (1.5pt);
		}	
	\end{tikzpicture}
	}\hskip 4mm
	\subfigure[][]{\label{fig:plus-inst-2}
	\begin{tikzpicture}
		\foreach \i in {2/3,1,4/3,5/3}{
			\draw[fill=black] (\i,0) ellipse (2pt and {20pt - 11*\i pt});
			\draw[fill=black] (-\i,0) ellipse (2pt and {20pt - 11*\i pt});
		}
		\foreach \i in {-3/8,-1/8,1/8,3/8}{
			\draw (-2/3,\i) -- (2/3,\i);
			\draw (\i,-5/3) -- (\i,5/3);
			\draw[fill=black] (\i,-5/3) circle (1.5pt);
			\draw[fill=black] (\i,5/3) circle (1.5pt);
			
		}
		\foreach \i in {-1/4,0,1/4}{
			\draw (-1,\i) -- (-2/3,\i);
			\draw (1,\i) -- (2/3,\i);
		}
	
		\foreach \i in {-1/8,1/8}{
			\draw (-4/3,\i) -- (-1,\i);
			\draw (1,\i) -- (4/3,\i);
		}
		\draw (-5/3,0) -- (-4/3,0);
		\draw (5/3,0) -- (4/3,0);		
	\end{tikzpicture}
	}\hskip 4mm
	\subfigure[][]{\label{fig:plus-inst-3}
	\begin{tikzpicture}
		\foreach \i in {2/3,1,4/3,5/3}{
			\draw[fill=black] (\i,0) ellipse (2pt and {20pt - 11*\i pt});
			\draw[fill=black] (-\i,0) ellipse (2pt and {20pt - 11*\i pt});
		}
		\foreach \i in {-3/8,-1/8,1/8,3/8}{
			\draw (\i,-5/3) -- (\i,5/3);

			\draw[fill=black] (\i,-5/3) circle (1.5pt);
			\draw[fill=black] (\i,5/3) circle (1.5pt);
			
		}
		\foreach \i in {-1/8,1/8}{
			\draw (-2/3,\i) -- (-3/8-0.08,\i);
			\draw (-3/8+0.08,\i) -- (3/8-0.08,\i);
			\draw (3/8+0.08,\i) -- (2/3,\i);
		}
		\foreach \i in {-3/8,3/8}{
			\draw (-2/3,\i) -- (2/3,\i);
		}
		\foreach \i in {-1/4,1/4}{
			\draw (-1,\i) -- (-2/3,\i);
			\draw (1,\i) -- (2/3,\i);
		}
	\end{tikzpicture}
	}\hskip 4mm
	\subfigure[][]{\label{fig:plus-inst-4}
	\begin{tikzpicture}
		\foreach \i in {-3/8,-1/8,1/8,3/8}{
			\draw[fill=black] (\i,-5/3) circle (1.5pt);
			\draw[fill=black] (\i,5/3) circle (1.5pt);
		}
		
		\draw[fill=black] (1/4,1/8) circle (1.5pt);
		\draw [thick,dashed] (1/4,1/8) -- (5/3,1/8);
		\draw[fill=black] (1/4,-1/8) circle (1.5pt);
		\draw [thick,dashed] (1/4,-1/8) -- (5/3,-1/8);
		\draw[fill=black] (-1/4,1/8) circle (1.5pt);
		\draw [thick,dashed] (-1/4,1/8) -- (-5/3,1/8);
		\draw[fill=black] (-1/4,-1/8) circle (1.5pt);
		\draw [thick,dashed] (-1/4,-1/8) -- (-5/3,-1/8);
		
		\draw (-1/8,5/3) -- (-1/8,1/8+3/8) -- (-1/8-3/8,1/8) -- (-1/8-3/8,-1/8) -- (-1/8,-1/8-3/8) -- (-1/8,-5/3);
		\draw (1/8,5/3) -- (1/8,1/8+3/8) -- (1/8+3/8,1/8) -- (1/8+3/8,-1/8) -- (1/8,-1/8-3/8) -- (1/8,-5/3);
		\draw (1/8,1/8+3/8) -- (-1/8,1/8+3/8);
		\draw (1/8,-1/8-3/8) -- (-1/8,-1/8-3/8);
		
		\draw (-3/8,5/3) -- (-3/8,1/8+5/8) -- (-3/8-5/8,1/8) -- (-3/8-5/8,-1/8) -- (-3/8,-1/8-5/8) -- (-3/8,-5/3);
		\draw (3/8,5/3) -- (3/8,1/8+5/8) -- (3/8+5/8,1/8) -- (3/8+5/8,-1/8) -- (3/8,-1/8-5/8) -- (3/8,-5/3);
		\draw (3/8,1/8+5/8) -- (-3/8,1/8+5/8);
		\draw (3/8,-1/8-5/8) -- (-3/8,-1/8-5/8);
	\end{tikzpicture}
	}
\caption{Transition from a brane web corresponding to a supergravity solution with four 5-brane poles (left) to a web corresponding to a solution with two poles and two punctures (right). 7-branes are shown as circles/ellipses; D5-branes extend horizontally and NS5-branes vertically.\label{fig:polarization-web}}
\end{figure}
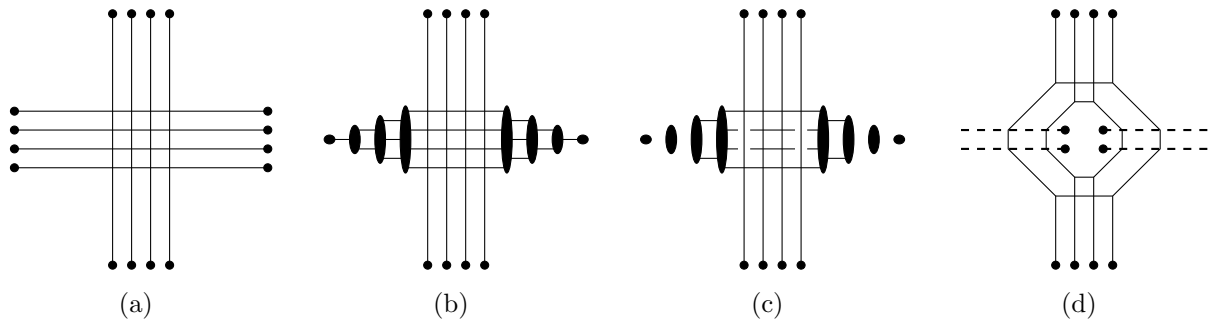

To illustrate the process we take an intersection of D5 and NS5 branes as an example (Fig.~\ref{fig:plus-inst-1}).
The SCFTs were called $+_{N,M}$ theories in \cite{Bergman:2018hin}.
The associated supergravity solution has 2 D5-brane poles and 2 NS5 poles, and is shown in Fig.~\ref{fig:plus-disc}. 
The semi-infinite D5 and NS5 branes can be terminated, respectively, on D7 and NS7 branes without changing the field theory engineered by the intersection. The 7-branes are indicated by the dots terminating the 5-branes in Fig.~\ref{fig:plus-inst-1}.
By rearranging the 7-branes one obtains an equivalent representation of the intersection in Fig.~\ref{fig:plus-inst-2}.
Since each 5-brane terminates on an individual 7-brane, there are no constraints from the $s$-rule \cite{Benini:2009gi}.

One can now move 5-brane segments stretching between 7-branes out of the plane of the brane web to infinity. In the supersymmetric context this is a relevant deformation which induces a flow onto the Higgs branch \cite{Aharony:1997bh}.
An example where 4 D5-brane segments on each side have been removed is shown in Fig.~\ref{fig:plus-inst-3}. Some D7-branes are left with no 5-branes attached and are irrelevant for the 5-brane junction, while the other D7-branes now have multiple 5-branes ending on them. As a result the junctions of D5 and NS5-branes in the center are constrained by the $s$-rule, leading to avoided intersections shown schematically as broken lines. 
One may move the 7-branes with 5-branes attached into the brane web by Hanany-Witten transitions, leading to  Fig.~\ref{fig:plus-inst-4} which is equivalent to Fig.~\ref{fig:plus-inst-3}.

The web in Fig.~\ref{fig:plus-inst-4} is a special case of the $+_{N,M,j,k}$ theories discussed in \cite{Chaney:2018gjc}. 
The supergravity solution corrsponding to Fig.~\ref{fig:plus-inst-4} has two NS5 poles and two D7 punctures in the interior of $\Sigma$ and is illustrated in Fig.~\ref{fig:plus-D7-disc}.
The geometry on $\Sigma$ precisely represents the structure of the brane web,  in the sense that coordinates labeling the faces of the brane web can be identified with a particular set of coordinates on $\Sigma$ \cite{Uhlemann:2020bek}.
The positions of the 7-brane punctures on $\Sigma$ encode which face of the brane web the 7-branes are located in.

The results of the previous section can be matched with this picture as follows. In the supersymmetric solutions we can add probe 5-branes to a corresponding 5-brane pole, where they are stable. This corresponds to increasing the 5-brane charge at the pole. We can also displace a group of 5-branes into the interior, by forming a non-commutative $S^2$ and polarizing them into 7-branes. 
Multiple 5-brane groups can be treated analogously.
These are the configurations studied in terms of the non-Abelian D5-brane action. In the brane web they describe the transition in Fig.~\ref{fig:polarization-web}, with the decomposition of $\sigma$ in (\ref{eq:sigma-decomp}) encoding how the 5-branes are grouped to form 7-branes.

In the non-supersymmetric solutions, probe 5-branes at the poles are unstable, signaling that the configuration of branes represented by the pole tends to disintegrate. For an Abelian 5-brane the instability amounts to moving individual 5-branes into $\Sigma$, which breaks the $SU(2)$ R-symmetry.
One may wonder whether this indicates that the 5-branes collectively move into $\Sigma$ to polarize into 7-branes, similar to the process discussed above for supersymmetric configurations.
Our results indicate that, though solutions with 5-branes polarized into 7-branes exist also in the non-supersymmetric backgrounds, and have lower potential than individual unpolarized 5-branes, they are again unstable. We found tachyons below the BF bound which transform non-trivially under $SU(2)$, signaling again that the configuration spreads out of the plane of the brane web.
The dynamical question whether the decay of a pole actually goes through (unstable) configurations with 5-branes polarized into 7-branes is not answered by our analysis. But the results either way show that non-supersymmetric solutions with 5-brane poles are generally unstable.

\section{Stability of probe 7-branes}\label{sec:probe-D7}

In the previous section we saw that 5-brane poles are unstable in the non-super\-symmetric solutions, with the 5-branes tending to move into the interior of $\Sigma$, and that configurations where the 5-branes are polarized into 7-branes near the boundary of $\Sigma$ are unstable as well.
In this section we study 7-branes more generally in the interior of $\Sigma$.
One might hope that configurations where all 5-branes are polarized into 7-branes deep in the interior of $\Sigma$, corresponding to `maximal Higgsing', could be stable.
We consider a sample of solutions, which includes solutions with fully backreacted 7-branes, and study embeddings of probe D7-branes.
We find a universal set of fluctuations, with identical mass spectrum regardless of the choice of background solution and the position of the 7-branes in $\Sigma$.
We find no instabilities among the $SU(2)_R$ singlet fluctuations, but tachyons below the BF bound among fluctuations that transform non-trivially under $SU(2)_R$. This shows that 7-branes in the sample solutions are unstable, and the universality of the spectrum suggests that solutions with 7-brane punctures are generally unstable.
The fluctuation spectrum smoothly connects to the non-Abelian D5-brane analysis of Sec.~\ref{sec:D5-nonAbelian}.

\subsection{Probe D7-branes}

The general action for a probe D7-brane, with $\cF=F-B_2$ and dilaton convention $\tau=\chi+ie^{-2\phi}$, reads
\begin{align}\label{eq:S-D7-gen}
	S_{D7}&=-T_7 \int d^8\xi e^{-2\phi}\sqrt{-\det\left(g_{ab}+\cF_{ab}\right)}
	+T_7\int e^\cF\wedge\sum_q C_{(q)}~.
\end{align}
The equations of motion in the $AdS_6\times S^2\times \Sigma$ backgrounds were discussed in detail in app.~B of \cite{Gutperle:2018vdd}, and we will borrow some results from there.
The relevant RR fields are
\begin{align}\label{eq:C6C8exp}
	C_{(6)}&= C_6 \,{\rm vol}_{AdS_6} ~,
	&
	C_{(8)}&=C_8 \, {\rm vol}_{AdS_6} \wedge {\rm vol}_{S^2} ~.
\end{align}
The worldvolume gauge field is restricted to $S^2$, such that $\cF=\mf \vol_{S^2}$ with $\mf=F-\Re\cC$.
We allow the embedding on $\Sigma$ to depend on the (Poincar\'e) AdS$_6$ radial coordinate $z$, such that the embedding is characterized by a function $w(z)$.
With $\tilde f_6$, $\tilde f_2$ and $\tilde \rho^2$ denoting the string-frame metric functions, the action becomes
\begin{align}\label{eq:S-D7-2}
	S_{\rm D7}&=\tilde T_7
	\int \frac{dz}{z^6} \left[-\: e^{-2\phi}\tilde f_6^5\sqrt{\tilde f_6^2+4\tilde\rho^2 z^2|w^\prime|^2} \sqrt{\tilde f_2^4+\mf^2}
	+C_8+C_6\mf\right],
\end{align}
where $\tilde T_7=T_7 \Vol_{\RR^{1,4}}\Vol_{S^2}$.
The resulting equation of motion for $w(z)$ is
\begin{align}\label{eq:D7-EOM}
	z^6\frac{d}{dz}\!\left[	\frac{2e^{-2\phi}\tilde f_6^5\tilde\rho^2\bar w^\prime\sqrt{\tilde f_2^4+\mf^2}}{z^{4}\sqrt{\tilde f_6^2+4\tilde\rho^2z^2|w^\prime|^2}}\right]
	-\partial_w\!\left[ e^{-2\phi}\tilde f_6^5\sqrt{\tilde f_6^2+4\tilde\rho^2 z^2|w^\prime|^2} \sqrt{\tilde f_2^4+\mf^2}
	-C_8-C_6\mf\right]&=0\,.
\end{align}	
For $w={\rm const}$ this reduces to eq.~(B.8) in \cite{Gutperle:2018vdd}.
We will be interested in small fluctuations around the $w={\rm const}$ embeddings.

\paragraph{Conformal D7-branes:}

For probe D7-branes preserving the AdS$_6$ symmetries (with $w={\rm const}$), we expect the embeddings in the supersymmetric backgrounds to also be solutions in the non-supersymmetric backgrounds:
The supersymmetric AdS$_6$ solutions with a probe D7-brane can be understood to arise from solutions with fully backreacted D7-branes in the limit where the monodromy at the puncture becomes infinitesimal.
Each supersymmetric solution with a backreacted D7-brane has a non-supersymmetric avatar, and taking the probe limit for the D7-brane puncture in this non-supersymmetric avatar leads to a non-supersymmetric AdS$_6$ solution with a probe D7-brane.

When $w={\rm const}$ the embedding preserves the AdS$_6$ isometries. 
The action reduces to
\begin{align}\label{eq:S-D7}
	\frac{S_{D7}}{T_7 \Vol_{AdS_6}\Vol_{S^2}}&=
	-\: e^{-2\phi}\tilde f_6^6 \sqrt{\tilde f_2^4+(F-\Re\cC)^2}
	+C_8+C_6 (F-\Re\cC)~.
\end{align}
The equation of motion for the worldvolume gauge field $A$ with $F=dA$ determines $F$ in terms of a constant $F_0$ and the location of the 7-brane on $\Sigma$ as follows,
\begin{align}\label{eq:eom-mf}
	F-\Re\cC&=\frac{(C_6+F_0) \tilde f_2^2}{\sqrt{e^{-4\phi}\tilde f_6^{12}-(C_6+F_0)^2}}~.
\end{align}
The equation of motion for the embedding is
\begin{align}
	\partial_w\Big[e^{-2\phi}\tilde f_6^6\sqrt{\tilde f_2^4+(F-\Re\cC)^2}-C_8-C_6(F-\Re\cC)\Big]&=0~,
\end{align}
where the derivative is to be evaluated before substituting the solution for $F$, i.e.\ $F$ is defined intrinsically on the worldvolume and does not depend on $w$.
Evaluating the derivative more explicitly, with the expressions for the 6- and 8-form fields of \cite{Gutperle:2018vdd}, leads to
\begin{align}\label{eq:eom-embedding}
	\partial_w\left[e^{-2\phi}\tilde f_6^6\sqrt{\tilde f_2^4+(F-\Re\cC)^2}\right]-i\tilde f_6^6\tilde f_2^2\partial_w\chi-(F-\Re\cC)\partial_w C_6&=0~.
\end{align}
This is one complex equation for one complex number, with one real free parameter in $F_0$.
We thus generically have a one-parameter family of solutions -- a curve in $\Sigma$.

\paragraph{\boldmath{$SU(2)$} singlet fluctuations:}

We start the discussion of fluctuations with the $SU(2)$ singlets, which are technically simpler.
We hold $\int_{S^2}F$ fixed while allowing the D7-brane to move on $\Sigma$, so that $w$ depends on $z$.
Expanding (\ref{eq:D7-EOM}) for small fluctuations around $w=w_0$ with fixed $F$ and keeping terms up to linear order leads to
\begin{align}\label{eq:fluct-eom}
	z^6\frac{d}{dz}(z^{-4}\,\bar w^\prime)
	-\cZ&=0~,
\end{align}
where
\begin{align}
	\cZ&\equiv \frac{\partial_w\left[e^{-2\phi}\tilde f_6^6\sqrt{\tilde f_2^4+(F-\Re\cC)^2}\right]-i\tilde f_6^6\tilde f_2^2\partial_w\chi-(F-\Re\cC)\partial_w C_6}{2e^{-2\phi}\tilde f_6^4\tilde\rho^2\sqrt{\tilde f_2^4+(F-\Re\cC)^2}}~.
\end{align}
The 8-form potential $C_{8}$ has been replaced using the relations in app.~C of \cite{Gutperle:2018vdd}. 

The mass matrix for the two real fluctuations in the real and imaginary parts of $w(z)$ is obtained from the second term in (\ref{eq:fluct-eom}).
Around a point $w_0$ where the zeroth-order equation of motion with $w=w_0$ constant is satisfied, we have
\begin{align}
	\mathcal Z &=\mathcal Z_w (w-w_0)+\mathcal Z_{\bar w}(\bar w-\bar w_0)+\ldots
\end{align}
Setting $w=x+iy$ the mass matrix for the fluctuations $\delta x$, $\delta y$ is given by
\begin{align}\label{eq:masss-matrix}
	M_{D7}&=\begin{pmatrix} \Re(\cZ_w+\cZ_{\bar w}) & -\Im(\cZ_w-\cZ_{\bar w})\\ -\Im(\cZ_w+\cZ_{\bar w}) & -\Re(\cZ_w-\cZ_{\bar w})  \end{pmatrix},
\end{align}
and the (squared) masses are given by the eigenvalues. Concrete examples will be discussed below.

\paragraph{Non-singlet fluctuations:}
We now spell out the ansatz for more general fluctuations, while being brief on the technical details.
For more general fluctuations which still respect 5d Poincar\'e symmetry, the embedding can in addition depend on the $S^2$ and new fluctuations are allowed in the gauge field.
We choose coordinates such that $ds_{S^2}^2=d\theta^2+\sin^2\!\theta\,d\phi^2$. The fluctuations come in representations of $SO(3)$ labeled by $\ell$, $m$. We choose the $m=0$ modes for simplicity, since the masses are identical within multiplets.
Then $w=w(z,\theta)$. We choose a gauge such that $A_z=0$, and then have gauge field components 
\begin{align}
	A&= -\hat F\cos\theta d\phi + A_\theta(z,\theta)d\theta+ A_\phi(z,\theta)d\phi~,
\end{align}
where we denote by the constant $\hat F$ the flux resulting from (\ref{eq:eom-mf}).
The induced (string frame) metric and worldvolume two-form then become
\begin{align}
	g&=\frac{\tilde f_6^2}{z^2}\,g_{\RR^{1,4}} + 
	\left(\tilde f_6^2+4z^2\tilde\rho^2 |\partial_z w|^2\right)\frac{dz^2}{z^2}
	+4\tilde\rho^2\left(\partial_z w\partial_\theta \bar w+\partial_z\bar w\partial_\theta w\right)dz\, d\theta
	\nonumber\\ &\hphantom{=}\,
	 +\left(\tilde f_2^2 + 4\tilde\rho^2 |\partial_\theta w|^2\right)d\theta^2+ \tilde f_2^2\sin^2\!\theta\, d\phi^2~,
	\nonumber\\
	\cF&=\partial_z  A_\theta\, dz \wedge d\theta + \partial_z A_\phi\,dz\wedge d\phi + ((\hat F-\Re \cC)\sin\theta+\partial_\theta A_\phi)\, d\theta \wedge d\phi~.
\end{align}
With these expressions the D7-brane action is given by
\begin{align}
	S_{D7}&=-T_7 \int d^8\xi e^{-2\phi}\sqrt{-\det\left(g_{ab}+\cF_{ab}\right)}
	+T_7\int C_8+\cF \wedge C_6~.
\end{align}
For a given embedding of a D7-brane at $w=w_0$, we have 	to expand this action to quadratic order in the fluctuations,
while expanding $w$, $A_\theta$ and $A_\phi$ in (vector) spherical harmonics.
Separating the fluctuations in the embedding into real and imaginary parts, we use
\begin{align}
	w(z,\theta)&= w_0+\sum_{\ell }\left(\delta x_{\ell}(z) +i\delta y_{\ell}(z)\right)Y_{\ell 0}~,
	\nonumber\\
	A_\theta(z,\theta)&=\sum_\ell A_\theta^\ell(z)\Psi_{\ell 0}~,
\hskip 14mm
	A_\phi(z,\theta)=-\cos\theta\, \hat F +  \sum_\ell A_\phi^\ell(z)\Phi_{\ell 0}~,
\end{align}
where $Y_{\ell m}$ are the scalar spherical harmonics and $\Psi_{\ell m}$, $\Phi_{\ell m}$ the transverse vector spherical harmonics.
With this ansatz and the expressions for the background supergravity solution we can then perform the expansion and read off the mass matrix.

\subsection{\texorpdfstring{$+_{N,M}$}{+[N,M]} solution}\label{sec:plus}

\begin{figure}
\centering
	\subfigure[][]{\label{fig:plus-web}
		\begin{tikzpicture}[scale=0.6]
			\foreach \i in {-1.2,-0.8,-0.4,0,0.4,0.8,1.2}{
				\draw[thick] (\i,-3) -- (\i,3);
			}
			\foreach \j in {-0.6,-0.2,0.2,0.6}{
				\draw[thick] (3.3,\j) -- (-3.3,\j);
			}
			\node at (0,-3.5) {$M$\,NS5};
			\node at (-4.2,0) {$N$\,D5};
			\node at (0,-3.7) {};
			
		\end{tikzpicture}
	}	\hskip 10mm
	\subfigure[][]{\label{fig:plus-disc}
		\begin{tikzpicture}
			\draw[fill=lightgray,opacity=0.2] (0,0) circle (1.5);
			\draw[thick] (0,0) circle (1.5);
			\draw[thick] (1.4,0) -- (1.6,0);
			\draw[thick] (-1.4,0) -- (-1.6,0);
			\draw[thick] (0,1.4) -- (0,1.6);
			\draw[thick] (0,-1.4) -- (0,-1.6);
			\node at (0.7,0.7) {$\mathbf \Sigma$};
			\node at (2.2,0) {$N$ D5};
			\node at (-2.2,0) {$N$ D5};
			\node at (0,1.85) {$M$ NS5};
			\node at (0,-1.85) {$M$ NS5};
			\draw[thick,blue, dashed] (-1.5,0) -- (1.5,0);
			\node at (-0.75,-0.25) {D7};
		\end{tikzpicture}
	}
	\caption{Left: Brane web for the $+_{N,M}$ theory. Right: Schematic representation of the associated supergravity solution with $\Sigma$ mapped to a disc. The dashed blue line indicates the curve along which probe D7-branes can be embedded.\label{fig:plus}}
\end{figure}
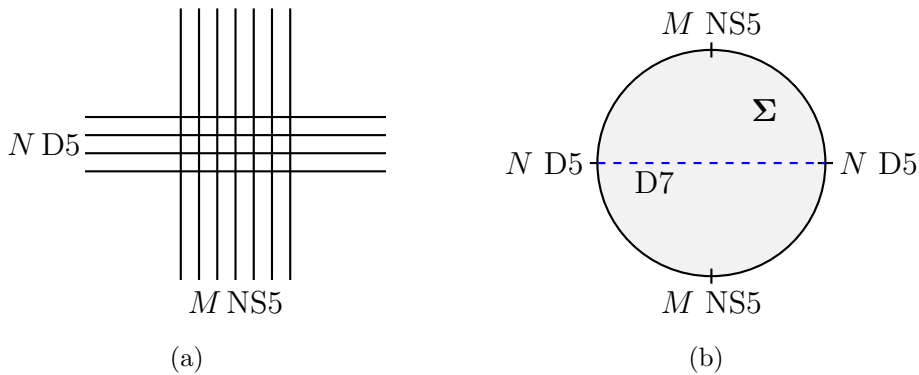

As a first example we take the $+_{N,M}$ solutions, which describe intersections of $N$ D5 and $M$ NS5 branes (Fig.~\ref{fig:plus-web}).
The field theories corresponding to the supersymmetric solutions were already discussed in \cite{Aharony:1997ju}; they admit deformations to quiver gauge theories with $M-1$ $SU(N)$ nodes, terminated by $N$ fundamental hypermultiplets at each end.
The expressions for $\cA_\pm$ and $\cG$ can be taken from \cite{Uhlemann:2020bek}.
It is convenient, though, to perform an $SL(2,\RR)$ transformation of the upper half plane to map the poles to $w\in\lbrace -1,0,1,\infty\rbrace$.
The functions $\cA_\pm$ and $\cG$ are then
\begin{align}
	\cA_\pm&=\frac{3}{8\pi}\left[iN\ln w+\pi N \pm M\ln(w-1)\mp M \ln(w+1)\right]~,
	\nonumber\\
	\cG&=\frac{9NM}{8\pi^2}\left[D_2\left(\frac{w-1}{w+1}\right)+D_2\left(\frac{w+1}{1-w}\right)\right]~,
\end{align}
where $D_2$ is the Bloch-Wigner function defined in  (\ref{eq:GBloch}).
The D5-brane poles are at $w=0$ and $w=\infty$; the NS5-brane poles are at $w=\pm 1$.
The schematic form of the solution after mapping the upper half plane to the disc is shown in Fig.~\ref{fig:plus-disc}.

The real part of the probe D7-brane equation of motion (\ref{eq:eom-embedding}) vanishes along the imaginary axis for symmetry reasons.
The imaginary part of (\ref{eq:eom-embedding}) determines the constant $F_0$.
The conformal embeddings are
\begin{align}
	w=i\xi
\end{align}
with $\xi\in \RR^+$.
For the supersymmetric solutions one can get an explicit expression for $F$ at each point from the BPS conditions. 
For the non-supersymmetric solutions we numerically solved the equation of motion for a sample of points on the imaginary axis.
The results for the supersymmetric and the non-supersymmetric backgrounds are both consistent with
\begin{align}
	F&=\frac{2}{3}\Im(\cA_++\bar \cA_-)=\frac{iM}{2\pi}\ln\left(\frac{\xi-i}{\xi+i}\right)~.
\end{align}
The right hand side is proportional to the D1 charge $N_{\rm D1}$ defined in \cite{Uhlemann:2020bek}. This is consistent with the identification of points on $\Sigma$ with faces in the brane web proposed there.

For each probe D7-brane	for given $\xi$ we can compute the mass matrix (\ref{eq:masss-matrix}) for $SU(2)$ singlet fluctuations.
This mass matrix takes the same form for all probe D7-branes on the imaginary axis.
For the susy backgrounds we find
\begin{align}
	M^{\rm susy}_{\rm D7}(\xi)&=\begin{pmatrix} -4 & 0 \\ 0 & 24\end{pmatrix}~.
\end{align}
The fluctuations $\delta x$, $\delta y$ in $w=x+iy$ are mass eigenstates.
The first entry on the diagonal corresponds to $\delta x$.
From $m^2=\Delta(\Delta-5)$ we conclude that $\delta x$ fluctuations are dual to relevant operators with scaling dimension $\Delta=4$, consistent with fermion mass terms.
The mass is the same as for the $\delta x$ fluctuations of D5 branes in (\ref{eq:D5-mass}).
The fluctuations $\delta y$ are dual to a irrelevant operators with $\Delta=8$. 
The mass is identical to that of the non-Abelian singlet D5-brane fluctuation discussed in Sec.~\ref{sec:D5-nonAbelian}.
The $\delta x$ and $\delta y$ masses are both above the BF bound.
The mass matrix for fluctuations in non-supersymmetric backgrounds is
\begin{align}
	M^{\rm non-susy}_{\rm D7}(\xi)&=\begin{pmatrix} 0 & 0 \\ 0 & 20 \end{pmatrix}~.
\end{align}
The mass matrix is again identical for all probe D7 branes along the imaginary axis, and $\delta x$, $\delta y$ are mass eigenstates.
The fluctuations $\delta x$ are now massless, corresponding to marginal operators with $\Delta=5$.
There are no supersymmetric marginal operators in 5d, and we have not imposed supersymmetry.
Both mass eigenvalues are above the BF bound.
The shift in $m^2$ from supersymmetric to non-supersymmetric solutions is $+4$ for the $\delta x$ fluctuation and $-4$ for the $\delta y$ fluctuation -- the same as for the D5 fluctuations in (\ref{eq:D5-mass}).

For the supersymmetric solutions one can understand the results from the brane web perspective: Vertical movements of the D7-branes in the web correspond to turning on a flavor mass term. 
Vertical moves in the brane web correspond in the supergravity solution to $\delta x$ fluctuations (this follows from the identification of the coordinate on $\Sigma$ with coordinates on the brane web in \cite{Uhlemann:2020bek}).
We thus find a consistent picture for flavor mass deformations between supergravity solutions and brane webs.
For non-supersymmetric solutions we do not see a reason to expect a particular mass, or the fluctuation masses to be independent of the location of the D7 probe. The latter is nevertheless the case.

For the fluctuations that transform non-trivially under $SU(2)$ we find again that the spectrum is independent of the position of the probe D7 brane on the imaginary axis. The $\delta x_\ell$ and $\delta A_\theta^\ell$ fluctuations both decouple, and we find
\begin{align}\label{eq:D7-SU2-mass0}
	\delta x_\ell:&& m^2_{\rm susy}&=9\ell(\ell+1)-4 &  m^2_{\rm non-susy}&=5\ell(\ell+1)~,
	\nonumber\\
	\delta A_\theta^\ell:&& m^2_{\rm susy}&=0 & m^2_{\rm non-susy}&=0~.
\end{align}
The masses are all above the BF bound. The $\delta A_\theta^\ell$ fluctuations match the $\mathbf{d}$ series in Table \ref{tab:masses}.
The $\delta y$ and $\delta A_\phi^\ell$ fluctuations mix, and we find the mass matrices
\begin{align}\label{eq:D7-SU2-mass}
	M^2_{\rm susy}&=3\begin{pmatrix} 8+3\ell(\ell+1) & 8\sqrt{\ell(\ell+1)}\\ 8 \sqrt{\ell(\ell+1)} & 3\ell(\ell+1)\end{pmatrix},
	\nonumber\\
	M^2_{\rm non-susy}&=5\begin{pmatrix} 4+\ell(\ell+1) & 4\sqrt{\ell(\ell+1)}\\ 4 \sqrt{\ell(\ell+1)} & \ell(\ell+1)\end{pmatrix},
\end{align}
where we used $(\delta y, \delta A_\phi)$ as basis.
The eigenvalues are given by
\begin{align}
	m^2_{\rm susy}&=\left\lbrace 3\ell(3\ell-5)\,,\, 3\ell(3\ell+11)+24\right\rbrace,
	\nonumber\\
	m^2_{\rm non-susy}&=\left\lbrace 5\ell(\ell-3)\,,\, 5\ell(\ell+5)+20\right\rbrace.
\end{align}
The masses in the supersymmetric solutions are all above the BF bound and do not signal instabilities.
In the non-supersymmetric solutions the modes with $\ell=1$ and $\ell=2$ have $m^2=-10$, which is below the BF bound and signals an instability.
Since the modes leading to this instability are a linear combination of $\delta y$ and $\delta A_\phi$, the 7-branes like to move along the imaginary axis while deforming the spherically symmetric worldvolume gauge field configuration.
This is a natural continuation of the stability discussed for 5-branes in the previous section to 7-branes in the interior of $\Sigma$.

The mass spectrum smoothly connects to the non-Abelian D5-brane results in Table \ref{tab:masses} for the case where the $\sigma$ matrices form a large irreducible representation of $SU(2)$.
To translate the $SU(2)$ representations in Table \ref{tab:masses} to the notation used here, we note that the $\mathbf{d-2}$ series corresponds to $d=2\ell+3$, while the $\mathbf{d+2}$ series corresponds to $d=2\ell-1$.
With these identifications the $SU(2)$ representations match and so do the masses. The multiplets of even dimension $d$ of Table \ref{tab:masses} cannot be observed in the probe D$7$-brane spectrum. We recall that the presence of the even-dimensional polarization modes relies on interactions between different brane stacks, which explains why they can not be seen in the Abelian D7-brane analysis.

\subsection{\texorpdfstring{$T_N$}{T[N]} solution}

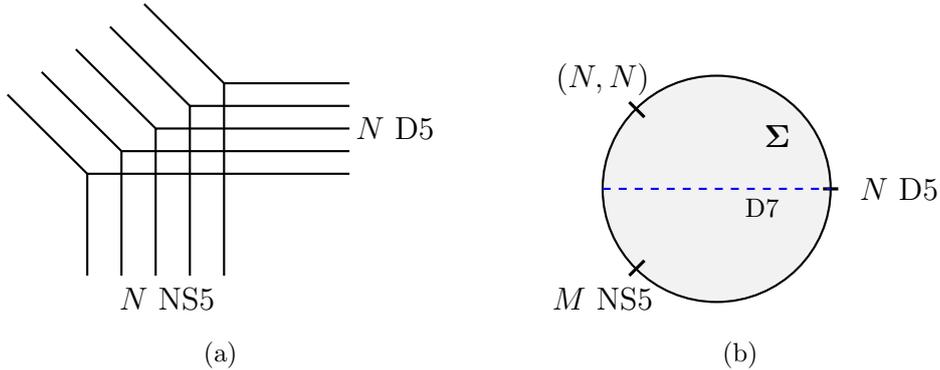
\begin{figure}
	\centering
	\subfigure[][]{\label{fig:TN-web}
		\begin{tikzpicture}[xscale=-0.6,yscale=0.6]
			\draw[thick] (-4,0.75) -- (-0.5,0.75) -- (-0.5,-3);
			\draw[thick] (-0.5,0.75) -- +(1.75,1.75);
			
			\draw[thick] (-4,0.25) -- (0.25,0.25) -- (0.25,-3);
			\draw[thick] (0.25,0.25) -- +(1.75,1.75);
			
			\draw[thick] (-4,-0.25) -- (1.0,-0.25) -- (1.0,-3);
			\draw[thick] (1.0,-0.25) -- +(1.75,1.75);
			
			\draw[thick] (-4,-0.75) -- (1.75,-0.75) -- (1.75,-3);
			\draw[thick] (1.75,-0.75) -- +(1.75,1.75);
			
			\draw[thick] (-4,1.25) -- (-1.25,1.25) -- (-1.25,-3);
			\draw[thick] (-1.25,1.25) -- +(1.75,1.75);
			
			\node at (0,-3.5) {$N$ NS5};
			\node at (-5,0.25) {$N$ D5};
		\end{tikzpicture}
	}\hskip 10mm
	\subfigure[][]{\label{fig:TN-disc}
		\begin{tikzpicture}
			\draw[fill=lightgray,opacity=0.2] (0,0) circle (1.5);
			\draw[thick] (0,0) circle (1.5);
			\draw[very thick] (1.4,0) -- (1.6,0);
			\draw[very thick] (-0.95,0.95) -- (-1.15,1.15);
			\draw[very thick] (-0.95,-0.95) -- (-1.15,-1.15);
			\node at (0.8,0.7) {$\mathbf \Sigma$};
			\node at (2.4,0) {$N$ D5};
			\node at (-1.5,1.45)  {$(N,N)$};
			\node at (-1.5,-1.45) {$M$ NS5};
			\draw[thick,blue,dashed] (-1.5,0) -- (1.5,0);
			\node at (0.6,-0.25) {\footnotesize D7};
		\end{tikzpicture}
	}
\caption{Left: brane web for the 5d $T_N$ theories. Right: schematic form of the supergravity solution; the dashed blue line is the curve along which probe D7-branes can be embedded.\label{fig:TN}}
\end{figure}

The next example is the $T_N$ solution, corresponding to a junction of $N$ D5, $N$ NS5 and $N$ $(1,1)$ 5-branes (Fig.\ref{fig:TN-web}).
The field theories corresponding to the supersymmetric solutions were discussed in \cite{Benini:2009gi} and admit quiver gauge theory deformations of the form $[2]-SU(2)-SU(3)-\ldots - SU(N-1)-[N]$.
For the $T_N$ solution,
\begin{align}
	\cA_\pm&=\frac{3N}{8\pi} \left[\pm \ln(w-1)+i\ln(2w)+(\mp 1-i)\ln(w+1)\right]~,
	\nonumber\\
	\cG_{T_N}&=\frac{9}{8\pi^2}N^2D\left(\frac{2w}{w+1}\right)~.
\end{align}
The D5 pole is at $w=0$, the NS5 pole at $w=1$ and the $(1,1)$ 5-brane pole at $w=-1$.
The schematic form of the solution is illustrated in Fig.~\ref{fig:TN-disc}.

The solutions for probe D7-branes in this coordinate take a similar form as the probe D7-brane embeddings in the $+_{N,M}$ solution,
\begin{align}
	w&=i\xi~, & F&=\frac{2}{3}\Im(\cA_++\bar \cA_-)=\frac{iN}{2\pi}\ln\left(\frac{\xi-i}{\xi+i}\right)~,
\end{align}
with $\xi\in\RR^+$.
This can be understood as follows: 
One can $SL(2,\RR)$ transform the $T_N$ solution to a solution with $(2,0)$, $(-1,1)$, $(-1,-1)$ 5-branes (this is the $Y_N$ theory of \cite{Bergman:2018hin}). 
The $SL(2,\RR)$ transformation has $c=0$ and $a=-1/b=1/d=\sqrt{2}$, and maps D5-branes into D5-branes and D7-branes into D7-branes.
The transformed solution has a $\ZZ_2$ symmetry with the imaginary axis as fixed line, similarly to the $+_{N,M}$ solution.
The mass matrices are identical as well:
\begin{align}\label{eq:D7-mass-TN}
	M^{\rm susy}_{\rm D7}(\xi)&=\begin{pmatrix} -4 & 0 \\ 0 & 24\end{pmatrix}~,
	&
	M^{\rm non-susy}_{\rm D7}(\xi)&=\begin{pmatrix} 0 & 0 \\ 0 & 20 \end{pmatrix}~.
\end{align}
This confirms the picture for mass deformations in the supersymmetric solutions and shows that both fluctuations have masses above the BF bound in the supersymmetric and non-supersymmetric solutions.
For the fluctuations transforming non-trivially under $SU(2)$ we find the same masses as in (\ref{eq:D7-SU2-mass0}) for the $\delta x_\ell$ and $\delta A_\theta^\ell$ fluctuations and the mass matrices (\ref{eq:D7-SU2-mass}) for $(\delta y, \delta A_\phi)$ fluctuations -- for all D7-branes along the imaginary axis.

The mass matrices for the D7-brane fluctuations are identical between the $+_{N,M}$ and $T_N$ solutions.
This suggests that they may, in fact, be universal for all solutions. As remarked above, this can be understood from the brane web picture for the supersymmetric solutions. The results here indicate that the same holds for the non-supersymmetric solutions. We will discuss one further example which includes fully backreacted 7-brane punctures to further support this conclusion.

\subsection{\texorpdfstring{$+_{N,M,j,k}$}{+[N,M,j,k]} solution}\label{sec:plus-D7}

\begin{figure}
	\centering
	\subfigure[][]{\label{fig:plus-D7-web}
		\begin{tikzpicture}[scale=0.7]
			\draw[dashed] (-5,0) -- (-0.75,0);
			\node at (-4.7,-0.4) {\footnotesize $T^4$};
			\draw[fill] (-0.75,0) circle (1.5pt);
			\draw[dashed] (5,0) -- (0.75,0);
			\node at (4.7,-0.4) {\footnotesize $T^2$};
			\draw[fill]  (0.75,0) circle (1.5pt) ;
			
			\draw (1.25,0) -- +(-0.25,0.25) -- (-1,0.25) -- (-1.5,0) -- (-1,-0.25) -- (1.0,-0.25) -- (1.25,0);
			\draw (2.25,0) -- +(-0.75,0.75) -- (-0.5,0.75) -- (-1,0.25);
			\draw (2.25,0) -- +(-0.75,-0.75) -- (-0.5,-0.75) -- (-1,-0.25);
			\draw (3.25,0) -- +(-1.25,1.25) -- (-1.5,1.25);
			\draw (3.25,0) -- +(-1.25,-1.25) -- (-1.5,-1.25);
			\draw (4.25,0) -- +(-1.75,1.75) -- (-1.0,1.75) -- (-1.5,1.25) -- (-4.0,0);
			\draw (4.25,0) -- +(-1.75,-1.75) -- (-1,-1.75) -- (-1.5,-1.25) -- (-4.0,0);
			
			\draw[fill] (2.5,1.75) -- +(0,0.5) circle (1.3pt);
			\draw[fill] (2.5,-1.75) -- +(0,-0.5) circle (1.3pt);
			
			\draw[fill] (2.0,1.25) -- +(0,1.0) circle (1.3pt);
			\draw[fill] (2.0,-1.25) -- +(0,-1.0) circle (1.3pt);
			
			\draw[fill] (1.5,0.75) -- +(0,1.5)  circle (1.3pt);
			\draw[fill] (1.5,-0.75) -- +(0,-1.5) circle (1.3pt);
			\draw[fill] (1,0.25) -- +(0,2) circle (1.3pt);
			\draw[fill] (1,-0.25) -- +(0,-2) circle (1.3pt);
			\draw[fill] (-0.5,0.75) -- +(0,1.5) circle (1.3pt);
			\draw[fill] (-0.5,-0.75) -- +(0,-1.5) circle (1.3pt);
			\draw[fill] (-1.0,1.75) -- +(0,0.5) circle (1.3pt);
			\draw[fill] (-1.0,-1.75) -- +(0,-0.5) circle (1.3pt);
			
			\draw[fill] (0,2.25) circle (1.3pt) -- (0,-2.25) circle (1.3pt);
			\draw[fill] (0.5,2.25) circle (1.3pt) -- (0.5,-2.25) circle (1.3pt);
			
			\draw[fill,white] (0,-3.16) circle (2pt);
		\end{tikzpicture}
	}\hskip 10mm
	\subfigure[][]{\label{fig:plus-D7-disc}
		\begin{tikzpicture}
			\draw[fill=lightgray,opacity=0.2] (0,0) circle (1.5);
			\draw[thick] (0,0) circle (1.5);
			\draw[thick] (0,1.4) -- (0,1.6);
			\draw[thick] (0,-1.4) -- (0,-1.6);
			\node at (0.7,0.8) {$\mathbf \Sigma$};
			\node at (0,1.87) {\small $M$ NS5};
			\node at (0,-1.92) {\small $M$ NS5};
			
			\draw[thick,blue] (-1,0) -- (0.5,0);
			
			\draw[fill=black] (0.5,0) circle (0.05);
			\draw[fill=black] (-1.0,0) circle (0.05);
			\draw[thick,dashed, black] (0.5,0) -- (1.5,0);
			\draw[thick,dashed, black] (-1.0,0) -- (-1.5,0);
			\node at (0.55,-0.3) {\footnotesize $T^k$};
			\node at (-0.95,-0.3) {\footnotesize $T^j$};
			\node at (0,0.25) {\footnotesize D7};
		\end{tikzpicture}
	}
\caption{Left: sample brane web for the $+_{N,M,j,k}$ theories. Right: disc representation of the supergravity solution with the branch cuts shown as black dashed lines; probe D7-branes can be embedded along the solid blue line and along the branch cuts.\label{fig:plus-D7}}
\end{figure}
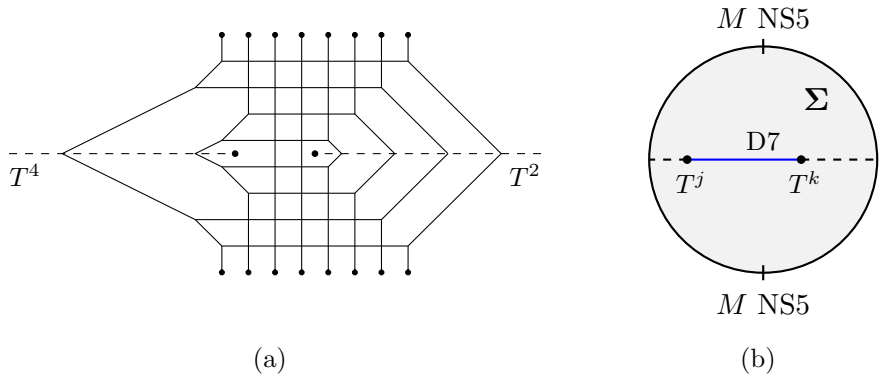

As a last example we discuss a family of solutions with backreacted D7-brane punctures.
Namely, the $+_{N,M,j,k}$ solutions discussed initially in \cite{Chaney:2018gjc}. 
They describe constrained junctions of $N$ D5 and $M$ NS5 branes, where the D5-branes terminate in $j$ groups on common D7-branes on one side and in $k$ groups on the other side (Fig.~\ref{fig:plus-D7-web} and \ref{fig:plus-inst-3}, \ref{fig:plus-inst-4}).

The general expressions for the functions $\cA_\pm$ can be taken from \cite{Uhlemann:2019lge}, which for this particular solution leads to
\begin{align}\label{eq:cA-plus-D7}
	\cA_\pm&=\frac{3M}{8\pi}\sum_{\ell=1}^2   c_\ell\left[\pm\ln(w-c_\ell)
	+\sum _{i=1}^2 \frac{n_i^2}{2\pi}L\left(c_i\frac{w-w_i}{w-\bar w_i},c_i\frac{c_\ell-w_i}{c_\ell-\bar w_i}\right)\right],
\end{align}
where $c_1=-c_2=1$, $n_1^2=j$, $n_2^2=k$, as well as 
\begin{align}
	L(z,t)&=\Li_2\left(2,\frac{z}{t}\right)+\ln z \ln\left(1-\frac{z}{t}\right)~.
\end{align}
The solution has poles representing $M$ NS5-branes each at $w=\pm 1$, and punctures on the imaginary axis at $w_1=i\tan\theta_1$ and $w_2=i\cot\theta_2$ with branch cuts extending in opposite directions along the imaginary axis. The punctures represent, respectively, $j$ and $k$ D7-branes.
The parameters $\theta_{1/2}$ are determined in terms of the brane web parameters by 
\begin{align}
k\theta_2&=j\theta_1~, &\pi N=2Mj\theta_1~.
\end{align}
The structure of the solutions is illustrated schematically in Fig.~\ref{fig:plus-D7-disc}.

The probe D7-brane equation of motion is satisfied on the imaginary axis, $w=i\xi$, corresponding to the real diameter of the disc in Fig.~\ref{fig:plus-D7-disc}, with
\begin{align}
	F&=\frac{2}{3}\Im(\cA_++\bar \cA_-)=\frac{iM}{2\pi}\ln\left(\frac{\xi-i}{\xi+i}\right)~.
\end{align}
This expression is insensitive to the branch cuts along the imaginary axis.
For the $SU(2)$ singlet fluctuations we find that the mass matrix is again given by (\ref{eq:D7-mass-TN}), including for probe D7 branes near the backreacted punctures. 
For the fluctuations transforming non-trivially under $SU(2)$ we similarly find again the masses as in (\ref{eq:D7-SU2-mass0}) and mass matrices as in (\ref{eq:D7-SU2-mass}).
This further supports the interpretation that the mass matrices in the supersymmetric and in the non-supersymmetric solutions are universal.

\section{Domain wall instabilities}\label{sec:DW}

In this section we study domain wall instabilities. 
The supersymmetric CFTs have a moduli space of vacua, which corresponds in the brane webs to opening up a brane intersection at a point to a brane web, without moving semi-infinite 5-branes.
In the supersymmetric solutions they lead to stable probe brane embeddings along Minkowski slices in Poincar\'e AdS, which describe displacements of individual brane segments.
In the non-supersymmetric solutions these probe brane embeddings can be destabilized, such that the branes are expelled towards the conformal boundary of AdS.
Following \cite{Gaiotto:2009mv} and the discussion in \cite{Apruzzi:2019ecr}, this can be used as a diagnostic for stability.
We start with a brief discussion of domain wall instabilities in the Brandhuber-Oz solution in massive Type IIA. We discuss Type IIB solutions corresponding to brane webs afterwards, focusing on Coulomb branch deformations.

\subsection{Domain walls in massive IIA}\label{sec:IIA-jets}

Type IIA  supergravity admits a supersymmetric AdS$_6$ solution named after the authors of \cite{Brandhuber:1999np} which is locally unique \cite{Passias:2012vp} (orbifolds were discussed in \cite{Bergman:2012kr}). This solution admits a consistent truncation to six-dimensional $F(4)$ gauged supergravity \cite{Romans:1985tw} and comes with a non-supersymmetric sibling. Both solutions have the form of a warped product AdS$_6\times M_4/\ZZ_2$ where $M_4$ has the topology of an $S^4$ with an O8 plane at the equator. The solutions are
\begin{align}\label{eq:BO-metric}
	&\drm s^2_{\text{BO}}\,=\, \frac{L^2\,\Delta^{1/2}}{X^{1/2}\,\sin \alpha^{1/3}}\left\{ \frac{9}{2}\,g^{-2}\drm s^2_{\AdS_6}\,+\,2g^{-2}\,X^2\left(\drm\alpha^2+\frac{\cos\alpha^2}{X^3\,\Delta}\drm s^2_{\Srm^3}\right)\right\}\,,\\
	& L^{-3}\,e^{\phi_0}\,F_4\,=\,-\frac{4\sqrt 2\,\Xi}{3\,g^3\,\Delta^2}\sin\alpha^{1/3}\,\cos\alpha^3\,\drm\alpha \wedge \text{vol}_{\Srm^3}\,,\quad\,L\,e^{\phi_0}\,F_0\,=\,\frac{\sqrt 2}{3} g\,,\\
	& e^{\phi}\,=\,e^{\phi_0}\frac{\Delta^{1/4}}{\sin\alpha^{5/6}\,X^{5/4}}\,,
\end{align}
where $g$ is a parameter and we introduced the functions
\begin{align}\label{eq:BO-Delta-Xi}
	\Delta&=\, X\,\cos\alpha^2\,+\,X^{-3}\,\sin\alpha^2\,,
	\nonumber\\
	\Xi&=\,X^{-6}\sin{\a}^2-3X^2\,\cos\alpha^2\,+\,4 X^{-2}\cos\alpha^2\,-\,6X^{-2}\,.
\end{align} 
$\phi_0$ and $L$ are free parameters. 
The supersymmetric solution is obtained for $X=1$, leading to $\Delta=1$ and $\Xi=-5$. In this case $M_4$ is an $\Srm^4$ of radius $\frac{2}{3m}$\,. The non-supersymmetric case instead corresponds to $X=3^{-1/4}$ and the $\Srm^4$ is deformed. $\Delta\,=\,3^{-1/4}(1+2\sin{\alpha}^2)$ is no longer a constant but is a (nowhere vanishing, positive) function. The function $\Xi$, on the other hand, is still a constant, $\Xi=-3\sqrt3$.
In the following, we use the rescaling
\begin{align} 
	g\,\,&\rightarrow\,\,\sqrt2\,m^{1/6}\,,& e^{\phi_0}\,\,&\rightarrow\,\, \frac{2}{3}m^{-5/6}\,g_s
\end{align} 
so that the solution takes the form
\begin{align}
	&\drm s^2_{\text{BO}}\,=\, \frac{L^2\,\Delta^{1/2}}{X^{1/2}\,(m\,\sin \alpha)^{1/3}}\left\{ \frac{9}{4}\drm s^2_{\AdS_6}\,+\,X^2\left(\drm\alpha^2+\frac{\cos\alpha^2}{X^3\,\Delta}\drm s^2_{\Srm^3}\right)\right\}\,,\\
	& L^{-3}\,g_s\,F_4\,=\,-\frac{\Xi}{\Delta^2}(m\,\sin\alpha)^{1/3}\,\cos\alpha^3\,\drm\alpha \wedge \text{vol}_{\Srm^3}\,,\quad\,L\,g_s\,F_0\,=\,m\,,\\
	& e^{\phi}\,=\,\frac{2\,g_s}{3}\frac{\Delta^{1/4}}{(m\,\sin\alpha)^{5/6}\,X^{5/4}}\,.
\end{align}
The coordinate $\alpha$ runs between $0$, corresponding to the equator of $S^4$ (the location of the $\O8$ plane), and $\pi/2$, where the $S^3$ shrinks smoothly, corresponding to the pole of the half-$S^4$. 

A diagnostic for instabilities of AdS backgrounds is provided by domain walls extended along Minkowski slices in AdS. For solutions sourced by a flux, it is natural to consider as domain walls the branes that couple to it. Indeed, since such AdS solutions typically arise as near horizon limits of brane configurations involving the same branes, the study of domain walls directly relates to the stability of the brane configuration. In particular, static domain wall embeddings are probes of the moduli space of vacua of the CFT, while non-static embeddings, instead, signal the presence of a potential, which can be dangerous. On the AdS side, such a danger is indicated by the domain wall feeling a force that pushes it towards the conformal boundary. This class of instabilities has been discussed in \cite[section 4.1.2]{Gaiotto:2009mv} and \cite[section 5]{Apruzzi:2019ecr}; see also \cite{Bena:2020xxb} for an M-theory example.

In the Brandhuber-Oz solution, the relevant domain walls are given by D4 branes, which couple to $F_6$. 
This decay channel was already discussed in \cite{Suh:2020rma}, but for completeness and to compare to the Type IIB analysis below we rederive it here.
In a supergravity solution with large flux, a single D4 brane,  which changes the flux only by a unit, has a small effect and can be studied in the probe approximation.
We start by writing the AdS$_6$ metric as
\begin{equation} 
	\drm s^2_{\AdS_6}\,=\,\frac{1}{r^2}(\drm r^2+\drm s^2_{\Mink_5})\,,
\end{equation} 
and we assume the probe brane to be initially localized at some fixed radius $r$.
The action of such an object reads
\begin{equation} 
	S_{\mathrm{D}4}\,=\,-\frac{3^6\,\mu_4}{2^6\,g_s}\,L^5\,\int \drm\,\frac{5\Delta\pm \Xi}{5\,r^5}\,\,,
\end{equation} 
where the two contributions come as usual from the DBI and WZ actions, and we used that
\begin{equation} 
	L^5\,g_s\,F_6\,=\,-\Xi\,\text{vol}_{\AdS_6}\,=\,-\frac{\Xi}{r^6}\,\drm r\wedge \text{vol}_{\Mink_5}\,.
\end{equation} 
For the supersymmetric solution, $\Delta=1\,,\Xi=-5$ so that the potential felt by the D4-brane vanishes identically. 
The supergravity background describes the near-horizon limit of a stack of D4-branes on a combination of D8 branes and an O8 plane, and the probe embeddings correspond to configurations with a single D4-brane separated from the stack.
Vanishing of the potential implies no force, and the supersymmetric solution is stable, as expected. 

\begin{figure}
	\centering
			\includegraphics[scale=0.18]{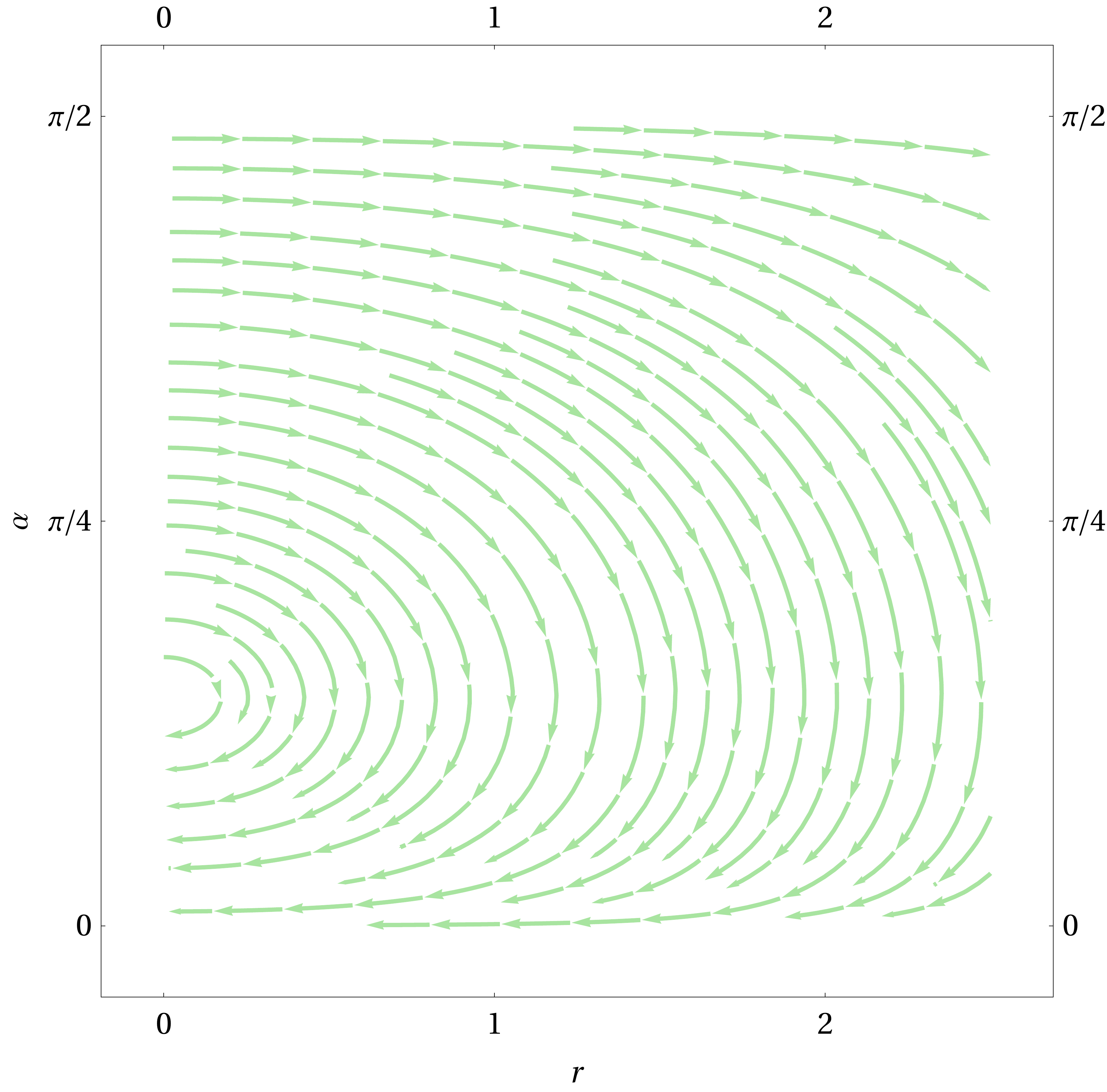}
			\caption{Force $\{F_\alpha, F_r\}$ on a probe D4 at $(\alpha,r)$, with $\alpha$/$r$ on the vertical/horizontal axis.	\label{DO-force}}
\end{figure}

In the non-supersymmetric case, however, $\Delta$ is no longer a constant and the $\mathrm{D}4$ brane experiences a potential of the form
\begin{equation}\label{eq:V-D4}
	V_{\mathrm{D}4}^{\rm non-susy}\,=\,\frac{1}{r^5}\left( 1+2 \sin\alpha^2-\frac{3^{7/4}}{5} \right)\,.
\end{equation} 
The corresponding force felt by the $\mathrm{D}4$ brane is illustrated in Fig.~\ref{DO-force}.
The brane is repelled towards the conformal boundary  of $\AdS_6$ and the $\O8$-plane. In fact, a brane sitting on top of the $\O8$-plane ($\alpha=0$) feels no force along the internal direction $\alpha$, and its only motion is along AdS$_6$, being repelled towards the conformal boundary.
Such $\mathrm{D}4$-branes destabilize the non-supersymmetric Brandhuber-Oz solution.

\subsection{Domain walls in Type IIB}

\begin{figure}[ht!]
	\centering
	\subfigure[][]{\label{fig:plus-CB}
		\begin{tikzpicture}[scale=0.8]
			\foreach \i in {-0.2,-0.1,0,0.1,0.2}{
				\draw[thick] (-2.5,\i) -- (2.5,\i);
			}
			\foreach \i in {-0.2,-0.066,0.066,0.2}{
				\draw[thick] (\i,-2.5) -- (\i,2.5);
			}
			\draw[thick] (-2.5,0.3) -- (-1.4,0.3) -- (-0.333,0.3+1.4-0.333) -- (-0.333,2.5);
			\draw[thick] (2.5,0.3) -- (1.4,0.3) -- (0.333,0.3+1.4-0.333) -- (0.333,2.5);
			\draw[thick] (-2.5,-0.3) -- (-1.4,-0.3) -- (-0.333,-0.3-1.4+0.333) -- (-0.333,-2.5);
			\draw[thick] (2.5,-0.3) -- (1.4,-0.3) -- (0.333,-0.3-1.4+0.333) -- (0.333,-2.5);
			
			\draw[thick] (-1.4,0.3) -- (-1.4,-0.3);
			\draw[thick] (1.4,0.3) -- (1.4,-0.3);
			\draw[thick] (0.333,0.3+1.4-0.333) -- (-0.333,0.3+1.4-0.333);
			\draw[thick] (0.333,-0.3-1.4+0.333) -- (-0.333,-0.3-1.4+0.333);
		\end{tikzpicture}
	}\hskip 15mm
	\subfigure[][]{\label{fig:plus-D7-CB}
		\begin{tikzpicture}[scale=0.8]
			\foreach \i in {-0.066,0.066}{
				\draw[thick] (\i,-2.5) -- (\i,2.5);
			}
			\draw[fill=black] (0.2,0) circle (2pt);
			\draw[fill=black] (-0.2,0) circle (2pt);
			\draw[thick,dashed] (0.2,0) -- (2.5,0);
			\draw[thick,dashed] (-0.2,0) -- (-2.5,0);
			
			\draw[thick] (-0.2,2.5) -- (-0.2,0.2) -- (-0.4,0);
			\draw[thick] (-0.333,2.5) -- (-0.333,0.4) -- (-0.733,0);
			
			\foreach \i in {-1,1}{\foreach \j in {-1,1}{
					\draw[thick] (-0.2*\i,2.5*\j) -- (-0.2*\i,0.2*\j) -- (-0.4*\i,0);
					\draw[thick] (-0.333*\i,2.5*\j) -- (-0.333*\i,0.4*\j) -- (-0.733*\i,0);
					
					\draw[thick] (-0.4666*\i,2.5*\j) -- (-0.4666*\i,1.4*\j) -- (-0.4666*\i-1.4*\i,0);
				}
				\draw[thick] (-0.2,0.2*\i) -- (0.2,0.2*\i);	
				\draw[thick] (-0.333,0.4*\i) -- (0.333,0.4*\i);	
				\draw[thick] (-0.4666,1.4*\i) -- (0.4666,1.4*\i);
			}
			\node at (2.4,-0.3) {\footnotesize $T^2$};
			\node at (-2.4,-0.3) {\footnotesize $T^2$};
		\end{tikzpicture}
	}
	\caption{Example Coulomb branch deformation for the $+_{N,M}$ theory with $N=7$ D5-branes and $M=6$ NS5 branes on the left. On the right for the $+_{N,M,j,k}$ theory with $j=k=2$, $M=8$ and $N=6$. 
		The external 5-branes have been separated slightly for illustration.
		The origin of the moduli space corresponds to the limit where all faces shrink to zero size.\label{fig:CB}}
\end{figure}
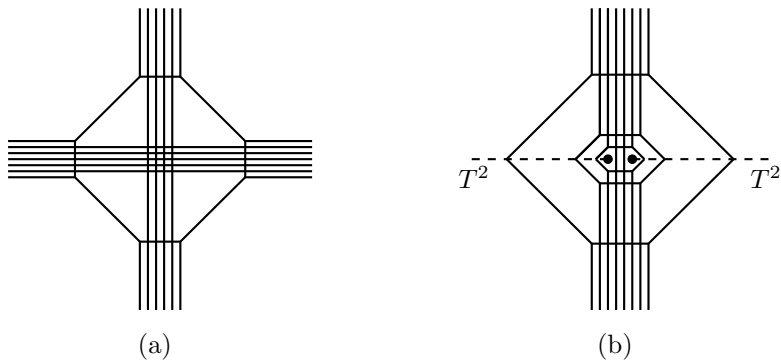

The engineering of 5d SCFTs in Type IIB differs from the D4-brane construction in that the SCFTs arise on intersections of 5-branes with 6-dimensional worldvolume, i.e.\ with an extra dimension compared to the field theory. The Coulomb branch, rather than displacing complete 5-branes, corresponds to opening up the intersection point and creating a brane web with closed faces, while leaving the external 5-branes of the intersection unchanged. 
Two examples, for the $+_{N,M}$ and $+_{N,M,j,k}$ theories, are shown in Fig.~\ref{fig:CB}.
A Coulomb branch deformation of the $+_{N,M}$ theory, for example, involves the displacement of $\pm (1,1)$ and $\pm (1,-1)$ 5-brane segments which form a quadrilateral.
The locations of the vertices are constrained by the positions of the external 5-branes.
More general deformations are possible, constrained by charge conservation at the 5-brane vertices.
Since the deformation is within the plane of the brane web, the $SU(2)$ R-symmetry is preserved.
The analogous deformation for the $+_{N,M,j,k}$ theories takes a similar form; the 5-brane quadrilateral is now suspended between the external NS5 branes and the branch cuts associated with 7-branes.

The nature of the Coulomb branch deformation in the brane web dictates the form of the associated probe brane embeddings in the supergravity solutions.
Namely, Coulomb branch deformations are described by probe 5-brane segments embedded along $\RR^{1,4}$ slices in AdS$_6$ and extending along a curve in $\Sigma$ (rather than being pointlike in the internal space). To preserve the $SU(2)$ R-symmetry they have to extend along the boundary of $\Sigma$, where the $S^2$ in the geometry collapses.
For the example deformation of the $+_{N,M}$ theory in Fig.~\ref{fig:CB}, we expect $\pm (1,1)$ and $\pm (1,-1)$ 5-brane segments which connect the poles representing the external 5-branes along the boundary of $\Sigma$. 
For the supersymmetric backgrounds we generally expect stable solutions.
For the non-supersymmetric backgrounds, an instability would arise if the probe 5-brane segments would be expelled from the brane junction. 
In the supergravity solutions this would correspond to the probe branes being expelled towards the conformal boundary of AdS$_6$.

In the following we will derive a general analytic solution for the Coulomb branch embeddings in the supersymmetric solutions, and then study concrete embeddings in supersymmetric and non-supersymmetric backgrounds for the $+_{N,M}$ and $+_{N,M,j,k}$ solutions.

\subsection{Coulomb branch 5-branes}\label{sec:CB-IIB}

To find the embeddings for Coulomb branch 5-branes in the supergravity solutions we start with D5-branes and then generalize the results using $SL(2,\ZZ)$.
We expand the probe brane action near the boundary of $\Sigma$ and, based on the expectation that Coulomb-branch 5-branes in the supersymmetric solutions describe supersymmetric states and should solve BPS equations, motivate a proposal for a general solution to the equations of motion. 
The validity of that proposal will be verified for examples in the next section.

We start with a D5-brane wrapping a curve in $\Sigma$ and an $\RR^{1,4}$ slice in Poincar\'e AdS$_6$, with coordinates such that
\begin{align}
	ds^2_{AdS_6}=d\xi^2+e^{2\xi}\,ds^2_{\RR^{1,4}}~.
\end{align}
Such a D5-brane does not capture $B_{(2)}$ and we set the worldvolume gauge field to zero.
The D5-brane action (with dilaton convention $\tau = \chi + i e^{-\phi}$) reads
\begin{align}
	S_{\rm D5}&=-T_{\rm D5}\int d^6\zeta e^{\phi/2}\sqrt{-g_{\rm E,ind}}-Q_{\rm D5}\int C_{(6)}~,
\end{align}
where $g_{\rm E,ind}$ is the metric induced from the Einstein-frame metric.
The 6-form potential is obtained from (\ref{eq:C6-cM}).
After a gauge transformation, and with real coordinates $w=x+i y$ on $\Sigma$,
\begin{align}
C_{(6)}&=-c(\xi)\left(dx\, \partial_x \Im\cM +dy\, \partial_y \Im\cM dy\right)\wedge \vol_{\RR^{1,4}}~,
\end{align}
where $c'(\xi)=e^{5\xi}$.

For embeddings along the boundary of $\Sigma$, i.e.\ along the real line with $w=x+i y$, the action simplifies.
In static gauge, with worldvolume coordinates $x$ and the coordinates on $\RR^{1,4}$, the embedding is parametrized by a function $\xi(x)$, and the action becomes
\begin{align}
	S_{\rm D5}&=-V_{\RR^{1,4}}\int dx \left[T_{\rm D5}c'(\xi)\,e^{\phi/2}f_6^6\sqrt{\xi'(x)^2+\frac{4\rho^2}{f_6^2}}-Q_{\rm D5}c(\xi) \partial_x \Im\mathcal M\right]~.
\end{align}
The ratio $4\rho^2/f_6^2$ can be evaluated using (\ref{eq:functionsExpl}), which leads to 
\begin{align}\label{eq:SD5-CB}
	S_{\rm D5}&=-V_{\RR^{1,4}}\int dx \left[T_{\rm D5}c'(\xi)\,e^{\phi/2}f_6^6\sqrt{\xi'(x)^2+\frac{V_0X^2}{20}\frac{2\kappa^2}{3\cG}}-Q_{\rm D5}c(\xi) \partial_x \Im\mathcal M\right].
\end{align}

We now evaluate the dilaton on the boundary.
From (B.21) of \cite{Uhlemann:2020bek}, with the appropriate insertions of $X$ to generalize the expression to non-supersymmetric solutions,
\begin{align}
	\Im(\tau)&=\frac{4\kappa^2 |\partial_w\cG|^2 T}{\left|(T/X^2+1)\partial_{\bar w}\cG \left(\partial_w\cA_+-\partial_w\cA_-\right)+(T/X^2-1)\partial_w\cG\left(\partial_{\bar w}\bar \cA_+ -\partial_{\bar w}\bar \cA_-\right)\right|^2 }~.
\end{align}
For the behavior on the boundary of $\Sigma$, where $\kappa^2,\cG\rightarrow 0$ with $\kappa^2/\cG$ finite and 
$T^2\rightarrow \frac{2|\partial_w\cG|^2}{3\kappa^2\cG}$, we find
\begin{align}
	\frac{1}{\Im(\tau)}\Big\vert_{\partial\Sigma}&=\frac{20\Re(\cA_+-\cA_-)^2}{V_0X^4 |\partial_w\cG|}\sqrt{\frac{3\cG}{2\kappa^2}}\left[
	\frac{V_0X^6}{20}\frac{\Re(\partial_w\cA_+-\partial_w\cA_-)^2}{\Re(\cA_+-\cA_-)^2}+\frac{V_0X^2}{20}\frac{2\kappa^2}{3\cG}
	\right].
\end{align}
This suggests the following solution to uniformize the square root in the DBI action in supersymmetric backgrounds with $X=1$, 
\begin{align}
	\xi'(x)&=\pm\frac{\Re(\partial_w\cA_+-\partial_w\cA_-)}{\Re(\cA_+-\cA_-)}~.
\end{align}
This can be integrated to
\begin{align}\label{eq:xi-CB-D5}
	\xi_{\rm D5}(x)&=\pm \ln \left(\cA_+(x)-\cA_-(x)+\bar\cA_+(x)-\bar \cA_-(x)\right)+\xi_0~,
\end{align}
where $\xi_0$ is a constant. In the examples considered below we verify that this is indeed a solution if the sign is chosen appropriately.

Since we have not made any assumption on the background, a candidate solution for general $(p,q)$ 5-branes can be obtained by $SL(2,\ZZ)$ transformations from (\ref{eq:xi-CB-D5}).
We follow the logic laid out in app.~B of \cite{Uhlemann:2020bek}, which leads to
\begin{align}\label{eq:xi-CB}
	\xi_{(p,q)}&=\pm \ln\Re \left((p-iq)\cA_+-(p+iq)\cA_-\right)+\xi_0~.
\end{align}
The boundary and regularity conditions associated with specific solutions will constrain admissible embeddings.
For example, for Coulomb branch embeddings we require that $\xi(x)$ does not reach to the conformal boundary along the segment of $\partial\Sigma$ at which a particular $(p,q)$ 5-brane is embedded.

\subsection{\texorpdfstring{$+_{N,M}$ and $+_{N,M,j,k}$}{+[N,M] and +[N,M,j,k]} solutions}\label{sec:dw-ex}

We now discuss the Coulomb branch embeddings in the supersymmetric  $+_{N,M}$ and $+_{N,M,j,k}$ solutions and in their non-supersymmetric siblings

\medskip

\begin{figure}[!ht]
	\centering
	\subfigure[][]{\label{fig:plus-CB-5-branes}
		\includegraphics[width=0.4\linewidth]{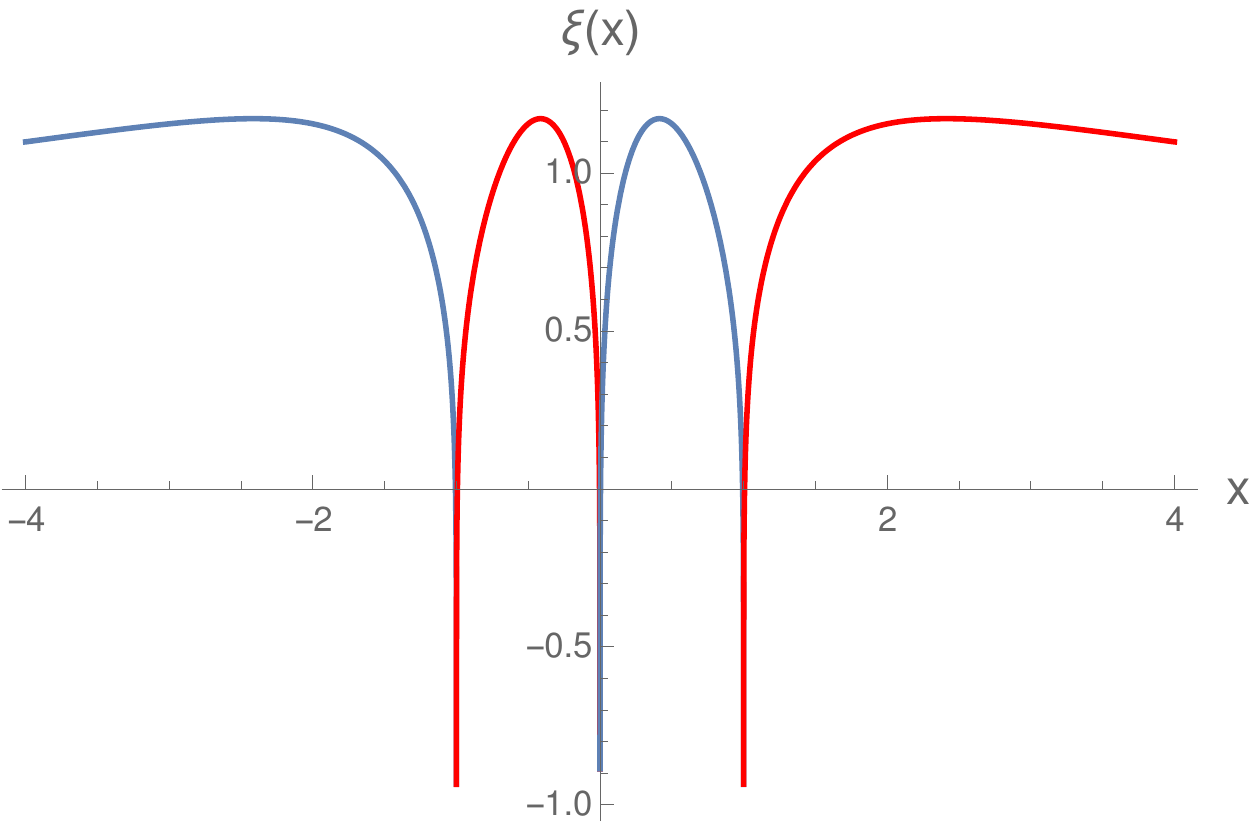}
	}
	\hskip 6mm
	\subfigure[][]{\label{fig:plus-CB-5-branes-non-susy}
		\begin{tikzpicture}
		\node at (0,0) {\includegraphics[width=0.4\linewidth]{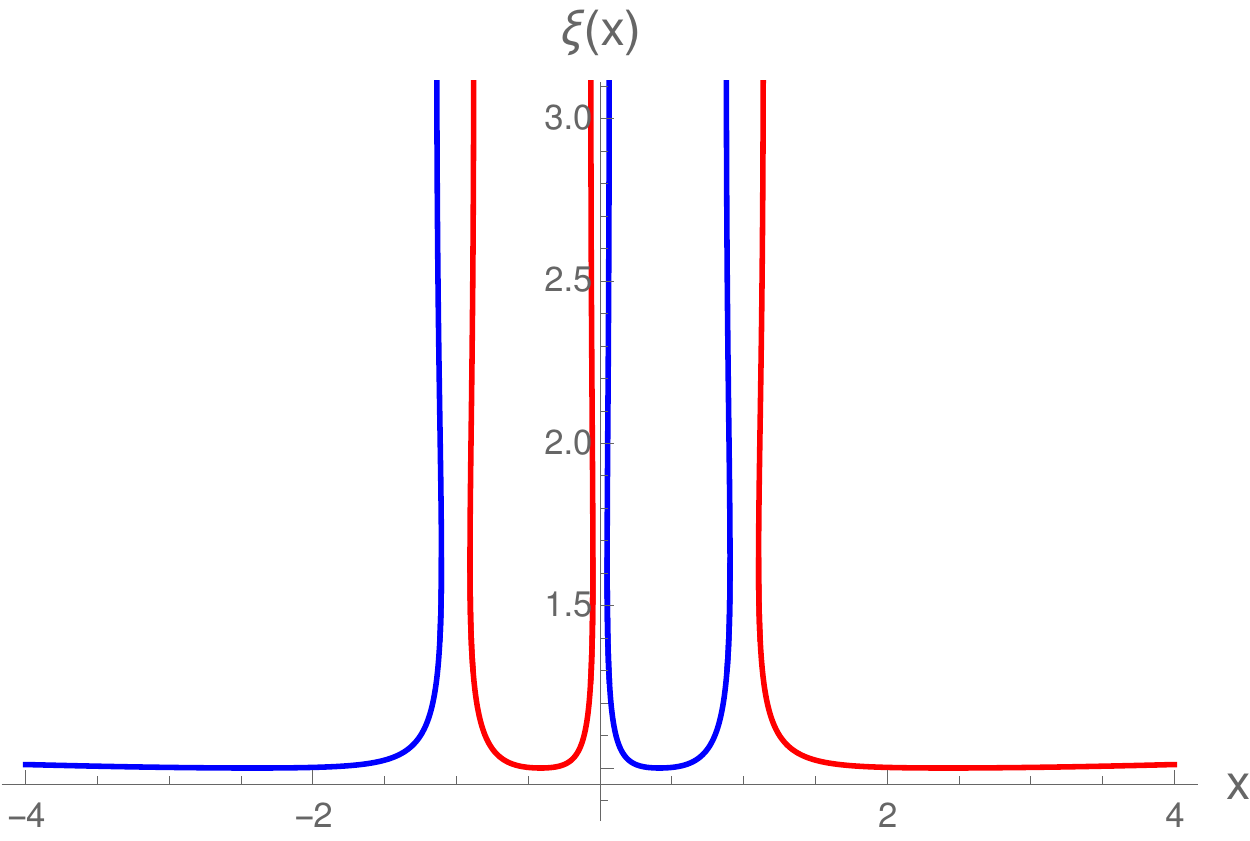}};
		\node at (2.8,0.65) {\includegraphics[width=0.18\linewidth]{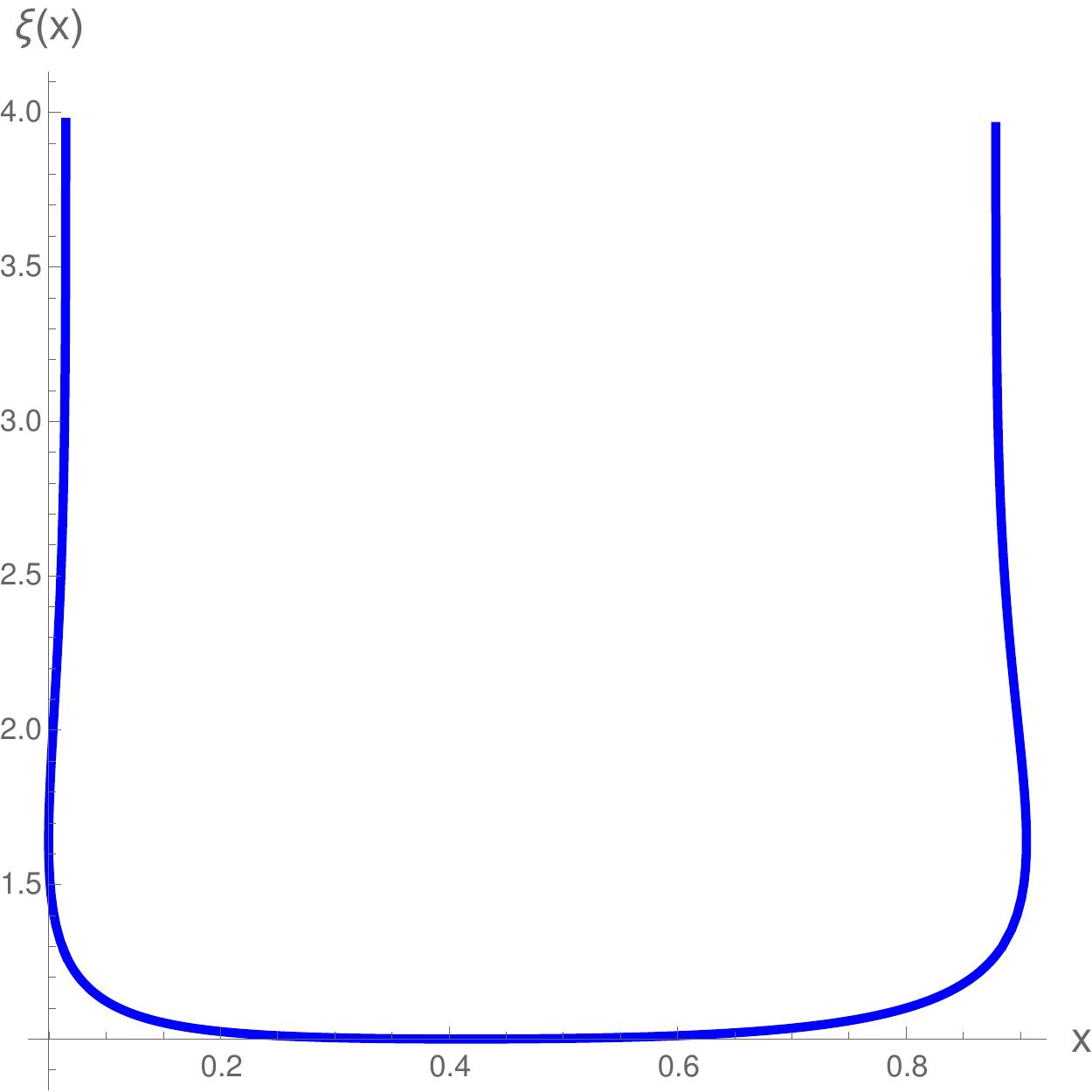}};
		\end{tikzpicture}
	}
	\caption{Left: Coulomb branch 5-branes corresponding to Fig.~\ref{fig:plus-CB} in the $+_{N,M}$ solution with $N=M$; $\pm (1,1)$ 5-branes in blue, $\pm (1,-1)$ 5-branes in red. 
		Right: embeddings in the non-supersymmetric sibbling,
		with one branch magnified to illustrate the turning points.
		Shifting $\xi(x)$ by a constant produces one-parameter families of solutions in both cases.}
\end{figure}

{\bf $\mathbf{+_{N,M}}$ solution:}
The functions $\cA_\pm$ for the $+_{N,M}$ solutions were discussed in Sec.~\ref{sec:plus}. 
The solution has poles representing $N$ D5-branes each at $w\in\lbrace 0,\infty \rbrace$ and poles representing $M$ NS5 branes each at $w=\pm 1$.
To realize the Coulomb branch deformation of Fig.~\ref{fig:CB}, we embed $\pm (1,1)$ 5-branes along the boundary segments $(-\infty,-1)$ and $(0,1)$ on the real line, and $\pm (1,-1)$ 5-branes along the boundary segments $(-1,0)$ and $(1,\infty)$ on the real line.
In the supersymmetric background the embeddings (\ref{eq:xi-CB}) indeed solve the equation of motion resulting from the $(p,q)$ 5-brane action (\ref{eq:(p,q)action}) with the appropriate charges.
The on-shell action vanishes.
The embeddings are shown in Fig.~\ref{fig:plus-CB-5-branes} for the $+_{N,M}$ theory with $N=M$.

For $N=M$ the brane web has a $\ZZ_4$ symmetry corresponding to rotation by $\pi/2$ combined with exchanging D5 and NS5 charges. As a result the brane embeddings along all 4 boundary segments take an equivalent form (they are related by $SL(2,\RR)$ transformations of the upper half plane). The points $x_f=\pm \sqrt{2}\pm 1$ where $\xi'(x_f)=0$ correpond in the brane web to the fixed points under reflection across the diagonals.
The brane embeddings do not extend to the conformal boundary of AdS$_6$ and describe non-conformal vacuum states of the SCFT, with no new sources turned on. Since the $SU(2)$ R-symmetry is preserved the states are on the Coulomb branch of the SCFT.
More general $(p,q)$ 5-brane embeddings with similar features are possible if the brane orientations are compatible with the brane webs for more general Coulomb branch deformations.
To sum up, in the supersymmetric AdS$_6$ backgrounds the embeddings (\ref{eq:xi-CB}) are static and not subject to any force.

For the non-supersymmetric backgrounds the embeddings (\ref{eq:xi-CB}) do not satisfy the equations of motion. This is analogous to the non-supersymmetric sibling of the Brandhuber-Oz solution, where the embeddings that describe the Coulomb branch D4-branes in the supersymmetric solution disappear in the non-supersymmetric background.
However, while the D4-brane in the non-supersymmetric sibling of the Brandhuber-Oz solution is pointlike in the internal space and can be treated like a particle moving in a potential, the D5-brane is a string in the internal space and as such has richer dynamics.
The D4-brane is expelled towards the conformal boundary by the potential in (\ref{eq:V-D4}); there are no solutions to the equation of motion for static embeddings. 
The 5-brane segments could in principle bend and adjust their profiles $\xi(x)$ along $\Sigma$ in such a way that the equations of motion would still be satisfied. 
This is indeed the case, as we show in Fig.~\ref{fig:plus-CB-5-branes-non-susy}. The figure shows solutions obtained by imposing $\xi'(x_f)=0$ as before and integrating the equations of motion.
Unlike the embeddings in the supersymmetric backgrounds in Fig.~\ref{fig:plus-CB-5-branes}, the non-supersymmetric embeddings do not reach the poles. Instead, they reach a turning point at a finite distance from the poles, where the tangent becomes vertical. Upon further following the curves, the embeddings then bend away from the poles slightly, back towards the center points $x_f$ (as shown in the magnified part of Fig.~\ref{fig:plus-CB-5-branes-non-susy}), and reach out to the conformal boundary.\footnote{The embeddings in Fig.~\ref{fig:plus-CB-5-branes-non-susy} each consist of three branches which are joined smoothly near the turning points; the central branch is obtained with the embedding parametrized by $\xi(x)$, the outer two using $x(\xi)$. This gives an overlapping regime of validity and allows to connect the branches smoothly}

As mentioned previously, the dynamics of the 5-branes is not as straightforward as for the pointlike D4-branes. However, a natural interpretation for the solutions in Fig.~\ref{fig:plus-CB-5-branes-non-susy} would be that the 5-brane segments have been partly expelled towards the conformal boundary of AdS$_6$ near the poles, signaling a similar instability. The embeddings in Fig.~\ref{fig:plus-CB-5-branes-non-susy} are akin to the end point of the motion of D4-brane domain walls in the non-supersymmetric background, which end up on the conformal boundary and for that reason also give rise to instabilities.
We will discuss a possible brane web interpretation at the end of this section.

\begin{figure}
	\centering
	\subfigure[][]{\label{fig:plus-D7-5-branes}
		\includegraphics[width=0.4\linewidth]{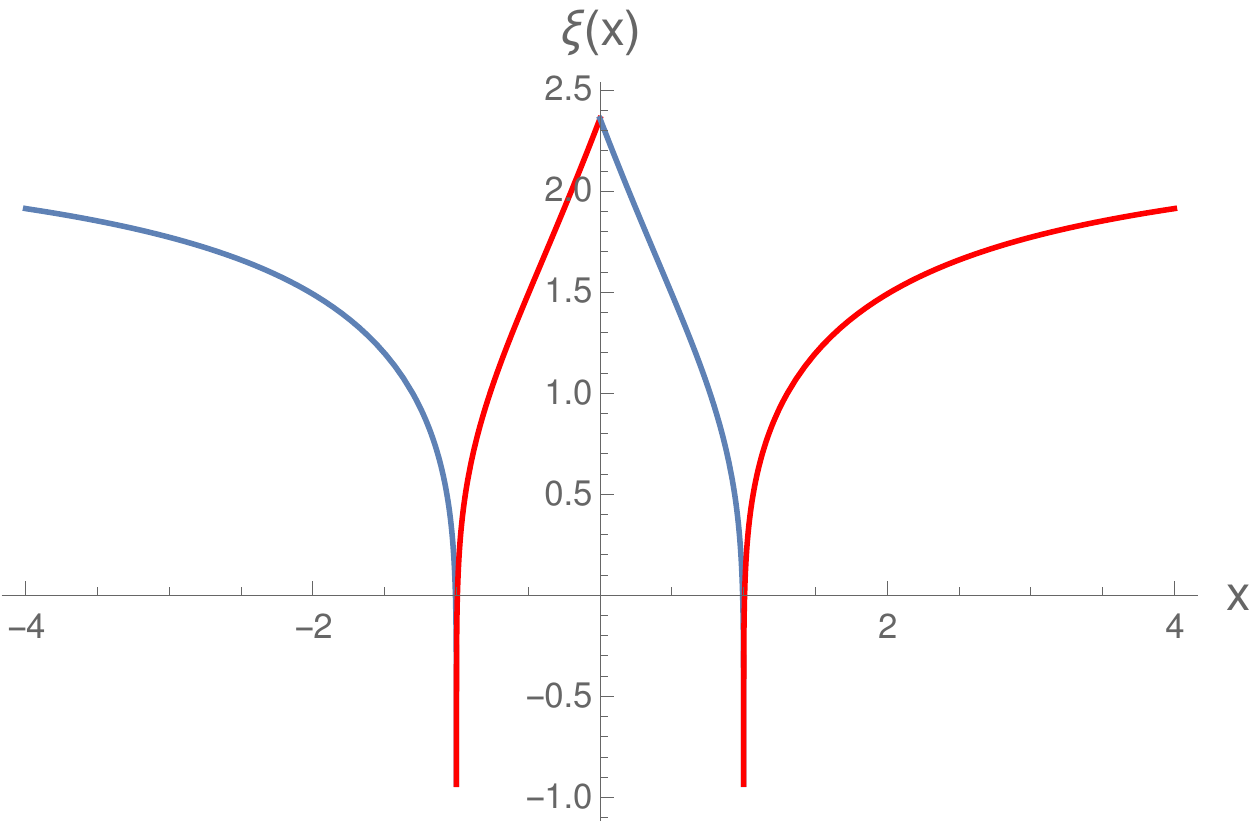}
	}
	\hskip 10mm
	\subfigure[][]{\label{fig:plus-D7-CB-5-branes-non-susy}
		\includegraphics[width=0.4\linewidth]{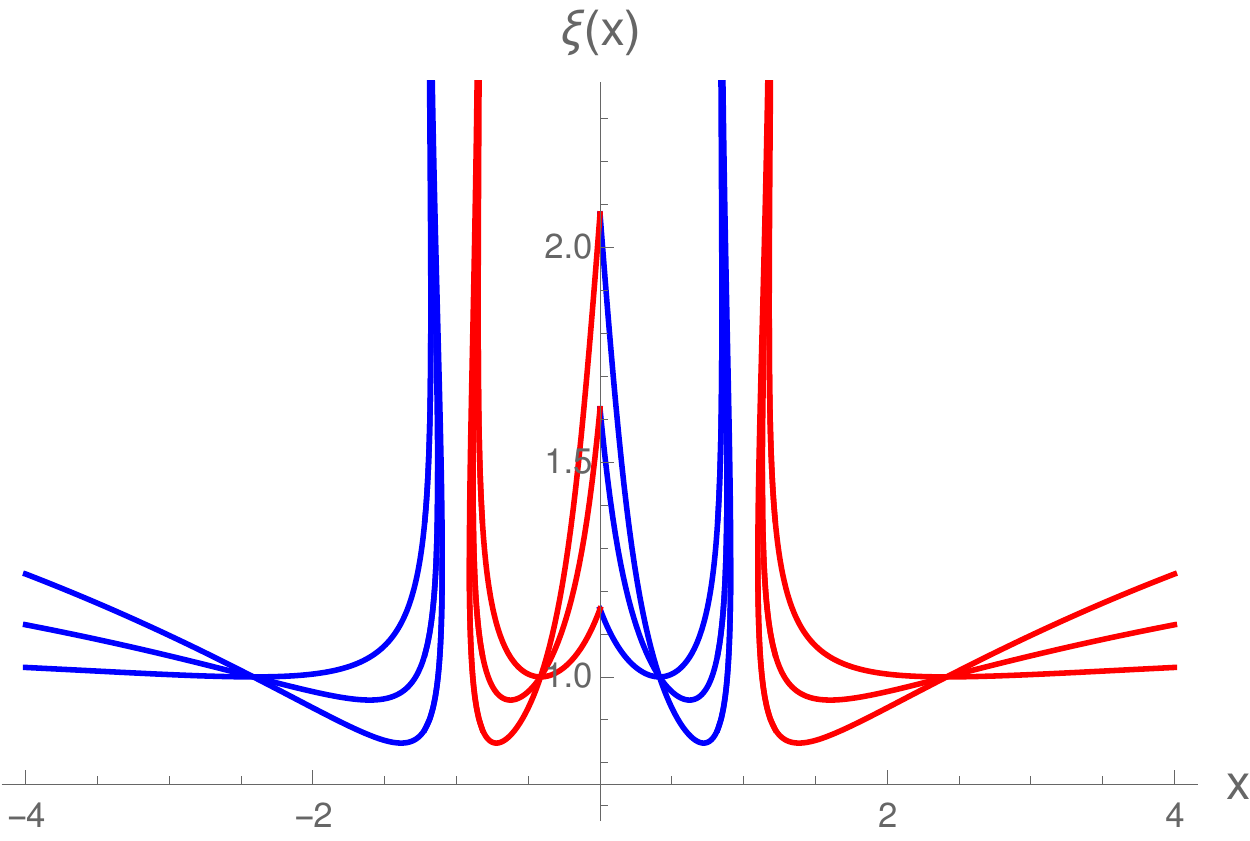}
	}
	\caption{Left: Coulomb branch 5-branes in the $+_{N,M,j,k}$ solution; $\pm (1,1)$ 5-branes in blue, $\pm (1,-1)$ 5-branes in red. 
		Right: analogous embeddings in the non-supersymmetric sibling. Shifting the embeddings by a constant again produces families of solutions.}
\end{figure}

\medskip
{\bf $\mathbf{+_{N,M,j,k}}$ solution:}
As a second example we take the $+_{N,M,j,k}$ solution.
The dual field theory arises as a Higgs-branch flow from the $+_{N,M}$ theories.
It is obtained by terminating the semi-infinite D5-branes in groups on 7-branes (Fig.~\ref{fig:plus-D7-web}, \ref{fig:plus-inst-3}).
The supergravity solution is illustrated schematically in Fig.~\ref{fig:plus-D7-disc} and the functions $\cA_\pm$ are in (\ref{eq:cA-plus-D7}).
The NS5-brane poles at $w=\pm 1$ remain; the D5-brane poles are replaced by two D7-brane punctures on the imaginary axis, with branch cuts extending in opposite directions along the imaginary axis.
The charges of the Coulomb-branch 5-branes we consider are the same as for the $+_{N,M}$ theories before (Fig.~\ref{fig:CB}). The embeddings $\xi(x)$ in (\ref{eq:xi-CB}) again solve the equations of motion resulting from the 5-brane action (\ref{eq:(p,q)action}) and are shown in Fig.~\ref{fig:plus-D7-5-branes}.

The branch cuts associated with the D7 branes cross the boundary of $\Sigma$ at $w=0$ and $w=\infty$. As seen in Fig.~\ref{fig:plus-D7-5-branes}, the brane embeddings do not drop into the Poincar\'e horizon when crossing the branch cuts. Instead, they reach the branch cuts at a finite value of $\xi$ and transform their charges appropriately when crossing the branch cut. The embeddings as shown are continuous across the branch cut but not smooth; this matches the brane web picture where the 5-branes change their angle when crossing the branch cut according to the change in their charge, as required by supersymmetry.

The analogous embeddings for the non-supersymmetric sibling are shown in Fig.~\ref{fig:plus-CB-5-branes-non-susy}. 
Since the $+_{N,M,j,k}$ solutions do not have the reflection symmetries across the diagonals of the brane web that were realized in the $+_{N,M}$ solutions with $N=M$, we show families of embeddings in the non-supersymmetric case. They are obtained by varying the derivative of $\xi$ at a reference point while holding the value of $\xi$ at that point fixed; the remaining free parameter in the embeddings amounts to shifts $\xi\rightarrow \xi+\xi_0$.
The behavior is modified compared to the $+_{N,M}$ solution in a similar way as in the supersymmetric context: the 5-branes are expelled towards the conformal boundary of AdS$_6$ at the NS5-brane poles at $w=\pm 1$, as for the non-super\-symmetric $+_{N,M}$ solution before, but when crossing the D7-brane branch cuts the embeddings remain finite and can be connected continuously.

\medskip

In summary, we find the expected Coulomb branch embeddings in the supersymmetric solutions, with 5-brane segments forming loops along the boundary of $\Sigma$ without reaching the conformal boundary. In the non-supersymmetric siblings we instead find embeddings where the 5-brane segments reach out to the conformal boundary near the 5-brane poles. 
This suggests an instability in the sense of \cite{Gaiotto:2009mv,Apruzzi:2019ecr,Bena:2020xxb}, in which 5-brane segments can be picked up at the Poincar\'e horizon and be partly expelled towards the conformal boundary by the 5-brane poles.
A brane web interpretation could be that the outer 5-brane quadrilateral in Fig.~\ref{fig:plus-CB} curls away from the brane intersection along the external 5-branes, effectively splitting up the external 5-brane stacks. This preserves the $SU(2)$ R-symmetry and acts in the plane of the brane web.
The effect would in a sense be opposite to the transition in Fig.~\ref{fig:polarization-web}, in which 5-branes are paired up in groups and terminated on the same 7-brane.

The discussion suggests a natural continuation to the non-supersymmetric siblings of solutions where all 5-brane poles are replaced by 7-brane punctures (which have not been constructed explicitly so far). 
In that case one would expect the probe 5-brane embeddings along the boundary of $\Sigma$ to not extend to the conformal boundary of AdS$_6$, and instead remain finite across the branch cuts associated with the 7-brane punctures in a similar way as for the D7 punctures in the $+_{N,M,j,k}$ solutions.
So the instability discussed above may not apply. We emphasize, however, that the discussion of probe 5-branes above is not exhaustive and there may be other instabilities.

\section{Bubble nucleation instabilities}\label{sec:D5-instanton}

In this section we discuss the nucleation of brane bubbles \cite{Maldacena:1998uz,Apruzzi:2019ecr} as a possible decay channel.
The relevant brane embeddings are technically similar to the domain wall embeddings of the previous section.
We first discuss the nucleation of D4-brane bubbles in the Brandhuber-Oz background.
The bubble nucleation instanton can be constructed as a configuration where the D4-brane wraps $S^5$ in Euclidean AdS$_6$. 
We then discuss the corresponding D5-brane instanton in the Hopf T-dual of the Brandhuber-Oz solution in Type IIB.\footnote{The brane interpretation of the T-dual is challenging -- see \cite{Lozano:2018pcp,Lozano:2018fvt} for a detailed discussion and proposals -- but it provides an instructive link in our discussion.}
In these cases the existence of bubble nucleation channels is closely tied to the existence of domain wall instabilities.
We then discuss bubble nucleation channels in 5-brane webs, and how they differ from the previous analyses, especially in their relation  to  domain wall instabilities. We summarize the outcome of a numerical scan for bubble nucleation channels in the sample Type IIB solutions of Sec.~\ref{sec:probe-D7}.

\subsection{D4 nucleation in massive IIA}\label{sub:D4SO5}

We show here the absence/existence of the explicit instanton solution responsible for the decay of the supersymmetric/non-supersymmetric Brandhuber-Oz solutions. The backgrounds are reviewed in Sec.~\ref{sec:IIA-jets}. We follow the strategy adopted in \cite{Apruzzi:2019ecr} and discuss the instanton brane both as a bounce solution in Euclidean time with $SO(5)$ symmetry in the spirit of \cite{Maldacena:1998uz} and as a thin wall with $SO(6)$ symmetry separating two vacua in the sense of Coleman-De Luccia \cite{Coleman:1980aw}. The Euclidean AdS$_6$ metric with manifest $SO(5)$ symmetry reads 
\begin{equation}\label{eq:glo-eads}
	ds^2_\mathrm{EAdS}= \cosh^2 (\rho) d\tau^2 +  d\rho^2 + \sinh^2 (\rho) ds^2_{S^{4}}\,.
\end{equation}
We embed the D4 brane probe such that it is characterized by two functions $\rho(\tau)$ and $\alpha(\tau)$. The D4-brane action reads
\begin{equation}\label{eq:S-D4}
	S_{\rm D4}= -\mathrm{Vol}_{S^{4}} \int d\tau \left(T_{\rm D4}K \sinh^{4} (\rho)\sqrt{\dot\rho^2 + \cosh^2 (\rho) + \frac{4X^2}{9}\dot{\alpha}^2} + \frac{Q_{\rm D4}}{5} f \sinh^{5} \rho\right) \,,
\end{equation}
where
\begin{align} \label{eq:Kf}
	K&= \frac{3^6 L^5 \Delta}{2^6 g_s}, & f &= \frac{3^6\Xi L^5}{2^6 g_s}\,, & Q_{\rm D4}&=T_{\rm D4}\, .
\end{align}
The equation of motion resulting from variation with respect to $\alpha$ reads
\begin{equation}
	\partial_{\tau} \left(   \frac{4 X^2\Delta \dot{\alpha} \sinh^{4} (\rho) }{9 \sqrt{\dot\rho^2 + \cosh^2 (\rho) + \frac{4X^2}{9}\dot{\alpha}^2}}  \right) = \sinh^{4} (\rho) \partial_{\alpha}(\Delta)\sqrt{\dot{\rho}^2 + \cosh^2 (\rho) + \frac{4X^2}{9}\dot{\alpha}^2} \,.
\end{equation}
This equation admits a simple solution when $\partial_{\alpha} \Delta(\alpha)=0$.
In the supersymmetric solution $\Delta$ is a constant and $\partial_\alpha\Delta =0$ for all $\alpha$; in the non-supersymmetric solution $\partial_\alpha\Delta$ vanishes for $\alpha=0$ and $\alpha=\frac{\pi}{2}$.
By using conservation of Euclidean energy \cite{Maldacena:1998uz}, it is possible to find an analytic solution when
\begin{equation}\label{eq:q-constr}
	q= \frac{Q_{\rm D4}f(\alpha_0)}{5T_{\rm D4}K(\alpha_0)}= -\frac{\Xi}{5 \Delta}>1~.
\end{equation}
This ratio has a natural interpretation as the ratio between the effective charge and tension of the brane. Indeed, if the charge is bigger than the tension, the electric force that wants to make the brane expand wins over the gravitational force that would make the bubble shrink. Thus, if $q>1$ the brane expands, eventually reaching the conformal boundary.

For the supersymmetric background $X=1$ leads to $q=1$, which implies that there are no solutions and an instanton does not exist. For the non-supersymmetric background $X=3^{-1/4}$, and eq.~(\ref{eq:q-constr}) can be satisfied for
\begin{equation}
	q= \frac{3^{7/4} }{5 (1 + 2 {\rm \sin}^2(\alpha_0))} >1\, .
\end{equation}
Combining this inequality with the constraint $\partial_\alpha\Delta = 0$ underlying our ansatz, which is satisfied for $\alpha_0=0$ and $\alpha_0=\pi/2$ in the non-supersymmetric solutions, we find a bubble nucleation channel for D4-branes at 
\begin{align}
	\alpha_0 =0~.
\end{align}
This is the same point at which the domain wall instability was found in Sec.~\ref{sec:IIA-jets}.

\paragraph{$SO(6)$ D4-brane Instanton}\label{sub:D4SO6}
In order to explicitly find bubble nucleating instanton solutions with $SO(6)$ symmetry in the spirit of Coleman-De Luccia \cite{Coleman:1980aw}, we use an AdS$_6$ metric with manifest $SO(6)$ symmetry. That is,
\begin{equation}\label{eq:ds2-EAdS5}
	ds^2_\mathrm{EAdS}= d\xi^2 + \sinh^2\!\xi\,  ds^2_{S^{5}}\,.
\end{equation}
To respect the $SO(6)$ symmetry the D4-brane is embedded at some fixed $\xi$ and a point in the internal space.
The D4-brane probe action reads
\begin{equation}
	S_{\rm D4} =  - T_{\rm D4} {\rm Vol}_5 \left( K \sinh^5\!\xi + f  \lambda(\xi) \right)\, ,
\end{equation}
where $K$ and $f$ are defined in \eqref{eq:Kf}, and $\lambda$ is defined such that $\lambda'(\xi) = \sinh^5(\xi)$. 
The equations of motion are
\begin{align}\label{eq:D4-SO6-EOM}
	&\partial_{\alpha} S_{\rm D4} = \Delta'(\alpha)=0~,\nonumber\\
	&\partial_{\xi} S_{\rm D4}= - 5 \Delta \cosh\xi\,\sinh^4\!\xi - \Xi \lambda'(\xi)=0~. 
\end{align}
The first equation, as before, leaves $\alpha$ unconstrained in the supersymmetric solution but fixes $\alpha_0\in\lbrace 0,\frac{\pi}{2}\rbrace$ in the non-supersymmetric solution.
The second equation implies
\begin{equation} \label{eq:finalinstconst}
	\tanh(\xi) = \frac{1}{q}= - \left. \frac{5  \Delta}{\Xi} \right|_{\alpha_0}~.
\end{equation}
In the supersymmetric case $q=1$ and there is no solution for finite $\xi$, whereas in the non-supersymmetric case $\alpha_0=0$ leads to a finite solution for $\xi$ and a bubble exists.
This is the same criterion as in \eqref{eq:q-constr}. Indeed, as shown in \cite{Apruzzi:2019ecr}, the two solutions are related by the fact that following the (Euclidean) time-evolution of the expanding $SO(5)$-symmetric configuration analyzed around (\ref{eq:S-D4}) produces the $SO(6)$-symmetric configuration described by \eqref{eq:D4-SO6-EOM}. This is true for objects that do not move along the internal space, which is indeed the case for a D4 brane at the equator of the (half) $S^4$ (cf.~Fig.~\ref{DO-force}).

Compared to the domain wall D4-branes, the existence of solutions is inverted: static domain wall D4-brane solutions exist in the supersymmetric background and do not exist in the non-supersymmetric sibling, while static (in the internal space) $SO(6)$ solutions do not exist in the supersymmetric background but exist in the non-supersymmetric one. The implications for stability, on the other hand, match: in the non-supersymmetric background the non-existence of domain wall solutions expells the D4-branes and causes an instability, and the existence of bubble nucleation solutions likewise has a destabilizing effect since it describes the nucleation of a brane that then expands and reaches the boundary in finite time. We will compare this to the Type IIB solutions in Sec.~\ref{sec:iib-bubbles}.

\subsection{D5 bubbles in the T-Dual of Brandhuber-Oz}\label{sec:bubble-Tdual}

We now turn to the Hopf T-dual of the Brandhuber-Oz solution \cite{Cvetic:2000cj},  where T-duality has been performed on the Hopf fiber of the $S^3$.
We first review the Type IIB AdS$_6$ supergravity solution. The metric in string frame reads,
\begin{equation}
	ds^2= f_6^2 ds_{{\rm AdS}_6}^2 + f_2^2 ds^2_{S^2} + f_{\alpha}^2 d\alpha^2 + f_{\varphi}^2 d\varphi^2\, ,
\end{equation}
where
\begin{align}
	f_6&= \frac{3wL\Delta^{1/4}}{2 X^{1/4}}, & f_2&=  \frac{w \cos(\alpha) L}{2 \Delta^{1/4}X^{3/4}}, & f_{\alpha}&= X^{3/4} w L \Delta^{1/4}, & f_{\varphi}&= f_{2}^{-1}, & w&= (m \sin(\alpha))^{-1/6},
\end{align}
in which $m$ corresponds to the Romans mass in Type IIA. The dilaton is 
\begin{equation}
	e^{-\phi}= \frac{3L^2X \cos(\alpha) }{4 w \Delta^{1/2}}\,.
\end{equation}
The non-vanishing fluxes read
\begin{align}
	F_3 &= \frac{\Xi L^4 \cos^3(\alpha)}{8w^2 \Delta^2} d\alpha \wedge {\rm vol}_{S^2}, & F_1 &= m d\varphi\, .
\end{align}
$\Delta$ and $\Xi$ were defined in (\ref{eq:BO-Delta-Xi}).
In the language of the general Type IIB solutions, $\alpha$ and $\varphi$ parametrize a surface $\Sigma$ which has a $U(1)$ isometry $\partial_\varphi$. 
The $S^2$ with size set by $f_2$ collapses at $\alpha=\frac{\pi}{2}$.
The functions $\cA_\pm$ reproducing this solution were given in \cite{DHoker:2016ujz}.

The $SO(6)$ symmetric D4-brane instanton in Type IIA corresponds to probe D5-branes with $SO(6)$ symmetry in the T-dual Type IIB background.
To study these D5-brane embeddings we again work in the AdS$_6$ metric which manifestly encodes this symmetry, (\ref{eq:ds2-EAdS5}).
The D5-branes wrap the $S^5$ in AdS$_6$ and extend along a curve in $\Sigma$. The embeddings can be parametrized by functions $\xi(\sigma), \alpha(\sigma), \varphi(\sigma)$ and the action reads
\begin{equation}
	S_{\rm D5}= -T_{\rm D5}\mathrm{Vol}_{S^{5}} \int d\sigma \left( f_6^6 e^{-\phi} \sinh^{5} (\xi)\sqrt{\dot\xi^2 + \frac{f_{\varphi}^2}{f_6^2}\dot{\varphi}^2+\frac{f_{\alpha}^2}{f_6^2}\dot{\alpha}^2} + f_7 \lambda(\xi) \dot{\varphi}  \right) \,,
\end{equation}
where $f_7$ is defined by $F_7=\ast_{10} F_3$ through
\begin{equation}
	F_7 = f_7 \sinh^5(\xi) d\xi \wedge {\rm Vol}_5\wedge d\varphi, \qquad f_7= \frac{3^6 L^6  \Xi}{2^6} \,.
\end{equation}
$\lambda$ is again defined such that $\lambda' = \sinh^5\!\xi$. The equations of motion are
\begin{align}
	& \partial_{\sigma} \left( \frac{f_6^4 f_{\varphi}^2e^{-\phi} \dot{\varphi} \sinh^5\!\xi }{\sqrt{\dot\xi^2 + \frac{f_{\varphi}^2}{f_6^2}\dot{\varphi}^2+\frac{f_{\alpha}^2}{f_6^2}\dot{\alpha}^2}} + f_7 \lambda\right)=0 \label{eq:TdBOeom1}\\
	& \partial_{\sigma} \left( \frac{f_6^4 f_{\alpha}^2e^{-\phi} \dot{\alpha} \sinh^5\!\xi }{\sqrt{\dot\xi^2 + \frac{f_{\varphi}^2}{f_6^2}\dot{\varphi}^2+\frac{f_{\alpha}^2}{f_6^2}\dot{\alpha}^2}} \right)=\partial_{\alpha}  \left( f_6^6 e^{-\phi} \sinh^{5}\!\xi\,\sqrt{\dot\xi^2 + \frac{f_{\varphi}^2}{f_6^2}\dot{\varphi}^2+\frac{f_{\alpha}^2}{f_6^2}\dot{\alpha}^2} \right) \dot{\varphi}+ \lambda \partial_{\alpha} f_7 \label{eq:TdBOeom2}\\
	& \partial_{\sigma} \left( \frac{f_6^6 e^{-\phi} \dot{\xi} \sinh^5\!\xi }{\sqrt{\dot\xi^2 + \frac{f_{\varphi}^2}{f_6^2}\dot{\varphi}^2+\frac{f_{\alpha}^2}{f_6^2}\dot{\alpha}^2}} \right)=5e^{-\phi} f_6^6  \cosh\xi \sinh^4\!\xi\, \sqrt{\dot\xi^2 + \frac{f_{\varphi}^2}{f_6^2}\dot{\varphi}^2+\frac{f_{\alpha}^2}{f_6^2}\dot{\alpha}^2}  + f_7 \dot{\varphi} \lambda'\label{eq:TdBOeom3}
\end{align}
To simplify the problem we embed the D5 such that $\dot{\alpha}=0$ and we choose a gauge such that $\sigma=\varphi$. Since $f_7$ is constant, \eqref{eq:TdBOeom2} is solved when $\dot{\xi}=0$, i.e.\ the size of the $S^5$ does not change along the embedding, and $\Delta'(\alpha)=0$. The condition $\Delta'(\alpha)=0$ analogously to the discussion in Type IIA implies that $\alpha=0$ or $\alpha=\pi/2$ in the non-supersymmetric background. \eqref{eq:TdBOeom1} is also solved for $\dot{\xi}=0$. The last condition comes from \eqref{eq:TdBOeom3}, that is 
\begin{equation}
\frac{729L^6\text{sinh}^4(\xi)(5X^3\text{cosh}(\xi)+(2+3X^3)\text{sinh}(\xi)}{64X^2}=0,
\end{equation}
This equation has solution only in the non-supersymmetric case and for $\alpha=0$.
This matches the discussion in Type IIA.

\subsection{5-brane bubbles in Type IIB}\label{sec:iib-bubbles}

We now discuss bubble nucleation in the Type IIB brane web solutions. For its technical simplicity we follow the perspective inspired by Coleman-De Luccia and seek $SO(6)$ symmetric brane bubble solutions in global Euclidean AdS$_6$.

For brane webs built out of 5-branes a natural decay channel is the nucleation of bubbles of new vacua with reduced 5-brane charge. As in the domain wall embeddings and in Sec.~\ref{sec:bubble-Tdual}, the relevant 5-brane embeddings extend along a curve in the internal space, rather than being pointlike. This leads to interesting differences with the D4-branes in Type IIA or the NS5 bubbles in AdS$_7$ solutions discussed in \cite{Apruzzi:2019ecr}.
One could study a variety of embeddings in Type IIB, but we focus on embeddings that are analogous to the Coulomb branch 5-branes studied in Sec.~\ref{sec:DW} in that they extend along the boundary of $\Sigma$.
This allows for a direct comparison to the D4-brane embeddings in Type IIA, where domain wall and bubble nucleation instabilities both originate from the same points in the internal space.

We conducted a numerical survey of 5-brane bubble embeddings along the boundary of $\Sigma$ in a set of sample solutions. This survey is not exhaustive, but we find a suggestive qualitative picture which matches with a brane web perspective that we will discuss. 
We illustrate the main points for the $+_{N,M}$ and $+_{N,M,j,k}$ solutions introduced in Sec.~\ref{sec:plus} and \ref{sec:plus-D7} (similar analyses for the $T_N$ and $X_{N,M}$ solutions led to qualitatively similar results).
Both solutions have NS5-brane poles at $w=\pm 1$. The $+_{N,M}$ solution in addition has D5-brane poles at $w\in\lbrace 0,\infty\rbrace$, while the $+_{N,M,j,k}$ solution has D7-brane punctures in $\Sigma$ whose branch cuts intersect the boundary of $\Sigma$ at $w\in\lbrace 0,\infty\rbrace$.
In line with the domain wall embeddings discussed in Sec.~\ref{sec:dw-ex}, we seek $\pm (1,1)$ 5-branes along the boundary segments $(-\infty,-1)$ and $(0,1)$ on the real line, and $\pm (1,-1)$ 5-branes along the boundary segments $(-1,0)$ and $(1,\infty)$.
The difference to the domain wall discussion is that the 5-branes now wrap $S^5$ in AdS$_6$ with metric (\ref{eq:glo-eads}) rather than Minkowski slices in Poincar\'e AdS$_6$.

\begin{figure}
	\centering
	\subfigure[][]{\label{fig:plus-bubble}
		\includegraphics[width=0.4\linewidth]{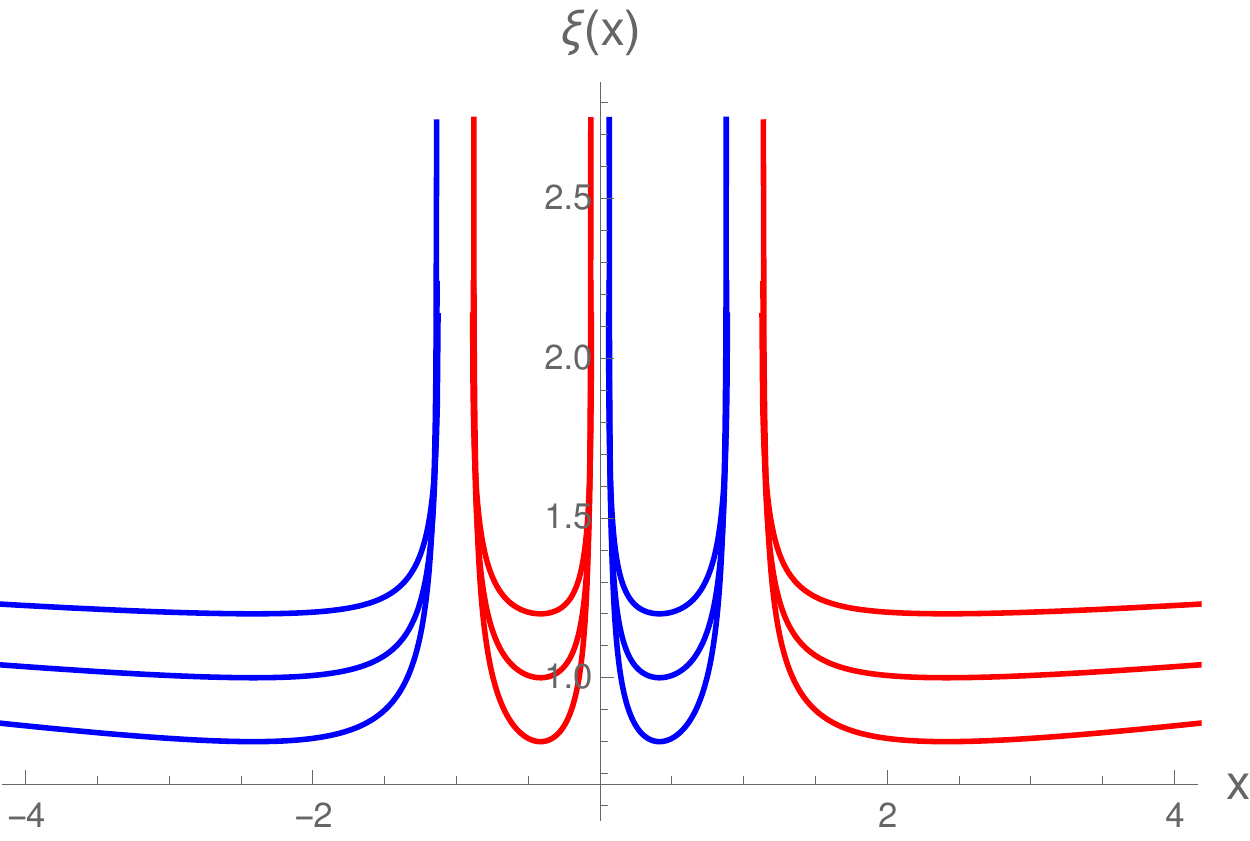}
	}
	\hskip 10mm
	\subfigure[][]{\label{fig:plus-bubble-D7}
		\includegraphics[width=0.4\linewidth]{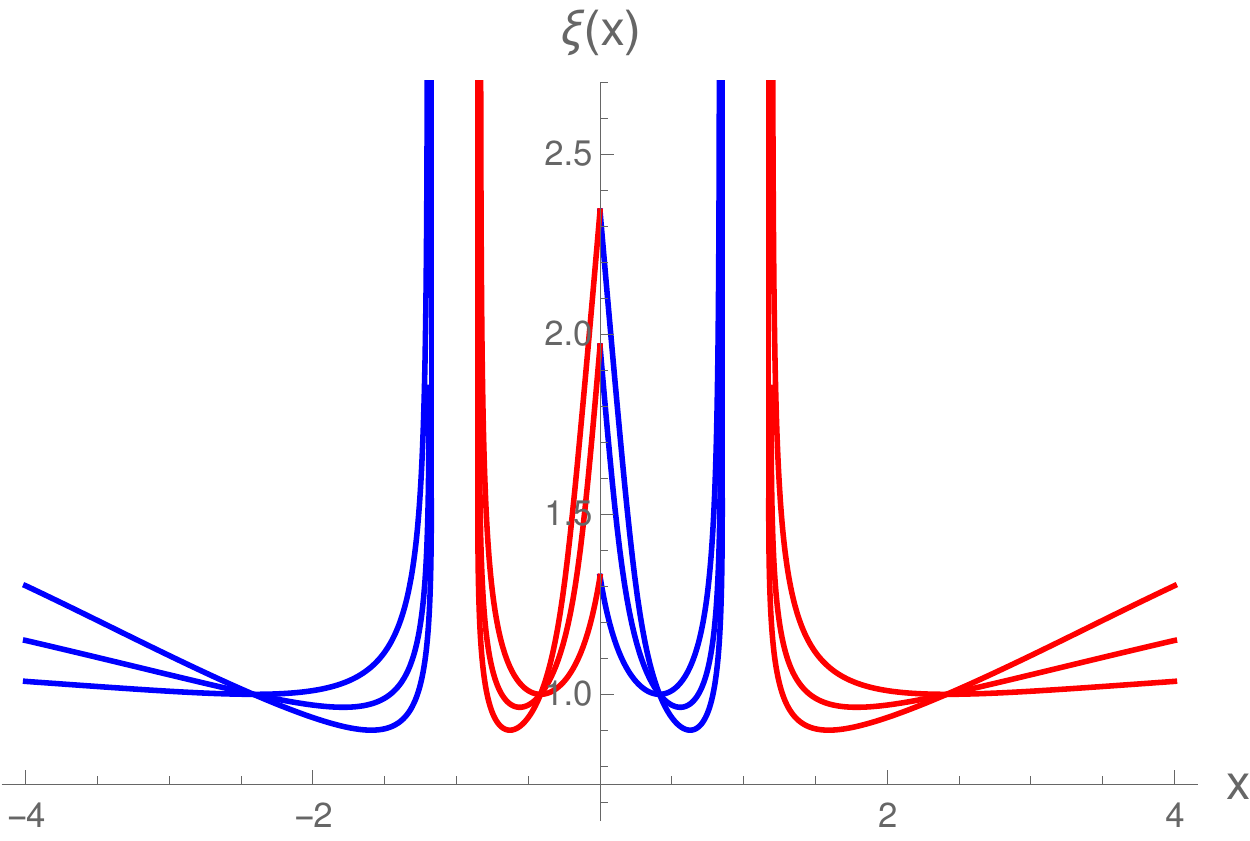}
	}
	\caption{Bubble nucleation 5-brane embeddings for the non-supersymmetric sibling of the $+_{N,M}$ solution with $N=M$ (left) and of the $+_{N,M,j,k}$ solution (right). The embeddings are qualitatively similar to the domain wall embeddings in Fig.~\ref{fig:plus-CB-5-branes-non-susy}, \ref{fig:plus-D7-CB-5-branes-non-susy}.}
\end{figure}

The embeddings are parametrized by a function $\xi(x)$ similar to the domain wall embeddings of Sec.~\ref{sec:CB-IIB}. For the $S^5$ embeddings shifts $\xi\rightarrow \xi+\xi_0$ are not a symmetry and the embeddings depend non-trivially on two parameters.
For the $+_{N,M}$ solutions with $N=M$ a natural starting point are embeddings that respect the reflection symmetries across the diagonals in the brane web (discussed in Sec.~\ref{sec:dw-ex}). This fixes $\xi'(x_f)=0$ at the fixed points of the corresponding symmetries in the supergravity solution, $x_f=\pm 1\pm\sqrt{2}$.
A sample of such embeddings in the non-supersymmetric backgrounds with $X=3^{-1/4}$ is shown in Fig.~\ref{fig:plus-bubble}.\footnote{Similar to the embeddings in Sec.~\ref{sec:CB-IIB}, the embeddings in Fig.~\ref{fig:plus-bubble} reach points of minimal distance to the poles at which they turn back towards the center of the boundary segment, and consist of 3 branches.}
The common qualitative feature is that $\xi(x)$ diverges towards the conformal boundary in both directions before reaching the pole terminating the boundary segment. 
The figure shows this feature for a special set of embeddings, but it applies to embeddings that do not respect the reflection symmetries as well. We do not find bubble nucleation embeddings that stay away from the boundary of AdS$_6$ and have finite action.
A sample of embeddings for the $+_{N,M,j,k}$ solutions is shown in Fig.~\ref{fig:plus-bubble-D7}. 
The embeddings are continuous across the 7-brane branch cuts, but similarly diverge before reaching the poles.

The bubble nucleation embeddings in the non-supersymmetric solutions are qualitatively similar to the domain wall embeddings (Fig.~\ref{fig:plus-CB-5-branes-non-susy}, \ref{fig:plus-D7-CB-5-branes-non-susy}), in the sense that both diverge towards the conformal boundary of AdS$_6$ before reaching the poles while remaining finite across branch cuts. The 5-brane poles, which drive the domain wall embeddings to the conformal boundary, now appear as an obstruction to bubble nucleation.

That the poles pose an obstruction to bubble nucleation can be understood from the brane web perspective. The bubble nucleation channel we seek would make the outer 5-brane quadrilateral in Fig.~\ref{fig:plus-CB} disappear. But since the $(p,q)$ 5-brane charges need to be conserved at the brane vertices, that implies a change in the numbers of semi-infinite 5-branes: 
to remove the outer quadrilateral of $\pm (1,1)$ and $\pm(1,-1)$ 5-branes in Fig.~\ref{fig:plus-CB}, one has to also remove semi-infinite D5 and NS5 branes.
The latter is not a finite energy change from the 5d perspective, which lines up with the absence of finite energy bubble solutions. On the other hand, when the external 5-branes represented by a particular pole are Higgsed to 7-branes, as done for the D5-branes in Fig.~\ref{fig:plus-D7-CB} (see also Fig.~\ref{fig:polarization-web}), one can now remove the 5-brane loop without having to attach semi-infinite D5-branes (though one still needs to include semi-infinite NS5-branes in Fig.~\ref{fig:plus-D7-CB}). 
This matches with Fig.~\ref{fig:plus-bubble-D7}, in which the 5-brane bubble embeddings are regular across the D7-brane branch cuts and only diverge near the NS5 poles.
If the NS5-branes in Fig.~\ref{fig:plus-D7-CB} were Higgsed as well, the outer 5-brane loop would be supported entirely by 7-brane branch cuts, and could be removed without altering semi-infinite 5-branes. 
This suggests that for the corresponding supergravity solutions this bubble nucleation channel may become active.

\subsection{Recap of instabilities}

To recapitulate, compared to the D4-brane case or the NS5-branes discussed in \cite{Apruzzi:2019ecr}, where bubble and domain wall instabilities are tied to each others existence, we find that the two decay channels appear to complement each other in the 5-brane web solutions (for the channels we discussed).

More precisely, as we have seen in Section \ref{sec:IIA-jets}, a D4 extended along the Poincaré slice of AdS$_6$ and sitting at the equator of the half $S^4$, only moves along AdS, reaching its conformal boundary while staying at the same internal location. This behavior is also captured by the existence of an SO(6) symmetric instanton as in Section \ref{sub:D4SO6}.

For 5-branes in Type IIB, we have identified as possible end-points of domain wall instabilities configurations that, while being extended along the Minkowski slice, are also extended along the boundary of $\Sigma$ (fig.~\ref{fig:plus-CB-5-branes-non-susy}  and \ref{fig:plus-D7-CB-5-branes-non-susy}).
We argued that these signal instabilities because, in the regions close to the 5-brane poles of the backgrounds, they are partially expelled towards the conformal boundary of AdS. However, embeddings along the boundary of $\Sigma$ do not give rise to SO(6) instantons, since the configurations we constructed in Section \ref{sec:iib-bubbles} also reach to the conformal boundary and do not have finite action.
This difference with the D4 case is not unexpected considering the differences in the brane constructions, as we discussed at the end of Section \ref{sec:iib-bubbles}.
It would be interesting to investigate the string-like dynamics of the 5-branes in the internal space in more detail.
Finally, although backgrounds without 5-brane poles have not been constructed explicitly, our analysis suggests that for such backgrounds the situation might be reversed, since the 5-brane poles that are acting as an obstruction to the existence of the instanton would not be present and 7-brane punctures do not obstruct this decay channel, as shown in Fig.~\ref{fig:plus-bubble-D7}.

\section{Spin-2 spectrum and scale separation}\label{sec:spin2}

In this section we study the spin 2 perturbations around the AdS$_6$ backgrounds \eqref{eq:metric-gen}, \eqref{eq:functionsExpl} with the aim of understanding whether they are scale separated. 
We will find that the answer is negative. This is a consequence of the existence of a universal class of low lying modes identified in \cite{spin2-susy}, whose mass, as we show here, is only mildly affected by the breaking of supersymmetry.

The starting point of our analysis is the spin 2 mass operator. As shown in \cite{bachas-estes,csaki-erlich-hollowood-shirman} this operator has a universal form only depending on the warping and the internal metric. There are various ways to write it,  and it has been recently noticed \cite{REC,deluca-deponti-mondino-tomasiello} that this operator is natural in the mathematical framework of Bakry-\'Emery geometry, where it arises as the weighted Laplacian:
\begin{equation}\label{eq:spin2Eq}
	\Delta_f \psi \equiv \Delta-\nabla f\cdot\nabla = -\frac{e^{-f}}{\sqrt{\bar g}}\partial_{m}\left(e^f \sqrt{\bar g}\, \bar{g}^{m n}\partial_n \psi  \right) \equiv M^2 \psi\;,
\end{equation}
where $e^f \equiv f_6^8$ is the weight function, and $\bar{ds}^2 \equiv f_6^{-2}(f_2^2 ds^2_{S^2} +4 \rho^2 |d w|^2)$ is the internal metric defined from $ds^2_{10} = f_6^2( ds^2_{\text{AdS}_6}+ \bar{ds}^2)$. 
The appearance of the weight $e^f$ in the definition of the Laplacian in \eqref{eq:spin2Eq} makes the operator self-adjoint with respect to the measure $e^f\sqrt{\bar{g}}$. 
This particular combination naturally appears in many of the internal integrals arising upon expanding the original 10-dimensional supergravity action around a warped vacuum compactification such as \eqref{eq:metric-gen}. In particular, it weights the internal part of the kinetic and mass terms of the spin 2 Kaluza-Klein modes, which are defined as transverse-traceless perturbation of the (unwarped) AdS$_6$ part of the full ten-dimensional metric.

All solutions for $\psi$ in equation \eqref{eq:spin2Eq}, with appropriate boundary conditions on $\Sigma$, give rise to spin-2 fields on AdS$_6$ with mass $M$.
We will now show that the masses in the non-supersymmetric backgrounds can be directly obtained from the masses in the supersymmetric backgrounds by a simple relation.
To see this, we will follow the analysis in \cite{spin2-susy} adapting it to our solutions involving the additional parameter $X$.

Spelling \eqref{eq:spin2Eq} out more explicitly leads to (cfr. \cite[Eq. 3.6]{spin2-susy})
\begin{equation}
	\frac{1}{f_6^4 f_2^2 \rho^2}\partial_a\left( f_6^6 f_2^2\eta^{a b} \partial_b\psi\right)+\frac{f_6^2}{f_2^2}\nabla^2_{S^2}\psi+M^2 \psi = 0
\end{equation}
with $a, b = \{w, \bar{w}\}$ and $\eta^{w \bar{w}} = \eta^{\bar{w} w} = \frac{1}{2}$.
Using the relations in  \eqref{eq:functionsExpl} we get
\begin{equation}
	\frac{f_6^2}{f_2^2} = \frac{2^2 3^2 5}{V_0 X^2} \left( X^4 + 2 \frac{| \partial_w \mathcal{G} |^2}{3 \kappa^2 \mathcal{G}} \right), \qquad f_6^6 f_2^2 =
	\frac{2^8 5^3}{V_0^3 X^2}  \mathcal{G}^2, \qquad \frac{1}{f_6^4 f_2^2 \rho^2} =
	\frac{3 V_0^2}{2^5 5^2  \mathcal{G} \kappa^2}~.
\end{equation}
Expanding $\psi$ in spherical harmonics on $S^2$ as
\begin{equation}
	\psi(y) \equiv \sum_{\ell, m} \phi_{\ell m }(w, \bar{w})Y_{\ell m}(S^2)	
\end{equation} 
we obtain for each level $\ell, m$ the equation
\begin{equation}
	\partial_a (\mathcal{G}^2 \eta^{a b} \partial_b \phi_{\ell, m}) - \ell (\ell + 1) \left(
	\frac{3}{2} \mathcal{G} \kappa^2 X^4 + | \partial_w \mathcal{G} |^2 \right) \phi_{\ell, m} +
	\frac{1}{120} \mathcal{G} \kappa^2 V_0 X^2 M^2 \phi_{\ell, m} = 0~.
\end{equation}
This can be further simplified with the redefinition
\begin{equation}
	\phi_{\ell m} \equiv\mathcal{G}^\ell \chi_{\ell m}
\end{equation}
and using the identity $\partial_w \partial_{\bar{w}}\mathcal{G} = -\kappa^2$ to give
\begin{equation}\label{eq:spin2V01}
	\partial_a (\mathcal{G}^{2 l + 2} \eta^{a b} \partial_b \chi_{l m}) + \frac{1}{6}
	\left( - 6 l \frac{M^2}{20} V_0 X^2 - 9 l (l + 1) X^4 \right) \mathcal{G}^{2 l + 1}
	\kappa^2 \chi_{l m} = 0\;.
\end{equation}
We can now rewrite \eqref{eq:spin2V01} as
\begin{equation}\label{eq:spin2V0}
	\partial_a (\mathcal{G}^{2 \ell + 2} \eta^{a b} \partial_b \chi_{\ell m}) + \frac{1}{6}
	\left( M^2_X-3 \ell (3\ell + 5)\right) \mathcal{G}^{2 \ell + 1} 
	\kappa^2 \chi_{\ell m} = 0
\end{equation}
with
\begin{equation}\label{eq:MX}
	M^2_X \equiv \frac{V_0}{20} M^2 X^2 - 9 \ell (\ell + 1) (X^4 - 1)~.
\end{equation}
The equation for spin-2 fluctuations, \eqref{eq:spin2V0}, is identical for the supersymmetric and non-supersymmetric backgrounds, aside from the shift in mass implied by the definition of $M_X$ in (\ref{eq:MX}).
We can therefore use the results on the spectrum in the super\-symmetric solutions from \cite{spin2-susy} and simply translate the masses to the non-supersymmetric context using \eqref{eq:MX}. The general relation, using that $X=1$ for the supersymmetric solutions and $X=3^{-1/4}$ for non-supersymmetric ones, is
\begin{equation}\label{eq:Mnonsusy}
	M^2_{\text{non-susy}} = \frac53 M^2_{\text{susy}}-10 \ell(\ell+1)~.
\end{equation}

As an example, the `minimal' solutions of \cite{spin2-susy} correspond to constant $\chi_{\ell m}$ and $M_X^2 = 3 \ell (3\ell+5)$.
From \eqref{eq:MX} this implies
\begin{align}
	M^2_{\text{susy, min}}& = 3\ell(3\ell+5)~,& M^2_{\text{non-susy, min}} &= 5\ell(3+\ell)~.
\end{align}
The corresponding eigefunctions only depend on the background through $\cA_\pm$ and $\mathcal{G}$, and are thus universally present in all AdS$_6$ vacua, both supersymmetric and non-supersymmetric. In particular, these modes are part of the spectrum also for backgrounds with O7-planes.
The first Kaluza-Klein state above the massless graviton (which corresponds to $\ell=0$) in this class is given by $\ell = 1$, for which
\begin{align}\label{eq:masses-scale}
	M^2_{\text{susy}} &= 24, & M^2_{\text{non-susy}} &= 20~.
\end{align}
These masses are identical to those of the D7-brane fluctuations along $\delta y$ discussed in Sec.~\ref{sec:probe-D7} -- for the supersymmetric as well as for the non-supersymmetric solutions. A similar coincidence between probe brane and graviton fluctuation masses was observed in the AdS$_7$ analysis in \cite{Apruzzi:2019ecr}.

The same logic applies to modes with general $\ell$, $m$, but \eqref{eq:masses-scale} suffices to show that there is no scale separation:
since we are working with unit-radius AdS$_6$ the masses  are of the same order as the cosmological constant.

\begin{acknowledgments}
We thank Minwoo Suh for collaboration in the initial stage of this work.
FA is supported by the European Union's Horizon 2020 Framework: ERC grants 682608, and the Swiss National Science Foundation. 
GBDL is supported in part by the Simons Foundation Origins of the Universe Initiative (modern inflationary cosmology collaboration) and by a Simons Investigator award.
GLM is supported by the Swedish Research Council grant number 2015-05333 and partially supported by ERC Consolidator Grant number 772408 ``String landscape''.
CFU is supported, in part, by the US Department of Energy under Grant No.~DE-SC0007859 and by the Leinweber Center for Theoretical Physics.
CFU completed part of this work at the Aspen Center for Physics, which is supported by National Science Foundation grant PHY-1607611.
\end{acknowledgments}

\appendix

\bibliographystyle{JHEP.bst}
\bibliography{ads6swamp}

\providecommand{\href}[2]{#2}\begingroup\raggedright\begin{thebibliography}{10}

\bibitem{Arkani-Hamed:2006emk}
N.~Arkani-Hamed, L.~Motl, A.~Nicolis and C.~Vafa, \emph{{The String landscape,
  black holes and gravity as the weakest force}},
  \href{https://doi.org/10.1088/1126-6708/2007/06/060}{\emph{JHEP} {\bfseries
  06} (2007) 060} [\href{https://arxiv.org/abs/hep-th/0601001}{{\ttfamily
  hep-th/0601001}}].

\bibitem{Ooguri:2016pdq}
H.~Ooguri and C.~Vafa, \emph{{Non-supersymmetric AdS and the Swampland}},
  \href{https://doi.org/10.4310/ATMP.2017.v21.n7.a8}{\emph{Adv. Theor. Math.
  Phys.} {\bfseries 21} (2017) 1787}
  [\href{https://arxiv.org/abs/1610.01533}{{\ttfamily 1610.01533}}].

\bibitem{Apruzzi:2019ecr}
F.~Apruzzi, G.~Bruno De~Luca, A.~Gnecchi, G.~Lo~Monaco and A.~Tomasiello,
  \emph{{On AdS$_{7}$ stability}},
  \href{https://doi.org/10.1007/JHEP07(2020)033}{\emph{JHEP} {\bfseries 07}
  (2020) 033} [\href{https://arxiv.org/abs/1912.13491}{{\ttfamily
  1912.13491}}].

\bibitem{Suh:2020rma}
M.~Suh, \emph{{The non-SUSY AdS$_{6}$ and AdS$_{7}$ fixed points are brane-jet
  unstable}}, \href{https://doi.org/10.1007/JHEP10(2020)010}{\emph{JHEP}
  {\bfseries 10} (2020) 010}
  [\href{https://arxiv.org/abs/2004.06823}{{\ttfamily 2004.06823}}].

\bibitem{Fei:2014yja}
L.~Fei, S.~Giombi and I.R.~Klebanov, \emph{{Critical $O(N)$ models in
  $6-\epsilon$ dimensions}},
  \href{https://doi.org/10.1103/PhysRevD.90.025018}{\emph{Phys. Rev. D}
  {\bfseries 90} (2014) 025018}
  [\href{https://arxiv.org/abs/1404.1094}{{\ttfamily 1404.1094}}].

\bibitem{Nakayama:2014yia}
Y.~Nakayama and T.~Ohtsuki, \emph{{Five dimensional $O(N)$-symmetric CFTs from
  conformal bootstrap}},
  \href{https://doi.org/10.1016/j.physletb.2014.05.058}{\emph{Phys. Lett. B}
  {\bfseries 734} (2014) 193}
  [\href{https://arxiv.org/abs/1404.5201}{{\ttfamily 1404.5201}}].

\bibitem{Bae:2014hia}
J.-B.~Bae and S.-J.~Rey, \emph{{Conformal Bootstrap Approach to O(N) Fixed
  Points in Five Dimensions}},
  \href{https://arxiv.org/abs/1412.6549}{{\ttfamily 1412.6549}}.

\bibitem{Chester:2014gqa}
S.M.~Chester, S.S.~Pufu and R.~Yacoby, \emph{{Bootstrapping $O(N)$ vector
  models in 4 $< d <$ 6}},
  \href{https://doi.org/10.1103/PhysRevD.91.086014}{\emph{Phys. Rev. D}
  {\bfseries 91} (2015) 086014}
  [\href{https://arxiv.org/abs/1412.7746}{{\ttfamily 1412.7746}}].

\bibitem{DeCesare:2021pfb}
F.~De~Cesare, L.~Di~Pietro and M.~Serone, \emph{{Five-dimensional CFTs from the
  $\epsilon$-expansion}},  \href{https://arxiv.org/abs/2107.00342}{{\ttfamily
  2107.00342}}.

\bibitem{Cordova:2018eba}
C.~C\'ordova, G.B.~De~Luca and A.~Tomasiello, \emph{{AdS$_{8}$ solutions in
  type II supergravity}},
  \href{https://doi.org/10.1007/JHEP07(2019)127}{\emph{JHEP} {\bfseries 07}
  (2019) 127} [\href{https://arxiv.org/abs/1811.06987}{{\ttfamily
  1811.06987}}].

\bibitem{BenettiGenolini:2019zth}
P.~Benetti~Genolini, M.~Honda, H.-C.~Kim, D.~Tong and C.~Vafa, \emph{{Evidence
  for a Non-Supersymmetric 5d CFT from Deformations of 5d $SU(2)$ SYM}},
  \href{https://doi.org/10.1007/JHEP05(2020)058}{\emph{JHEP} {\bfseries 05}
  (2020) 058} [\href{https://arxiv.org/abs/2001.00023}{{\ttfamily
  2001.00023}}].

\bibitem{Bertolini:2021cew}
M.~Bertolini and F.~Mignosa, \emph{{Supersymmetry breaking deformations and
  phase transitions in five dimensions}},
  \href{https://arxiv.org/abs/2109.02662}{{\ttfamily 2109.02662}}.

\bibitem{Maldacena:1998uz}
J.M.~Maldacena, J.~Michelson and A.~Strominger, \emph{{Anti-de Sitter
  fragmentation}},
  \href{https://doi.org/10.1088/1126-6708/1999/02/011}{\emph{JHEP} {\bfseries
  02} (1999) 011} [\href{https://arxiv.org/abs/hep-th/9812073}{{\ttfamily
  hep-th/9812073}}].

\bibitem{Gaiotto:2009mv}
D.~Gaiotto and A.~Tomasiello, \emph{{The gauge dual of Romans mass}},
  \href{https://doi.org/10.1007/JHEP01(2010)015}{\emph{JHEP} {\bfseries 01}
  (2010) 015} [\href{https://arxiv.org/abs/0901.0969}{{\ttfamily 0901.0969}}].

\bibitem{Narayan:2010em}
P.~Narayan and S.P.~Trivedi, \emph{{On The Stability Of Non-Supersymmetric AdS
  Vacua}}, \href{https://doi.org/10.1007/JHEP07(2010)089}{\emph{JHEP}
  {\bfseries 07} (2010) 089} [\href{https://arxiv.org/abs/1002.4498}{{\ttfamily
  1002.4498}}].

\bibitem{Antonelli:2019nar}
R.~Antonelli and I.~Basile, \emph{{Brane annihilation in non-supersymmetric
  strings}}, \href{https://doi.org/10.1007/JHEP11(2019)021}{\emph{JHEP}
  {\bfseries 11} (2019) 021}
  [\href{https://arxiv.org/abs/1908.04352}{{\ttfamily 1908.04352}}].

\bibitem{Bena:2020xxb}
I.~Bena, K.~Pilch and N.P.~Warner, \emph{{Brane-Jet Instabilities}},
  \href{https://doi.org/10.1007/JHEP10(2020)091}{\emph{JHEP} {\bfseries 10}
  (2020) 091} [\href{https://arxiv.org/abs/2003.02851}{{\ttfamily
  2003.02851}}].

\bibitem{Bena:2020qpa}
I.~Bena, G.B.~De~Luca, M.~Gra\~na and G.~Lo~Monaco, \emph{{Oh, wait, O8 de
  Sitter may be unstable!}},
  \href{https://doi.org/10.1007/JHEP03(2021)168}{\emph{JHEP} {\bfseries 03}
  (2021) 168} [\href{https://arxiv.org/abs/2010.05936}{{\ttfamily
  2010.05936}}].

\bibitem{Basile:2021vxh}
I.~Basile, \emph{{Supersymmetry Breaking and Stability in String Vacua: brane
  dynamics, bubbles and the swampland}},
  \href{https://doi.org/10.1007/s40766-021-00024-9}{\emph{Riv. Nuovo Cim.}
  {\bfseries 1} (2021) 98} [\href{https://arxiv.org/abs/2107.02814}{{\ttfamily
  2107.02814}}].

\bibitem{Guarino:2020jwv}
A.~Guarino, J.~Tarrio and O.~Varela, \emph{{Brane-jet stability of
  non-supersymmetric AdS vacua}},
  \href{https://doi.org/10.1007/JHEP09(2020)110}{\emph{JHEP} {\bfseries 09}
  (2020) 110} [\href{https://arxiv.org/abs/2005.07072}{{\ttfamily
  2005.07072}}].

\bibitem{Guarino:2020flh}
A.~Guarino, E.~Malek and H.~Samtleben, \emph{{Stable Nonsupersymmetric
  Anti\textendash{}de Sitter Vacua of Massive IIA Supergravity}},
  \href{https://doi.org/10.1103/PhysRevLett.126.061601}{\emph{Phys. Rev. Lett.}
  {\bfseries 126} (2021) 061601}
  [\href{https://arxiv.org/abs/2011.06600}{{\ttfamily 2011.06600}}].

\bibitem{Aharony:1997ju}
O.~Aharony and A.~Hanany, \emph{{Branes, superpotentials and superconformal
  fixed points}},
  \href{https://doi.org/10.1016/S0550-3213(97)00472-0}{\emph{Nucl. Phys. B}
  {\bfseries 504} (1997) 239}
  [\href{https://arxiv.org/abs/hep-th/9704170}{{\ttfamily hep-th/9704170}}].

\bibitem{Aharony:1997bh}
O.~Aharony, A.~Hanany and B.~Kol, \emph{{Webs of (p,q) five-branes,
  five-dimensional field theories and grid diagrams}},
  \href{https://doi.org/10.1088/1126-6708/1998/01/002}{\emph{JHEP} {\bfseries
  01} (1998) 002} [\href{https://arxiv.org/abs/hep-th/9710116}{{\ttfamily
  hep-th/9710116}}].

\bibitem{DeWolfe:1999hj}
O.~DeWolfe, A.~Hanany, A.~Iqbal and E.~Katz, \emph{{Five-branes, seven-branes
  and five-dimensional E(n) field theories}},
  \href{https://doi.org/10.1088/1126-6708/1999/03/006}{\emph{JHEP} {\bfseries
  03} (1999) 006} [\href{https://arxiv.org/abs/hep-th/9902179}{{\ttfamily
  hep-th/9902179}}].

\bibitem{Benini:2009gi}
F.~Benini, S.~Benvenuti and Y.~Tachikawa, \emph{{Webs of five-branes and N=2
  superconformal field theories}},
  \href{https://doi.org/10.1088/1126-6708/2009/09/052}{\emph{JHEP} {\bfseries
  09} (2009) 052} [\href{https://arxiv.org/abs/0906.0359}{{\ttfamily
  0906.0359}}].

\bibitem{Bergman:2015dpa}
O.~Bergman and G.~Zafrir, \emph{{5d fixed points from brane webs and
  O7-planes}}, \href{https://doi.org/10.1007/JHEP12(2015)163}{\emph{JHEP}
  {\bfseries 12} (2015) 163}
  [\href{https://arxiv.org/abs/1507.03860}{{\ttfamily 1507.03860}}].

\bibitem{Zafrir:2015ftn}
G.~Zafrir, \emph{{Brane webs and $O5$-planes}},
  \href{https://doi.org/10.1007/JHEP03(2016)109}{\emph{JHEP} {\bfseries 03}
  (2016) 109} [\href{https://arxiv.org/abs/1512.08114}{{\ttfamily
  1512.08114}}].

\bibitem{DHoker:2016ujz}
E.~D'Hoker, M.~Gutperle, A.~Karch and C.F.~Uhlemann, \emph{{Warped $AdS_6\times
  S^2$ in Type IIB supergravity I: Local solutions}},
  \href{https://doi.org/10.1007/JHEP08(2016)046}{\emph{JHEP} {\bfseries 08}
  (2016) 046} [\href{https://arxiv.org/abs/1606.01254}{{\ttfamily
  1606.01254}}].

\bibitem{DHoker:2016ysh}
E.~D'Hoker, M.~Gutperle and C.F.~Uhlemann, \emph{{Holographic duals for
  five-dimensional superconformal quantum field theories}},
  \href{https://doi.org/10.1103/PhysRevLett.118.101601}{\emph{Phys. Rev. Lett.}
  {\bfseries 118} (2017) 101601}
  [\href{https://arxiv.org/abs/1611.09411}{{\ttfamily 1611.09411}}].

\bibitem{DHoker:2017mds}
E.~D'Hoker, M.~Gutperle and C.F.~Uhlemann, \emph{{Warped $AdS_6\times S^2$ in
  Type IIB supergravity II: Global solutions and five-brane webs}},
  \href{https://doi.org/10.1007/JHEP05(2017)131}{\emph{JHEP} {\bfseries 05}
  (2017) 131} [\href{https://arxiv.org/abs/1703.08186}{{\ttfamily
  1703.08186}}].

\bibitem{DHoker:2017zwj}
E.~D'Hoker, M.~Gutperle and C.F.~Uhlemann, \emph{{Warped $AdS_6\times S^2$ in
  Type IIB supergravity III: Global solutions with seven-branes}},
  \href{https://doi.org/10.1007/JHEP11(2017)200}{\emph{JHEP} {\bfseries 11}
  (2017) 200} [\href{https://arxiv.org/abs/1706.00433}{{\ttfamily
  1706.00433}}].

\bibitem{Apruzzi:2014qva}
F.~Apruzzi, M.~Fazzi, A.~Passias, D.~Rosa and A.~Tomasiello, \emph{{AdS$_{6}$
  solutions of type II supergravity}},
  \href{https://doi.org/10.1007/JHEP11(2014)099}{\emph{JHEP} {\bfseries 11}
  (2014) 099} [\href{https://arxiv.org/abs/1406.0852}{{\ttfamily 1406.0852}}].

\bibitem{Kim:2015hya}
H.~Kim, N.~Kim and M.~Suh, \emph{{Supersymmetric AdS$_6$ Solutions of Type IIB
  Supergravity}},
  \href{https://doi.org/10.1140/epjc/s10052-015-3705-1}{\emph{Eur. Phys. J. C}
  {\bfseries 75} (2015) 484}
  [\href{https://arxiv.org/abs/1506.05480}{{\ttfamily 1506.05480}}].

\bibitem{Hong:2018amk}
J.~Hong, J.T.~Liu and D.R.~Mayerson, \emph{{Gauged Six-Dimensional Supergravity
  from Warped IIB Reductions}},
  \href{https://doi.org/10.1007/JHEP09(2018)140}{\emph{JHEP} {\bfseries 09}
  (2018) 140} [\href{https://arxiv.org/abs/1808.04301}{{\ttfamily
  1808.04301}}].

\bibitem{Malek:2018zcz}
E.~Malek, H.~Samtleben and V.~Vall~Camell, \emph{{Supersymmetric AdS$_{7}$ and
  AdS$_6$ vacua and their minimal consistent truncations from exceptional field
  theory}}, \href{https://doi.org/10.1016/j.physletb.2018.09.037}{\emph{Phys.
  Lett. B} {\bfseries 786} (2018) 171}
  [\href{https://arxiv.org/abs/1808.05597}{{\ttfamily 1808.05597}}].

\bibitem{Malek:2019ucd}
E.~Malek, H.~Samtleben and V.~Vall~Camell, \emph{{Supersymmetric AdS$_7$ and
  AdS$_6$ vacua and their consistent truncations with vector multiplets}},
  \href{https://doi.org/10.1007/JHEP04(2019)088}{\emph{JHEP} {\bfseries 04}
  (2019) 088} [\href{https://arxiv.org/abs/1901.11039}{{\ttfamily
  1901.11039}}].

\bibitem{Gursoy:2002tx}
U.~Gursoy, C.~Nunez and M.~Schvellinger, \emph{{RG flows from spin(7), CY 4
  fold and HK manifolds to AdS, Penrose limits and pp waves}},
  \href{https://doi.org/10.1088/1126-6708/2002/06/015}{\emph{JHEP} {\bfseries
  06} (2002) 015} [\href{https://arxiv.org/abs/hep-th/0203124}{{\ttfamily
  hep-th/0203124}}].

\bibitem{Karndumri:2012vh}
P.~Karndumri, \emph{{Holographic RG flows in six dimensional F(4) gauged
  supergravity}}, \href{https://doi.org/10.1007/JHEP01(2013)134}{\emph{JHEP}
  {\bfseries 01} (2013) 134} [\href{https://arxiv.org/abs/1210.8064}{{\ttfamily
  1210.8064}}].

\bibitem{Vafa:2005ui}
C.~Vafa, \emph{{The String landscape and the swampland}},
  \href{https://arxiv.org/abs/hep-th/0509212}{{\ttfamily hep-th/0509212}}.

\bibitem{Chaney:2018gjc}
A.~Chaney and C.F.~Uhlemann, \emph{{On minimal Type IIB AdS$_{6}$ solutions
  with commuting 7-branes}},
  \href{https://doi.org/10.1007/JHEP12(2018)110}{\emph{JHEP} {\bfseries 12}
  (2018) 110} [\href{https://arxiv.org/abs/1810.10592}{{\ttfamily
  1810.10592}}].

\bibitem{Uhlemann:2019lge}
C.F.~Uhlemann, \emph{{AdS$_6$/CFT$_5$ with O7-planes}},
  \href{https://doi.org/10.1007/JHEP04(2020)113}{\emph{JHEP} {\bfseries 04}
  (2020) 113} [\href{https://arxiv.org/abs/1912.09716}{{\ttfamily
  1912.09716}}].

\bibitem{Bergman:2018hin}
O.~Bergman, D.~Rodr\'\i{}guez-G\'omez and C.F.~Uhlemann, \emph{{Testing
  AdS$_{6}$/CFT$_{5}$ in Type IIB with stringy operators}},
  \href{https://doi.org/10.1007/JHEP08(2018)127}{\emph{JHEP} {\bfseries 08}
  (2018) 127} [\href{https://arxiv.org/abs/1806.07898}{{\ttfamily
  1806.07898}}].

\bibitem{Fluder:2018chf}
M.~Fluder and C.F.~Uhlemann, \emph{{Precision Test of AdS$_6$/CFT$_5$ in Type
  IIB String Theory}},
  \href{https://doi.org/10.1103/PhysRevLett.121.171603}{\emph{Phys. Rev. Lett.}
  {\bfseries 121} (2018) 171603}
  [\href{https://arxiv.org/abs/1806.08374}{{\ttfamily 1806.08374}}].

\bibitem{Uhlemann:2019ypp}
C.F.~Uhlemann, \emph{{Exact results for 5d SCFTs of long quiver type}},
  \href{https://doi.org/10.1007/JHEP11(2019)072}{\emph{JHEP} {\bfseries 11}
  (2019) 072} [\href{https://arxiv.org/abs/1909.01369}{{\ttfamily
  1909.01369}}].

\bibitem{Uhlemann:2020bek}
C.F.~Uhlemann, \emph{{Wilson loops in 5d long quiver gauge theories}},
  \href{https://doi.org/10.1007/JHEP09(2020)145}{\emph{JHEP} {\bfseries 09}
  (2020) 145} [\href{https://arxiv.org/abs/2006.01142}{{\ttfamily
  2006.01142}}].

\bibitem{Gutperle:2020rty}
M.~Gutperle and C.F.~Uhlemann, \emph{{Surface defects in holographic 5d
  SCFTs}}, \href{https://doi.org/10.1007/JHEP04(2021)134}{\emph{JHEP}
  {\bfseries 04} (2021) 134}
  [\href{https://arxiv.org/abs/2012.14547}{{\ttfamily 2012.14547}}].

\bibitem{Legramandi:2021uds}
A.~Legramandi and C.~Nunez, \emph{{Electrostatic Description of
  Five-dimensional SCFTs}},  \href{https://arxiv.org/abs/2104.11240}{{\ttfamily
  2104.11240}}.

\bibitem{Gutperle:2021nkl}
M.~Gutperle and N.~Klein, \emph{{A Penrose limit for type IIB AdS$_{6}$
  solutions}}, \href{https://doi.org/10.1007/JHEP07(2021)073}{\emph{JHEP}
  {\bfseries 07} (2021) 073}
  [\href{https://arxiv.org/abs/2105.10824}{{\ttfamily 2105.10824}}].

\bibitem{Roychowdhury:2021oiq}
D.~Roychowdhury, \emph{{Non-integrability for $ \mathcal{N}=1 $ SCFTs in $ 5d
  $}},  \href{https://arxiv.org/abs/2106.10646}{{\ttfamily 2106.10646}}.

\bibitem{Alencar:2021ljc}
G.~Alencar and M.O.~Tahim, \emph{{Non-Integrability of Strings in
  $AdS_{6}\times S^{2}\times\Sigma$ Background and its 5D Holographic Duals}},
  \href{https://arxiv.org/abs/2106.11288}{{\ttfamily 2106.11288}}.

\bibitem{Legramandi:2021aqv}
A.~Legramandi and C.~Nunez, \emph{{Holographic description of SCFT$_5$
  compactifications}},  \href{https://arxiv.org/abs/2109.11554}{{\ttfamily
  2109.11554}}.

\bibitem{Brandhuber:1999np}
A.~Brandhuber and Y.~Oz, \emph{{The D-4 - D-8 brane system and five-dimensional
  fixed points}},
  \href{https://doi.org/10.1016/S0370-2693(99)00763-7}{\emph{Phys. Lett. B}
  {\bfseries 460} (1999) 307}
  [\href{https://arxiv.org/abs/hep-th/9905148}{{\ttfamily hep-th/9905148}}].

\bibitem{Coleman:1980aw}
S.R.~Coleman and F.~De~Luccia, \emph{{Gravitational Effects on and of Vacuum
  Decay}}, \href{https://doi.org/10.1103/PhysRevD.21.3305}{\emph{Phys. Rev. D}
  {\bfseries 21} (1980) 3305}.

\bibitem{lust-palti-vafa}
D.~{L\"ust}, E.~Palti and C.~Vafa, \emph{{AdS and the Swampland}},
  \href{https://doi.org/10.1016/j.physletb.2019.134867}{\emph{Phys. Lett.}
  {\bfseries B797} (2019) 134867}
  [\href{https://arxiv.org/abs/1906.05225}{{\ttfamily 1906.05225}}].

\bibitem{malek-samtleben-kk}
E.~Malek and H.~Samtleben, \emph{{Kaluza--Klein Spectrometry for
  Supergravity}},
  \href{https://doi.org/10.1103/PhysRevLett.124.101601}{\emph{Phys. Rev. Lett.}
  {\bfseries 124} (2020) 101601}
  [\href{https://arxiv.org/abs/1911.12640}{{\ttfamily 1911.12640}}].

\bibitem{malek-nicolai-samtleben}
E.~Malek, H.~Nicolai and H.~Samtleben, \emph{{Tachyonic Kaluza--Klein modes and
  the AdS swampland conjecture}},
  \href{https://doi.org/10.1007/JHEP08(2020)159}{\emph{JHEP} {\bfseries 08}
  (2020) 159} [\href{https://arxiv.org/abs/2005.07713}{{\ttfamily
  2005.07713}}].

\bibitem{csaki-erlich-hollowood-shirman}
C.~Csaki, J.~Erlich, T.J.~Hollowood and Y.~Shirman, \emph{{Universal aspects of
  gravity localized on thick branes}},
  \href{https://doi.org/10.1016/S0550-3213(00)00271-6}{\emph{Nucl. Phys.}
  {\bfseries B581} (2000) 309}
  [\href{https://arxiv.org/abs/hep-th/0001033}{{\ttfamily hep-th/0001033}}].

\bibitem{bachas-estes}
C.~Bachas and J.~Estes, \emph{{Spin-2 spectrum of defect theories}},
  \href{https://doi.org/10.1007/JHEP06(2011)005}{\emph{JHEP} {\bfseries 06}
  (2011) 005} [\href{https://arxiv.org/abs/1103.2800}{{\ttfamily 1103.2800}}].

\bibitem{REC}
G.B.~De~Luca and A.~Tomasiello, \emph{{Leaps and bounds towards scale
  separation}},  \href{https://arxiv.org/abs/2104.12773}{{\ttfamily
  2104.12773}}.

\bibitem{deluca-deponti-mondino-tomasiello}
G.B.~De~Luca, N.~De~Ponti, A.~Mondino and A.~Tomasiello, \emph{{Cheeger bounds
  on spin-two fields}},  \href{https://arxiv.org/abs/2109.11560}{{\ttfamily
  2109.11560}}.

\bibitem{spin2-susy}
M.~Gutperle, C.F.~Uhlemann and O.~Varela, \emph{{Massive spin 2 excitations in
  $AdS_6\times S^2$ warped spacetimes}},
  \href{https://doi.org/10.1007/JHEP07(2018)091}{\emph{JHEP} {\bfseries 07}
  (2018) 091} [\href{https://arxiv.org/abs/1805.11914}{{\ttfamily
  1805.11914}}].

\bibitem{polchinski-silverstein}
J.~Polchinski and E.~Silverstein, \emph{{Dual Purpose Landscaping Tools: Small
  Extra Dimensions in AdS/CFT}},  in \emph{Strings, gauge fields, and the
  geometry behind: The legacy of Maximilian Kreuzer}, A.~Rebhan, L.~Katzarkov,
  J.~Knapp, R.~Rashkov and E.~Scheidegger, eds., pp.~365--390 (2009),
  \href{https://doi.org/10.1142/9789814412551_0018}{DOI}
  [\href{https://arxiv.org/abs/0908.0756}{{\ttfamily 0908.0756}}].

\bibitem{gautason-schillo-vanriet-williams}
F.F.~Gautason, M.~Schillo, T.~Van~Riet and M.~Williams, \emph{{Remarks on scale
  separation in flux vacua}},
  \href{https://doi.org/10.1007/JHEP03(2016)061}{\emph{JHEP} {\bfseries 03}
  (2016) 061} [\href{https://arxiv.org/abs/1512.00457}{{\ttfamily
  1512.00457}}].

\bibitem{dgkt}
O.~DeWolfe, A.~Giryavets, S.~Kachru and W.~Taylor, \emph{Type {IIA} moduli
  stabilization}, {\emph{JHEP} {\bfseries 07} (2005) 066}
  [\href{https://arxiv.org/abs/hep-th/0505160}{{\ttfamily hep-th/0505160}}].

\bibitem{Petrini:2013ika}
M.~Petrini, G.~Solard and T.~Van~Riet, \emph{{AdS vacua with scale separation
  from IIB supergravity}},
  \href{https://doi.org/10.1007/JHEP11(2013)010}{\emph{JHEP} {\bfseries 11}
  (2013) 010} [\href{https://arxiv.org/abs/1308.1265}{{\ttfamily 1308.1265}}].

\bibitem{Cribiori:2021djm}
N.~Cribiori, D.~Junghans, V.~Van~Hemelryck, T.~Van~Riet and T.~Wrase,
  \emph{{Scale-separated AdS$_4$ vacua of IIA orientifolds and M-theory}},
  \href{https://arxiv.org/abs/2107.00019}{{\ttfamily 2107.00019}}.

\bibitem{Demirtas:2021nlu}
M.~Demirtas, M.~Kim, L.~McAllister, J.~Moritz and A.~Rios-Tascon, \emph{{Small
  Cosmological Constants in String Theory}},
  \href{https://arxiv.org/abs/2107.09064}{{\ttfamily 2107.09064}}.

\bibitem{Romans:1985tw}
L.J.~Romans, \emph{{The F(4) Gauged Supergravity in Six-dimensions}},
  \href{https://doi.org/10.1016/0550-3213(86)90517-1}{\emph{Nucl. Phys. B}
  {\bfseries 269} (1986) 691}.

\bibitem{Gutperle:2018vdd}
M.~Gutperle, A.~Trivella and C.F.~Uhlemann, \emph{{Type IIB 7-branes in warped
  AdS$_{6}$: partition functions, brane webs and probe limit}},
  \href{https://doi.org/10.1007/JHEP04(2018)135}{\emph{JHEP} {\bfseries 04}
  (2018) 135} [\href{https://arxiv.org/abs/1802.07274}{{\ttfamily
  1802.07274}}].

\bibitem{Corbino:2017tfl}
D.~Corbino, E.~D'Hoker and C.F.~Uhlemann,
  \emph{{$\text{AdS}_{2}\times\text{S}^{6}$ versus
  $\text{AdS}_{6}\times\text{S}^{2}$ in Type IIB supergravity}},
  \href{https://doi.org/10.1007/JHEP03(2018)120}{\emph{JHEP} {\bfseries 03}
  (2018) 120} [\href{https://arxiv.org/abs/1712.04463}{{\ttfamily
  1712.04463}}].

\bibitem{Cordova:2016emh}
C.~Cordova, T.T.~Dumitrescu and K.~Intriligator, \emph{{Multiplets of
  Superconformal Symmetry in Diverse Dimensions}},
  \href{https://doi.org/10.1007/JHEP03(2019)163}{\emph{JHEP} {\bfseries 03}
  (2019) 163} [\href{https://arxiv.org/abs/1612.00809}{{\ttfamily
  1612.00809}}].

\bibitem{Apruzzi:2016rny}
F.~Apruzzi, G.~Dibitetto and L.~Tizzano, \emph{{A new 6d fixed point from
  holography}}, \href{https://doi.org/10.1007/JHEP11(2016)126}{\emph{JHEP}
  {\bfseries 11} (2016) 126}
  [\href{https://arxiv.org/abs/1603.06576}{{\ttfamily 1603.06576}}].

\bibitem{Bergshoeff:2006gs}
E.A.~Bergshoeff, M.~de~Roo, S.F.~Kerstan, T.~Ortin and F.~Riccioni,
  \emph{{SL(2,R)-invariant IIB Brane Actions}},
  \href{https://doi.org/10.1088/1126-6708/2007/02/007}{\emph{JHEP} {\bfseries
  02} (2007) 007} [\href{https://arxiv.org/abs/hep-th/0611036}{{\ttfamily
  hep-th/0611036}}].

\bibitem{Myers:1999ps}
R.C.~Myers, \emph{{Dielectric branes}},
  \href{https://doi.org/10.1088/1126-6708/1999/12/022}{\emph{JHEP} {\bfseries
  12} (1999) 022} [\href{https://arxiv.org/abs/hep-th/9910053}{{\ttfamily
  hep-th/9910053}}].

\bibitem{McGuirk:2012sb}
P.~McGuirk, G.~Shiu and F.~Ye, \emph{{Soft branes in supersymmetry-breaking
  backgrounds}}, \href{https://doi.org/10.1007/JHEP07(2012)188}{\emph{JHEP}
  {\bfseries 07} (2012) 188} [\href{https://arxiv.org/abs/1206.0754}{{\ttfamily
  1206.0754}}].

\bibitem{DeLuca:2018zbi}
G.B.~De~Luca, A.~Gnecchi, G.~Lo~Monaco and A.~Tomasiello, \emph{{Holographic
  duals of 6d RG flows}},
  \href{https://doi.org/10.1007/JHEP03(2019)035}{\emph{JHEP} {\bfseries 03}
  (2019) 035} [\href{https://arxiv.org/abs/1810.10013}{{\ttfamily
  1810.10013}}].

\bibitem{Passias:2012vp}
A.~Passias, \emph{{A note on supersymmetric AdS$_6$ solutions of massive type
  IIA supergravity}},
  \href{https://doi.org/10.1007/JHEP01(2013)113}{\emph{JHEP} {\bfseries 01}
  (2013) 113} [\href{https://arxiv.org/abs/1209.3267}{{\ttfamily 1209.3267}}].

\bibitem{Bergman:2012kr}
O.~Bergman and D.~Rodriguez-Gomez, \emph{{5d quivers and their AdS(6) duals}},
  \href{https://doi.org/10.1007/JHEP07(2012)171}{\emph{JHEP} {\bfseries 07}
  (2012) 171} [\href{https://arxiv.org/abs/1206.3503}{{\ttfamily 1206.3503}}].

\bibitem{Lozano:2018pcp}
Y.~Lozano, N.T.~Macpherson and J.~Montero, \emph{{AdS$_{6}$ T-duals and type
  IIB AdS$_{6} \times$ S$^{2}$ geometries with 7-branes}},
  \href{https://doi.org/10.1007/JHEP01(2019)116}{\emph{JHEP} {\bfseries 01}
  (2019) 116} [\href{https://arxiv.org/abs/1810.08093}{{\ttfamily
  1810.08093}}].

\bibitem{Lozano:2018fvt}
Y.~Lozano, N.T.~Macpherson and J.~Montero, \emph{{AdS$_6$ T-duals and AdS$_6
  \times S^2$ geometries in Type IIB}},
  \href{https://doi.org/10.22323/1.347.0122}{\emph{PoS} {\bfseries CORFU2018}
  (2019) 122} [\href{https://arxiv.org/abs/1811.00054}{{\ttfamily
  1811.00054}}].

\bibitem{Cvetic:2000cj}
M.~Cvetic, H.~Lu, C.N.~Pope and J.F.~Vazquez-Poritz, \emph{{AdS in warped
  space-times}}, \href{https://doi.org/10.1103/PhysRevD.62.122003}{\emph{Phys.
  Rev. D} {\bfseries 62} (2000) 122003}
  [\href{https://arxiv.org/abs/hep-th/0005246}{{\ttfamily hep-th/0005246}}].

\end{thebibliography}\endgroup
\end{document}